\documentclass[a4paper,12pt]{article}
\usepackage{amsfonts}
\usepackage{amsmath}
\usepackage{physics}
\usepackage{amssymb}
\usepackage{jheppub}
\usepackage{lmodern}
\usepackage[section]{placeins}
\usepackage{xcolor}
\usepackage{bm}
\usepackage[utf8]{inputenc}
\usepackage[T1]{fontenc}
\definecolor{refkey}{gray}{0.45}
\definecolor{labelkey}{RGB}{155,48,48}
\hypersetup{
  colorlinks=true,
  citecolor=magenta,
  linkcolor=blue,
  urlcolor=violet
 }

\newcommand{\del}{\partial}
\newcommand{\mi}{\mathrm{i}}
\def\be{\begin{equation}}\def\ee{\end{equation}}

\title{Near-Extremal Fluid Mechanics}

\author{Upamanyu Moitra,}
\author{Sunil Kumar Sake,}
\author{Sandip P. Trivedi}
\affiliation{Department of Theoretical Physics, Tata Institute of Fundamental Research, \\1 Homi Bhabha Road, Colaba, Mumbai -- 400005, India}

\emailAdd{upamanyu@theory.tifr.res.in}
\emailAdd{sunil.sake@tifr.res.in}
\emailAdd{sandip@theory.tifr.res.in}

\preprint{{TIFR/TH/20-12}}

\abstract{

We analyse near-extremal black brane configurations in asymptotically $\mathrm{AdS}_4$ spacetime with the temperature $T$, chemical potential $\mu$, and three-velocity  $u^\nu$, varying slowly.
We consider a low-temperature limit where the rate of variation is much slower than $\mu$, but much bigger than $T$. This limit  is different from the one considered for conventional fluid-mechanics in which the rate of variation is much smaller than both $T$, $\mu$. We find that in our limit, as well, the Einstein-Maxwell equations can  be solved in a systematic perturbative expansion. At first order, in the rate of variation,  the resulting constitutive relations for the stress tensor and charge current are local in the boundary theory and can be easily calculated. At higher orders, we show that these relations become non-local in time but the perturbative expansion is still valid. We find  that   there are four linearised modes in this limit; these   are  similar  to the hydrodynamic modes found  in conventional fluid mechanics with the same dispersion relations.  We also study some linearised time independent perturbations exhibiting  attractor behaviour  at the horizon --- these  arise  in the presence of external driving forces in the boundary theory.
}

\begin{document}
\maketitle
\flushbottom

\section{Introduction}\label{sec-in}

Physical systems often behave like  fluids. The underlying circumstances when this happens  are as  follows. 
 Once local equilibrium has set in,  a sufficiently big, but not too big,  part of the system which has locally equilibrated can be assigned a local temperature $T(x^\mu)$ 
 and a local velocity  $u^\nu(x^\mu)$ with which its centre of mass  moves.  

The subsequent dynamics is then determined by the  evolution of these local quantities and is governed by the celebrated Navier--Stokes (NS) equations of fluid mechanics. Our interest here will be  in  relativistic field theories in $2+1$ dimensions and  $u^\nu$   hereafter will denote  the three-velocity with    the index $\nu$ taking values $0, 1, 2$.  Also, we will  be interested in  systems with a conserved particle number or charge;  for such systems the local state is also specified by a chemical potential $\mu(x^\nu)$. 

The NS equations provide  a good description of the system when $T$, $\mu$,  $u^\nu$, vary slowly compared to a characteristic scale  $\ell$  over which the system  has locally equilibrated; often this scale  is determined by the mean free path of the constituents. Corrections to  the  equations are  suppressed in powers of $\ell/d$ where $d$ is the characteristic scale of variation of $T$, $\mu$, $u^\nu$, and the NS equations are a good approximation provided $\ell/d \ll 1$ . The NS equations take a universal form; only a few parameters of the specific system enter in them and  these are determined by the equilibrium properties of the system and  some dissipative properties which give rise to its  viscosity, conductivity, etc. The NS equations are, in fact, an effective theory --- higher order corrections   can be organised in terms of a derivative expansion in spatial and temporal derivatives.  These corrections   can be systematically included with the introduction of only a few additional parameters at each order. 

Conformal field theories  have no underlying scale. For an uncharged state in such a system,  one expects the role of the mean free path to be  played by the temperature $T$,  with  fluid mechanics arising as  a good approximation in  situations which are slowly varying    compared to $T$. It is well known by  now that strongly coupled  conformal theories often have weakly coupled gravity duals \cite{Aharony:1999ti}. 
 In a series of beautiful developments starting with \cite{Bhattacharyya:2008jc}, it has been shown that the fluid mechanics approximation in   field theory  has an exact  parallel in gravity,  with  the   complicated and non-linear Einstein equations admitting  a systematic derivative approximation  exactly for situations which  are slowly varying compared to $T$. In fact, in these circumstances,  it has been shown that   solutions of the Einstein equations map  directly to solutions of the relativistic Navier-Stokes equations order by order in the derivative expansion.  For related works, see \cite{Bhattacharyya:2007vs, Loganayagam:2008is,  Bhattacharyya:2008xc, Bhattacharyya:2008ji, Banerjee:2008th, Bhattacharyya:2008mz,   Bhattacharyya:2008kq, Hubeny:2009zz, Bhattacharya:2011eea, Bhattacharya:2011tra};  see also the reviews  \cite{Rangamani:2009xk, Hubeny:2011hd}  and references therein. These results have proved to be of considerable interest in the study of strongly coupled conformal and non-conformal systems, \cite{Kanitscheider:2009as, David:2009np},  and have also been extended to charged fluids, \cite{Erdmenger:2008rm},  \cite{Banerjee:2008th} where it was shown that the solutions on the gravity side  map to those of the charged relativistic NS equations. For an account of  many of the exciting developments in the field, see  the reviews  \cite{Son:2007vk, Erdmenger:2007cm, Gubser:2009md, Herzog:2009xv, Hartnoll:2009sz, McGreevy:2009xe,  Horowitz:2010gk, CasalderreySolana:2011us, Hartnoll:2016apf} and references therein. 

Here we will be interested in near-extremal black holes/branes in gravity. These   are solutions   which  arise  in  theories of gravity  that  also   
contain  gauge fields; in particular,    the black holes/branes of interest carry charges under  these gauge fields. The simplest such example, which we will focus on here,  consists  of  Einstein gravity coupled to a  Maxwell field.  The $\mathrm{U}(1)$ gauge symmetry in the bulk maps to a conserved $\mathrm{U}(1)$ global charge  in the boundary 
theory in this case. 

 As was mentioned above, for a field theory with such a conserved charge the equilibrium state is specified by  a temperature $T$ and a chemical potential  $\mu$.  
 In the earlier work mentioned above, extending the gravity analysis to charged fluids, the map of the Einstein equations to    NS equations of  charged fluid mechanics was shown to arise when one considered departures from equilibrium which were slowly varying compared to $T$. In the near-extremal limit where 
 \be
 \label{condaa}
 T\ll \mu,
 \ee
 this means, 
 denoting the frequency and momentum scale characterising the temporal and spatial variations by $\omega$ and $k$ respectively, that the earlier analysis applied  when the condition
 \be
 \label{condusual}
 \omega,k \ll T,\mu
 \ee
 is met, so that the rate of variation is the smallest scale compared to both $T$ and $\mu$.

 For a near-extremal black brane, since condition (\ref{condaa}) is valid,  $T$ and $\mu$  are two very different scales, 
  (in fact,  $T=0$ in the extremal case). 
 So one can also consider the regime where  $\omega$ and $k$  satisfy the following condition instead, 
 \be
 \label{condtwo}
 T \ll \omega, k \ll \mu.
 \ee
 That is when, in contrast to eq.(\ref{condusual}),   $T$ is the smallest scale and  the rate of spatial and temporal variation is bigger than it, but smaller than $\mu$.  
   We would like to  ask: what is the behaviour of the system in such  situations? In particular, we wish to explore whether there is a systematic approximation in which one can solve the Einstein equations in such circumstances and whether the resulting behaviour can be described in the boundary theory by solutions to a set of  equations,  local in space and time, analogous to the conventional NS equations,  with corrections which can be incorporated by including successively higher derivatives in space and time.

 This will be the question we investigate here. For simplicity, as mentioned above, we will consider a theory of Einstein gravity  with a Maxwell field 
 and a negative cosmological constant. The near-extremal  Reissner-Nordstr{\"o}m black brane solution in this system is among the simplest instances in which   a near-extremal black brane solution can arise. We will also restrict ourself to $3+1$ dimensions on the gravity side and therefore $2+1$ dimensions in the field theory. Our results can be easily extended  to include additional gauge fields, scalars, etc, and also in other dimensions.

 The motivation for our investigation comes partly from the renewed recent interest in studying extremal and near-extremal systems, \cite{Maldacena:2016upp}. 
 The near-horizon region in such systems often have an   $\mathrm{AdS}_2 $ factor in their geometry, and it has been found that the thermodynamics and response to probes at low frequencies can often be described in terms of a two dimensional theory of gravity, called Jackiw-Teitelboim (JT) gravity, \cite{Teitelboim:1983ux, Jackiw:1984je} coupled to extra degrees of freedoms. In fact,  it turns out that  two-dimensional JT gravity can     equivalently be replaced by a theory in one dimension --- time alone \cite{Sachdev:1992fk, Kitaev, Maldacena:2016hyu}. The degrees of freedom in this one-dimensional theory are time reparametrisations and  
 they, along with a phase mode \cite{Davison:2016ngz}, correctly reproduce the near-extremal thermodynamics quite generally in  near-extremal systems with an $\mathrm{AdS}_2$ near horizon geometry,  \cite{Nayak:2018qej, Moitra:2018jqs, Moitra:2019bub, Moitra:2019xoj}. After coupling to additional fields,  the time reparametrisation modes and phase mode also correctly reproduce the response to probes of the black hole, at small frequencies, $\omega \ll \mu$.  See \cite{Sarosi:2017ykf, Rosenhaus:2018dtp} for reviews.

 The $\mathrm{AdS}_2$ near-horizon region of these  near-extremal branes, in fact, corresponds to a  long ``throat''  in their geometry. What we mean more precisely by  this  is that the spacetime  in these solutions can be foliated by a set of spatial hypersurfaces orthogonal to the timelike Killing field
 and the proper distance along these hyper surfaces to the horizon grows big and diverges in the extremal limit. The one-dimensional theory mentioned above lives at the boundary of this throat region, where it glues into the outside region ultimately opening out into  the asymptotic $\mathrm{AdS}_4$ region.

 By solving the Einstein equations in the $\mathrm{AdS}_2$ near-horizon region and in the outside region, and matching the solutions  together at the boundary of the throat region, we find that the Einstein-Maxwell  equations, in fact, do admit a systematic approximation in the limit eq.(\ref{condtwo}). The starting point in finding the solution, like in the conventional case, eq.(\ref{condusual}),  is a  boosted near-extremal black brane configuration with local values  of $T$, $\mu$, $u^\nu$, but now these are varying at time and length scales satisfying condition (\ref{condtwo}). Corrections to this starting configuration, we find, can be found by carrying out a  systematic double expansion in the parameters  
  \be
 \label{valeps}
 \epsilon \sim {\omega \over \mu}, \,\, {k\over \mu} \ll 1,
 \ee
 and
  
 \be
 \label{valttil}
 {\widetilde T}  \equiv \frac{T}{\omega} \ll 1.
 \ee
 We carry out this expansion by first working to a given order in $\epsilon$, and then at that order in $\epsilon$, carrying out an expansion  in ${\widetilde T}$. 
 
 At $\mathcal{O}(\epsilon^1)$  we find   that the solutions to Einstein equations can be described by a fluid on the boundary with a constitutive relation that is local, including a viscosity term in the stress tensor and also dissipative terms in the charge current which give rise to charge diffusion.

 However, beyond first order in $\epsilon$, while the corrections continue to be small, their effects in the boundary stress tensor and charge current  can no longer be incorporated by adding local terms in the  constitutive relations. 
 For example, at second order, there are effects which go like $\mathcal{O}(\epsilon^2 \log {\omega /\mu})$, and also correction at $\mathcal{O}(\epsilon^2 )$ in an asymptotic expansion
 in ${\widetilde T}$. 
 These terms are non-local in time, and their effects cannot be obtained by adding local higher derivative terms to the stress tensor and charge current. Note that although 
 some of the  non-local terms which goes like  $\mathcal{O}(\epsilon^2 \log \epsilon)$ are  logarithmically enhanced, they are  still smaller than the $\mathcal{O}(\epsilon)$ terms at first order, and the presence of such terms therefore does not invalidate the systematic expansion in powers of $\epsilon$. 
 At higher orders, we expect there to be  further logarithmically  enhanced terms but again their presence will not invalidate the perturbative expansion described above;
 despite the logarithmically  enhanced terms, the contribution  at  $\mathcal{O}(\epsilon^n)$ will continue to be suppressed compared to terms which arise  up to $\mathcal{O}(\epsilon^{n-1})$. It is worth emphasising that the non-locality is only in time, the effects of spatial variation can be incorporated in local corrections to  the constitutive relations involving spatial derivatives to the required order. The non-locality is,  in fact, tied to the scale invariant nature of the near horizon $\mathrm{AdS}_2$ region; under the scaling symmetry involved only time and not the spatial directions transform non-trivially.

 The fact that a systematic approximation scheme exists for the situations considered here  is  of interest in increasing our understanding of the behaviour of the Einstein equations which are,  in general,  notoriously difficult to solve. 
 In fact,  finding such an additional approximation scheme was one of the   motivations behind our work.  In particular, it is our fond  hope  that the analysis presented here can be applied in the future  for studying near-extremal Kerr black holes, in asymptotically flat spacetime.   These are  known to occur in nature and of considerable observational interest.

 It is also worth contrasting the behaviour we find with that in conventional fluid mechanics, which is valid when condition (\ref{condusual}) holds. 
 In the conventional case, close to extremality, where $T\ll \mu$, one can carry  out a double  expansion in powers of $\epsilon^n$ and $(\omega/T)$ by,  first working at $\mathcal{O}(\epsilon^n)$ and then at this order calculating  the corrections to all orders in an expansion in $(\omega/ T)$.  The resulting constitutive relations
 are then manifestly local to all orders in this double   expansion. In the linearised approximation, it is well known that there are $4$ hydrodynamic modes in 
 the fluid mechanics regime, one shear mode, two sound modes and one charge diffusion mode. 
 We find in the limit (\ref{condtwo}) considered here, that there are also four modes in the linearised approximation; these map to the four modes in the fluid mechanics regime, however, in order to find the dispersion relation for the charge diffusion mode correctly we need to go beyond the $\mathcal{O}(\epsilon)$ corrections in the constitutive relations and also include terms at $\mathcal{O}(\epsilon^2)$, While these $\mathcal{O}(\epsilon^2)$ corrections are complicated and non-local as emphasised above in general, they simplify for the charge diffusion mode and can be obtained quite easily. The result we get then is quite interesting: the dispersion relations for all four modes, the shear, two sound and charge diffusion modes agree in the conventional case and the limit we consider here\footnote{We are deeply grateful to Richard Davidson for a discussion on this point.}. 
 
 More generally, one finds that working with the local constitutive relations  only  up to $\mathcal{O}(\epsilon)$ imposes some important limitations. There are four parameters which determine the state of the fluid, $T$, $\mu$, and two components of the velocity $\beta^i, i=1,2$. Corrections in the constitutive relation beyond the perfect fluid part, which we obtain only to $\mathcal{O}(\epsilon)$, 
 do not allow the time derivatives of $T$, $\mu$, in the local rest frame of the fluid, to be obtained beyond $\mathcal{O}(\epsilon)$ (more precisely $\mathcal{O}(\epsilon/{\widetilde T})$).  This is not enough to incorporate the leading effect of dissipation reliably in the time evolution. Higher order corrections, at least to $\mathcal{O}(\epsilon^2)$ in the constitutive relations need to be retained for a full understanding of the dynamics, these can be obtained through a systematic approximation as mentioned above and discussed in more detail in the case of a prototypical scalar field in section \ref{sec-toy}.

 This paper is structured as follows. In section \ref{sec-fgr}, we begin with a review of near-extremal RN branes and the conventional fluid mechanics-gravity correspondence. 
 In section \ref{sec-toy}, we study a massless scalar field in a near-extremal background which is a prototype for the subsequent analysis. In section \ref{sec-deo}, we  turn to gravity and gauge field perturbations and show how they can be analysed systematically in the double expansion mentioned above. In section \ref{sec-hdm} we consider linearised perturbations and also obtain the full non-linear constitutive relations at $\mathcal{O}(\epsilon)$ --- these include corrections up to $\mathcal{O}(\epsilon {\widetilde T})$ in our double expansion. 
 In section \ref{sec-hdm},  we consider some time-independent situations, some static and others stationary, which arise when external forces are turned on in the boundary theory. We find that the resulting  perturbations die out at the extremal horizon exhibiting attractor behaviour. We end with some discussion in section \ref{sec-co}. Appendices \ref{app-leq}--\ref{app-constime} contain important details. 
 
 Before proceeding let us note some additional references. In approaches different from ours (e.g, study of quasi-normal modes) from ours, hydrodynamics of low-temperature systems was  explored in, for example, \cite{Denef:2009yy, Edalati:2009bi, Edalati:2010hk, Edalati:2010pn, Davison:2011uk, Davison:2013bxa, Erdmenger:2016wyp}; see also \cite{Oh:2010jp}. The holographic correspondence has been of considerable interest in describing strongly interacting systems and has been applied to describe  the behaviour of a  wide range of systems --- from cold atom systems to heavy ion collisions at the Relativistic Heavy Ion Collider (RHIC). Some key references in this regard are: \cite{Policastro:2001yc, Policastro:2002se, Son:2002sd, Herzog:2002fn, Policastro:2002tn, Herzog:2002pc, Nunez:2003eq, Herzog:2003ke, Kovtun:2003wp, Buchel:2003tz, Kovtun:2004de, Buchel:2004hw, Buchel:2004di, Buchel:2004qq, Kovtun:2005ev, Benincasa:2005iv, Shuryak:2005ia, Janik:2005zt, Benincasa:2006fu, Mas:2006dy, Son:2006em, Saremi:2006ep, Maeda:2006by, Herzog:2006gh, Liu:2006ug, Gubser:2006bz, Janik:2006gp, Nakamura:2006ih, Janik:2006ft, Friess:2006kw, Nakamura:2006xk, Liu:2006he, Herzog:2007ij, Heller:2007qt, Hartnoll:2007ai, Gubser:2007ga, Kats:2007mq, Brigante:2007nu, Baier:2007ix, Gubser:2008px, Brigante:2008gz, Hartnoll:2008vx, Gubser:2008yx, Son:2008ye, Balasubramanian:2008dm, Gubser:2008pc, Gubser:2008sz, Myers:2008yi, Karch:2008fa, Haack:2008cp, Herzog:2008wg, Adams:2008wt, Buchel:2008ae, Iqbal:2008by, Herzog:2008he, Hartnoll:2008kx, Rangamani:2008gi, Buchel:2008vz, Cai:2009zv, Myers:2009ij, Chesler:2009cy, deBoer:2009pn, Goldstein:2009cv, Rebhan:2011vd, Bredberg:2011jq, Davison:2011ek, Donos:2014cya}.

\section{A Review of Nearly Extremal Black Branes and the  Conventional Fluid-Gravity Correspondence}\label{sec-fgr}

\subsection{Basic Set-Up and Conventions}\label{ss-c}

Our starting point is the Einstein-Maxwell action in four-dimensional asymptotically AdS spacetime,
\begin{equation}\label{ema}
I = \frac{1}{16\pi G} \int \mathrm{d}^4 x \, \sqrt{-g} \pqty{ R - 2 \Lambda - F_{MN} F^{MN}   }.
\end{equation}
Throughout this paper, we work in units in which $L_{\mathrm{AdS}} = 1$, i.e.,
\begin{equation}\label{defcc}
\Lambda = -3.
\end{equation}
The uppercase Latin letters $M, N, \ldots$ correspond to four-dimensional bulk spacetime coordinates, the Greek letters $\mu, \nu, \ldots$ correspond to three-dimensional boundary spacetime coordinates (2 spacelike and 1 timelike) and the lowercase Latin letters $i, j, \ldots$ correspond to boundary spacelike coordinates.

The Maxwell equations read,
\begin{equation}\label{em}
\mathsf{M}^N \equiv \nabla_M F^{MN} = 0,
\end{equation}
while the Einstein equations are given by,
\begin{equation}\label{ein}
\mathsf{E}_{MN} \equiv R_{MN} - \frac{1}{2} R g_{MN} + \Lambda g_{MN} - \pqty{ 2 F_{M S} F_N {}^S - \frac{1}{2} g_{MN} F_{RS} F^{RS}} = 0.
\end{equation}

The electrically charged Reissner-Nordstr\"{o}m black brane is described by the metric and gauge field,
\begin{align}
\bar{g}_{MN} \, \mathrm{d} x^M \, \mathrm{d} x^N &= - r^2 f(r) \, \mathrm{d} t^2 + \frac{\mathrm{d} r^2}{r^2 f(r) } + r^2 \pqty{ \mathrm{d} x^2 + \mathrm{d} y^2 }, \label{metrt}  \\
\bar{A}_M \, \mathrm{d} x^M &= g(r) \, \mathrm{d} t.  \label{grt}
\end{align}
Here,
\begin{align}
f(r) &= 1 - \frac{2GM}{r^3} + \frac{Q^2}{r^4}, \label{fdef} \\
g(r) &= - \frac{Q}{r}, \label{gdef}
\end{align}
where $M, Q$ are proportional to the mass density and charge density respectively.

The ingoing Eddington-Finkelstein coordinate system is well-suited for describing the fluid-gravity correspondence. In order to go to this coordinate system, we make the change of variables,
\begin{equation}
\label{defv}
v = t + r^*,
\end{equation}
where,
\begin{equation}
\label{defrw}
r^* \equiv  \int \frac{\mathrm{d} r}{r^2 f(r)}.
\end{equation}
In the new coordinates $(v,r,x,y)$, the metric and gauge field read
\begin{align}
\mathrm{d} s^2 = \bar{g}_{MN} \, \mathrm{d} x^M \, \mathrm{d} x^N &= - r^2 f(r) \, \mathrm{d} v^2 + 2 \, \mathrm{d} v \, \mathrm{d} r + r^2 ( \mathrm{d} x^2 + \mathrm{d} y^2), \label{metv} \\
A = \bar{A}_M \, \mathrm{d} x^M &= g(r) \, \mathrm{d} v. \label{gfv}
\end{align}
Note that we have eliminated the component $A_r$ of the gauge field by a suitable gauge transformation.\footnote{{The gauge chosen here, eq.(\ref{gfv}) with $g$ given by (\ref{gdef}) and the coordinate system are both non-singular on the future event horizon. We continue to work in this gauge and coordinate system through the rest of the paper. However, note that (\ref{grt}) with $g(r)$ given by eq.(\ref{gfv}) corresponds to a singular gauge choice on the event horizon. This can be made non-singular easily by subtracting a constant from $A_t$ in the $(t,r,x,y)$ coordinate system: $A_t = g(r) - g(r_+)$, where $r=r_+$ is the location of the outer horizon. }}

The event horizon of the black brane is denoted to be located at $r=r_+$:  $f(r_+)  =0$ and $f(r)>0$ for all $r>r_+$. The Hawking temperature (physical temperature) of the black brane is given by,
\begin{equation}\label{defth}
\widehat{T} = \frac{3r_+^4 - Q^2}{4\pi r_+^3}.
\end{equation}
Throughout this paper, we find it convenient to use the symbol $T$ for the rescaled temperature, this is related to the  actual temperature   ${\hat T}$  by,
\begin{equation}\label{deft}
T \equiv \frac{2\pi}{3} \widehat{T}.
\end{equation} 
The chemical potential $\mu$ of the system is given by,
\begin{equation}
\label{defcp}
\mu = \frac{Q}{r_+}.
\end{equation}

We are interested in the  regime where the black brane temperature is zero or very small compared to the chemical potential, i.e., the black brane being extremal or nearly extremal,
\be
\label{next}
T\ll \mu.
\ee
At extremality, the locations of the event horizon and the inner horizon coincide at $r=r_h$ and $f(r)$ has a double zero at $r=r_h$ ($f(r_h) = 0 = f'(r_h)$). In this paper, $r_h$ will refer to the location of the extremal horizon. At extremality, the  mass density $\mathcal{E}$ and charge density $\rho$ get related as,
\begin{equation}
\label{extmq}
4\pi G \, \mathcal{E} = G M = 2 r_h^3, \quad 4\pi G \, \rho = Q = \sqrt{3} r_h^2,
\end{equation}
while the chemical potential is given by,
\begin{equation}
\label{extmu}
\mu = \sqrt{3}r_h.
\end{equation}
In this extremal case, the function $f(r)$, eq.(\ref{fdef}) and $g(r)$, eq.(\ref{gdef}) are given respectively by,
\begin{align}
\label{f0r}
f_0 (r) &= 1 - \frac{4r_h^3}{r^3} + \frac{3r_h^4}{r^4} = \frac{(r-r_h)^2 (r^2 + 2r r_h +3 r_h^2)}{r^4}, \\
g_0 (r) &= -\frac{\sqrt{3}r_h^2}{r}. \label{g0r}
\end{align}

For a black brane at arbitrary temperature $T$  and chemical potential $\mu$, the location of the horizon, energy density and charge density can be expressed as,
\begin{align}
r_+ &= \sqrt{\frac{\mu^2}{3} +T^2 } + T, \label{crp} \\
4\pi G \, \mathcal{E} = GM &= 2\pqty{ \frac{\mu^2}{3} +T^2}^{3/2} + T(\mu^2 +2T^2) , \label{cm} \\
4\pi G \, \rho = Q &= \mu \pqty{ \sqrt{ \frac{\mu^2}{3} +T^2 } + T }\label{cq}.
\end{align}

We are interested in studying near-extremal black branes meeting the condition, eq.(\ref{next}).  It is easy to see that the metric and gauge field up to linear order in $T$  (at fixed $\mu$) take the form,
\begin{align}
\bar{g}_{MN} \, \mathrm{d} x^M \, \mathrm{d} x^N &= - F(r) \, \mathrm{d} v^2 + 2 \, \mathrm{d} v \, \mathrm{d} r + r_h^2 \pqty{ \mathrm{d} x^2 + \mathrm{d} y^2 }, \label{metnh} \\
\bar{A}_M \, \mathrm{d} x^M &= G(r) \mathrm{d} v, \label{gfnh}
\end{align}
with,
\begin{align}
F(r) &= 6(r-r_h) (r- r_h - T) \label{Fdef} \\
G(r) &= -\sqrt{3} (r_h +T) + \frac{\sqrt{3} (r_h +T)}{r_h} (r-r_h)  \label{Gdef},
\end{align}
where, in our notation,   $r_h$ continues to be  related to the chemical potential of the finite temperature black brane by the relation eq.(\ref{extmu}). We see that the metric (\ref{metnh}) in the $v$-$r$ plane describes an $\mathrm{AdS}_2$ spacetime; when  $(T\neq 0)$ it is actually thermal $\mathrm{AdS}_2$.

\subsection{A Review of the Conventional  Fluid-Gravity Correspondence}\label{ss-rfg}

The starting point for the fluid-gravity correspondence is the black brane metric and gauge field in the ingoing Eddington-Finkelstein coordinates, given by eqs.(\ref{metv}--\ref{gfv}). The first step is to perform a three-dimensional Lorentz boost in the $x^i$ directions, with velocity $\beta_i$, which transforms the metric and gauge field into the form,
\begin{align}
\mathrm{d} s^2  &= - r^2 f(r; \mu, T) \, u_\mu u_\nu \, \mathrm{d}  x^\mu \, \mathrm{d} x^\nu - 2 \, u_\mu \, \mathrm{d}  x^\mu\, \mathrm{d} r + r^2 P_{\mu \nu} \, \mathrm{d} x^\mu \, \mathrm{d} x^\nu , \label{metvb} \\
A  &= -g(r; \mu, T) \, u_\mu \, \mathrm{d}  x^\mu. \label{gfvb}
\end{align}
The functions $f$ and $g$ depend parametrically on the chemical potential and the temperature, which we have made explicit above.
Here,
\begin{equation}
\label{defumu}
u_\mu = \frac{1}{\sqrt{1-\bm{\beta}^2}} (-1, \beta_i)
\end{equation}  
is a normalized ($u^\mu u_\mu =-1$) timelike Lorentz 3-vector and 
\begin{equation}
\label{defPr}
P_{\mu \nu} = \eta_{\mu \nu} + u_\mu u_\nu
\end{equation}
is the projector orthogonal to the velocity $u^\mu$. Here $\eta_{\mu \nu} = \mathrm{diag}(-1,1,1)$ is the metric on the flat boundary spacetime. The next step is to make the chemical potential, temperature and the boost parameters functions of the boundary coordinates $x^\sigma$. We call the resulting metric and gauge field the ``zeroth order'' ones,
\begin{align}
g^{(0)}_{MN} ( x^\sigma) \, \mathrm{d} x^M \, \mathrm{d} x^N &= - r^2 f \pqty{r; \mu(x^\sigma) , T(x^\sigma) } \, u_\mu (x^\sigma) u_\nu (x^\sigma)  \, \mathrm{d}  x^\mu \, \mathrm{d} x^\nu - 2 \, u_\mu (x^\sigma) \, \mathrm{d}  x^\mu\, \mathrm{d} r \nonumber\\
&\quad + r^2 P_{\mu \nu}  (x^\sigma) \, \mathrm{d} x^\mu \, \mathrm{d} x^\nu, \label{met0} \\
A^{(0)}_{M} \, \mathrm{d} x^M &= -g\pqty{ r; \mu(x^\sigma), T(x^\sigma) } u_\mu (x^\sigma)  \, \mathrm{d}  x^\mu \label{gauge0}. 
\end{align}
The equations (\ref{met0}--\ref{gauge0}) above do not, in general, meet the system of Maxwell and Einstein equations (\ref{em}, \ref{ein}). In order to meet the equations of motion, we must correct the metric in a systematic manner. The strategy of solving for the corrections is the well-known ``derivative expansion'', in which the zeroth order metric parameters are assumed to be slowly varying functions of $x^\mu$. The perturbation theory is formulated as follows: We expand the full metric and gauge field in a power series in $\epsilon$, where $\epsilon$ is a small parameter describing the ``slowness' of the variation of the parameters,
\begin{align}
g_{MN} &= \sum_{n=0} \epsilon^n g^{(n)}_{MN} \big( \mu( \epsilon x^\sigma),  T(\epsilon x^\sigma), \beta_i (\epsilon x^\sigma)  \big), \label{sermet}\\
A_M &= \sum_{n=0} \epsilon^n A^{(n)}_{M} \big( \mu( \epsilon x^\sigma),  T(\epsilon x^\sigma), \beta_i (\epsilon x^\sigma)  \big). \label{serA}
\end{align}
Here, the functions $g^{(n)}_{MN}$, $A^{(n)}_{M}$ are the corrections to the metric and gauge field, in the $\epsilon$-expansion. In fact, the parameters $\mu$, $T$ and $\beta_i$ also get corrected order by order in $\epsilon$.

In the standard fluid-gravity correspondence, the fluid parameters, $\mu$, $T$ and $\beta_i$  vary slowly compared to $T$. A solution can be obtained to the coupled Einstein-Maxwell equations by starting with the zeroth order solution eqs.(\ref{met0}, \ref{gauge0}) and correcting it order by order in $\epsilon$. 
The normalisable components of the metric and gauge field to order $n$  then determine the stress tensor and charge current in the boundary theory  to that order. 
The equations of motion for the metric and gauge field ensure that the resulting stress tensor and current satisfy the conservation equations,
\begin{align}
\partial^\mu T_{\mu \nu} &= 0, \label{nsft} \\
\partial^\nu J_\nu &= 0 . \label{nsfj}
\end{align}
Eq. (\ref{nsft}) is  the relativistic Naiver-Stokes equation (with higher derivative corrections for $n\ge 2$). 

A few more details are  worth giving.
It is convenient to work in the gauge 
\begin{align}
g_{rr} &= 0 = A_r,  \nonumber \\ g_{r\mu} &\propto u_{\mu}, \label{gauge} \\
 g^{(0)\, MN}g^{(n)}_{MN} &= 0, \quad n\geq 1. \nonumber
\end{align}
Having obtained the solution up to order $n-1$ one finds at the $n$\textsuperscript{th} order, i.e. the next order in the derivative expansion, that the 
Einstein equations obtained from varying $g_{r} {}^\mu$ do not involve perturbations at the next order, i.e. $g_{MN}^{(n)}$, $A_{M}^{(n)}$. Instead, they simply impose the Naiver-Stokes equations eq.(\ref{nsft}) on the $(n-1)$\textsuperscript{th} order solution. 
Similarly the Maxwell equation obtained by varying $A_r$ leads to the current conservation equation, eq.(\ref{nsfj}) for the $(n-1)$\textsuperscript{th} order solution. 
The remaining Einstein and Maxwell equations then take the form

\begin{equation}
\label{diffop}
\mathbb{D} \pqty{ g_{MN}^{(n)}, A_{M}^{(n)} } = s^{(n)} \pqty{ g_{MN}^{(n-1)}, \ldots , A_{M}^{(n)}, \ldots },
\end{equation}
where $\mathbb{D}$ is a differential operator acting on the metric and gauge field at the $n$\textsuperscript{th} order and $s^{(n)}$ is a source term generated by the action of  additional derivatives along the boundary directions acting on lower order terms. The operator $\mathbb{D}$ has several noteworthy features. First, it is an operator involving derivatives only in $r$. Second, the operator remains unchanged to every order in the $\epsilon$-expansion; in particular, it is determined by the zeroth order metric. The specific form of the operator, however, depends on the kind of metric and gauge field component under consideration. A solution to eq.(\ref{diffop}) requires one to fix the constants of integration; these are constrained  by imposing ingoing boundary conditions at the horizon, normalisable fall-off at the boundary, and extra conditions corresponding to a choice of Landau frame as described below.  We will discuss in detail the specific forms of the operator and the solutions in section \ref{sec-deo}.

The fluid-gravity correspondence, in fact, has been generalised \cite{Bhattacharyya:2008ji} to include the scenario in which the boundary metric is not flat $\eta_{\mu \nu}$, but weakly curved with a metric $\widetilde{\gamma}_{\mu \nu}$. In addition, a non-trivial electromagnetic field strength $\widetilde{F}_{\mu \nu}$ can be turned on. In such a case, the conservation equations (\ref{nsft}--\ref{nsfj}) are modified to,
\begin{align}
\widetilde{\nabla}^\mu T_{\mu \nu} &= \widetilde{F}_{\nu \sigma} J^\sigma, \label{nsct} \\
\widetilde{\nabla}^\nu J_{\nu} &= 0 . \label{nscj}
\end{align}
Here the raising and lowering of indices and covariant differentiation is assumed to be done with the boundary metric $\widetilde{\gamma}_{\mu \nu}$.

\subsection{Determining the Boundary Stress Tensor and Charge Current}\label{ss-stress}

Using the methods of holographic renormalisation,(see, for example, \cite{Henningson:1998gx, Balasubramanian:1999re, Kraus:1999di}, and, for a pedagogical review, \cite{Skenderis:2002wp}, and references therein) ,  we can find the expression for the stress tensor,
\begin{equation}
\label{tmndef}
8 \pi G T^\mu {}_\nu =  -\lim_{r\to \infty} r^3 \pqty{ K^{\mu} {}_\nu - (K-2) \delta^\mu {}_\nu - \pqty{ {R^{(3)} }^\mu {}_\nu  - \frac{1}{2} R^{(3)} \delta^\mu {}_\nu } }.
\end{equation}
Here $K_{\mu \nu}$ is the extrinsic curvature tensor for a hypersurface defined by a constant value of $r$. The Ricci tensor $R^{(3)}_{\mu \nu}$ is constructed out of the induced metric $\gamma_{\mu \nu}$ on the boundary. The form of the metric on the boundary manifold is given by,
\begin{equation}
\label{bmetdef}
\widetilde{\gamma}_{\mu \nu} = \lim_{r\to \infty} \frac{1}{r^2} \gamma_{\mu \nu}.
\end{equation}
We can similarly find out the boundary gauge field $\widetilde{A}_\mu$ from the large-$r$ expansion of the bulk gauge field
\begin{equation}
\label{bgdef}
\widetilde{A}_\mu = \lim_{r\to \infty} A_\mu .
\end{equation}
The boundary electromagnetic field strength is defined, as usual,
\begin{equation}
\label{fmnb}
\widetilde{F}_{\mu \nu} = \del_\mu \widetilde{A}_\nu - \del_\nu. \widetilde{A}_\mu,
\end{equation}
while the boundary charge current is given by,
\begin{equation}
\label{jmudef}
4\pi G J^\nu =  \lim_{r\to \infty} \sqrt{-g} F^{\nu r}.
\end{equation}

When the boundary conditions are such that no non-normalisable components of the metric and the gauge field are present, we would have, $\widetilde{\gamma}_{\mu \nu} = \eta_{\mu \nu}$ and $\widetilde{A}_\mu = 0$. This is the case we would deal with in section \ref{sec-hdm}. In section \ref{sec-hs}, we shall encounter cases in which the boundary manifold has a non-trivial curvature as well as a non-zero electromagnetic field strength.

Before we end this section, let us discuss the Landau frame conditions mentioned previously. 
Suppose we have   corrections at the $n$\textsuperscript{th} order to the stress tensor and charge current. Then some of these corrections can be incorporated in the zeroth order stress tensor and current themselves, by changing the temperature, etc.\ at the $n$\textsuperscript{th} order, i.e. making suitable changes,
 $T \rightarrow T + \epsilon^n \delta T$, $\mu \rightarrow \epsilon^n \delta \mu, \beta^i\rightarrow \beta^i + \epsilon^n \delta \beta^i$. 
 After doing so the  remaining $n$\textsuperscript{th} order corrections in the stress tensor and current  can be taken to  satisfy,  
\begin{align}
u^{\mu} T^{(n)}_{\mu \nu} &= 0, \nonumber\\
u^{\nu} J^{(n)}_{\nu} &= 0. \label{landau}
\end{align}
These are called the Landau frame conditions (sometimes also referred to as the Landau-Lifshitz frame conditions, see \cite{LANDAU}). 
We will impose these conditions in our discussion of near-extremal fluid mechanics in section \ref{sec-hdm}  as well.

\section{Massless Scalar in Extremal and Near-Extremal Background} \label{sec-toy}
In this section, we study the behaviour of a massless scalar field in an extremal or near-extremal black brane background. The scalar field satisfies non-normalisable boundary conditions which are slowly varying,  along the boundary directions, compared to the chemical potential. We will see that this is a good prototype for the study of fluid mechanics in such backgrounds.

The scalar field $\phi$  satisfies the equation of motion,
\begin{equation}
\label{phim}
\nabla^2 \phi \equiv \frac{1}{\sqrt{-g}} \del_M \pqty{ \sqrt{-g} g^{M N} \del_N \phi } = 0.
\end{equation}
A constant value of $\phi = \phi^{(0)}$ obviously meets this equation. Now, in the spirit of the fluid-gravity correspondence, we want to make $\phi$ to the zeroth order, a slowly-varying function of the boundary coordinates $x^\mu$,
\begin{equation}
\label{phi0}
\phi^{(0)} = \phi^{(0)} ( \epsilon x^\mu). 
\end{equation}
In order that the equation (\ref{phim}) above is still met, we have to add a correction to $\phi$.

To be concrete, let us work in a basis of plane waves in the $x^\mu$ directions and write $\phi$ as a perturbative series,
\begin{equation}
\label{phic}
\phi = e^{- \mathrm{i} \omega v + \mathrm{i} k_x x} \phi_c (r) \equiv e^{- \mathrm{i} \omega v + \mathrm{i} k_x x} \sum_{n=0} \phi^{(n)} (r) .
\end{equation}
We take 
\be 
\label{defepsilon}
\frac{\omega}{\mu}, \frac{k_x}{\mu} \sim \epsilon \ll 1.
\ee
  We work order by order in perturbation theory in $\epsilon$ and  
the number $n$ within the parentheses labels the order in perturbation theory. The perturbation $\phi^{(n)}$ is sourced by the corrections at lower orders, $\phi^{(n-1)}, \cdots, \phi^{(0)}$.

We take $\phi$ to meet the boundary condition at the asymptotic $\mathrm{AdS}_4$ boundary, $r\rightarrow \infty$,
\be
\label{brass}
\phi\rightarrow \phi^{(0)}(\epsilon x^\mu),
\ee
so that the non-normalisable behaviour we are imposing for  $\phi$ is already met by $\phi^{(0)}$; the higher order corrections $\phi^{(n)}$, $n>0$ must then be purely normalisable meeting the condition
\be
\label{bcnn}
\phi^{(n)}\rightarrow 0
\ee
as $r\rightarrow \infty$. 

  We easily find that the equation (\ref{phim}) leads to,
\begin{equation}
\label{radp}
\frac{\mathrm{d} }{\mathrm{d} r} \pqty{ r^4 f(r) \frac{\mathrm{d} \phi_c }{\mathrm{d} r} } - 2 \mathrm{i} \omega r \frac{\mathrm{d} \pqty{ r \phi_c } }{\mathrm{d} r} - k_x^2 \phi_c = 0.
\end{equation}
We take $\phi^{(0)}$ to be a constant (with the exponential dependence $e^{- \mathrm{i} \omega v + \mathrm{i} k_x x}$ peeled off) and write down the equation determining $\phi^{(1)}$,
\begin{equation}
\label{radp1}
\frac{\mathrm{d} }{\mathrm{d} r} \pqty{ r^4 f(r) \frac{\mathrm{d} \phi^{(1)} }{\mathrm{d} r} } - 2 \mathrm{i} \omega r \frac{\mathrm{d} \pqty{ r \phi^{(1)} } }{\mathrm{d} r} - k_x^2 \phi^{(1)} = (2 \mathrm{i} \omega r + k_x^2) \phi^{(0)}.
\end{equation}  

In the first order calculation, analogous to what is done in the standard fluid-gravity correspondence, we would neglect all terms involving $\omega$ and $k_x$ on the left hand side and retain only the term linear in $\omega$ on the right hand side. 

Let us now specialise to  the extremal case. 
It is immediately clear in this case  that the approximation mentioned above  is not a good one. This follows from noting that  the near-horizon region is  $\mathrm{AdS}_2\times \mathbb{R}^2$,
\be
\label{geomads2}
\mathrm{d}s^2= -6(r-r_h)^2 \, \mathrm{d} v^2 +  2 \, \mathrm{d} v \, \mathrm{d} r + r_h^2 \pqty{ \mathrm{d}x^2 + \mathrm{d}y^2 },
\ee
 and it follows  from the scaling symmetry of this geometry,
 \begin{align}
 v &\rightarrow \lambda v, \nonumber  \\
 (r-r_h) &\rightarrow \frac{1}{\lambda} (r-r_h), \label{scalads2}
 \end{align} 
  that one can rescale $\omega $ to be unity. This  shows that the effects of 
  $\omega$, no matter how small, always become important sufficiently close to the horizon. Thus, the $\omega$--dependent terms on the LHS of eq.(\ref{radp1}) cannot be neglected. 
  In contrast, the $k_x$--dependent terms continue to be small as long as $\epsilon\ll 1$ and they can be neglected to first order in $\epsilon$, in eq.(\ref{radp1}).

  For temperatures close to extremality, with $\omega$ meeting the condition, 
  \be
  \label{condomega}
   T\ll \omega \ll  \mu,
   \ee
   these considerations continue to be true since the effects of the  temperature ``die away'' while one is still in the near-horizon region,
   \be
   \label{nhr}
   {r-r_h\over r_h}\ll 1.
   \ee
   The scaling argument above  then shows that  in this region the effects of $\omega $ on the LHS in eq.(\ref{radp1})  are important. 
   
 In contrast to the near horizon  region, once one is sufficiently far from the horizon, 
 \be
 \label{suffar}
 \omega\ll {r-r_h},
 \ee
 the $\omega$-- and $k_x$--dependent terms on the LHS of eq.(\ref{radp1}) can be neglected since they are second order in $\epsilon$. It is therefore convenient to separate the analysis of eq.(\ref{radp1}) into the far and near regions, $r>r_B$ and $r<r_B$ respectively with the boundary $r_B$ where they meet satisfying the conditions, 
 \begin{align}
 \omega  & \ll   {r_B-r_h}, \label{condrba}\\
 {r_B-r_h} &\ll  r_h. \label{condrba2}
 \end{align}
 Note that these two conditions are compatible since $\omega\ll \mu = \sqrt{3} r_h$.
 At $r=r_B$, the inside solution and the outside one must satisfy the continuity conditions for the function $\phi^{(1)}$ and its first derivative.

Equation (\ref{radp1}) in the far region can therefore be approximated as,
\begin{equation}
\label{radp2}
\frac{\mathrm{d} }{\mathrm{d} r} \pqty{ r^4 f(r) \frac{\mathrm{d} \phi^{(1)}_{\mathrm{out}} }{\mathrm{d} r} }  = 2 \mathrm{i} \omega r \phi^{(0)} .
\end{equation}
This differential equation has the immediate solution,
\begin{equation}
\label{sphi}
\phi^{(1)}_{\mathrm{out}} = B^{(1)}_{\mathrm{out}} -\int_{r}^\infty \frac{\mathrm{d} r'}{{r'}^4 f(r')} \pqty{ A^{(1)}_{\mathrm{out}} +  \int_{r_B}^{r'} \mathrm{d} r'' \, 2\mathrm{i} \omega   r'' \phi^{(0)} }  .
\end{equation}
Note that in the solution above, we have chosen the limit on the inner integral at the boundary of the $\mathrm{AdS}_2$, $r=r_B$ and that of the outer integral at the asymptotic infinity of $\mathrm{AdS}_4$, $r\to \infty$.  Note that we use $f(r)=f_0(r)$, eq.(\ref{f0r}) for the extremal geometry, we will consider the effects of non-zero temperature in \S\ref{ss-ft}. 

Imposing the boundary condition eq.(\ref{bcnn}) at the asymptotic infinity sets,
\begin{equation}
\label{bout}
B^{(1)}_{\mathrm{out}} = 0.
\end{equation} 
This leaves one undetermined constant $A^{(1)}_\mathrm{{out}}$ which will be determined by matching with the inner region solution.

We can actually write down a closed-form expression for $\phi^{(1)}_{\mathrm{out}}$ above,
\begin{align}
\phi^{(1)}_{\mathrm{out}} &= - \frac{1}{6r_h^2} \frac{A^{(1)}_{\mathrm{out}} - \mathrm{i} \omega \phi^{(0)} (r_B^2 - r_h^2) }{r-r_h} - \frac{A^{(1)}_{\mathrm{out}}  - \mathrm{i} \omega \phi^{(0)} (r_B^2 + 2r_h^2 ) }{18 r_h^3} \log \frac{(r-r_h)^2}{r^2 + 2r r_h + 3 r_h^2 } \nonumber \\
&\quad -\frac{A^{(1)}_{\mathrm{out}}  - \mathrm{i} \omega \phi^{(0)} (r_B^2 - 7 r_h^2 ) }{36 \sqrt{2} r_h^3} \pqty{ \pi - 2 \tan^{-1}\pqty{ \frac{r+r_h}{\sqrt{2}r_h} }  }. \label{phex1}
\end{align}
We can look at the behaviour of this solution near $r=r_B$   where we can expand the function in terms of $(r-r_h)/r_h$.  From eq.(\ref{phex1}) above,  we get the behaviour,
\begin{equation}
\label{phinh}
\phi^{(1)}_{\mathrm{out}}  = - \frac{1}{6r_h^2} \frac{A^{(1)}_{\mathrm{out}} - \mathrm{i} \omega \phi^{(0)} (r_B^2 - r_h^2) }{r-r_h} - \frac{A^{(1)}_{\mathrm{out}}  - \mathrm{i} \omega \phi^{(0)} (r_B^2 + 2r_h^2 ) }{9 r_h^3} \log \frac{r-r_h}{r_h} + \cdots ,
\end{equation}
where the ellipsis denotes terms of order $\mathcal{O}\bqty{ ( ( r-r_h)/ r_h )^0 }$. 

Suppose  we tried to directly extrapolate eq.(\ref{phinh}) to the horizon $r_h$ and obtain a non-singular solution.  
Cancelling the leading behaviour  which goes like $1/(r-r_h)$, one finds  the condition, 
\begin{equation}
\label{a1out}
A^{(1)}_{\mathrm{out}} = \mathrm{i} \omega \phi^{(0)} (r_B^2 - r_h^2),
\end{equation}
but this  still leaves a sub-leading logarithmic term which  would have diverged at  $r_h$. 
We will see below  that including the $\omega$--dependent terms in the inner region, in fact, gets rid of this logarithmic divergence, and matching the inner solution to the outer  one  fixes $A^{(1)}_{\mathrm{out}}$ to be exactly the value in eq.(\ref{a1out}) thereby removing the leading term in eq.(\ref{phinh}).

Let us now study the near-horizon behaviour in more detail. Neglecting the $k_x^2$ terms, the near-horizon form of the equation (\ref{radp}) is given by,
\begin{equation}
\label{phi2}
 \frac{\mathrm{d} }{\mathrm{d} r} \pqty{ F(r) \frac{\mathrm{d} \phi^{(1)}_{\mathrm{in}} }{\mathrm{d} r} } - 2 \mathrm{i} \omega \frac{\mathrm{d} \phi^{(1)}_{\mathrm{in}} }{\mathrm{d} r} = \frac{2 \mathrm{i} \omega \phi^{(0)} }{r_h}.
\end{equation}
It is worth drawing attention to a feature of the above equation. Had we started with the near-horizon metric of the form (\ref{metnh}) with $T=0$, we would have got the same form for the left-hand side. However, we would not have any source term that appears on the right-hand side of eq.(\ref{phi2}) above. This term comes from a correction to the $\mathrm{AdS}_2 \times \mathbb{R}^2$ metric --- more specifically, the variation of the transverse two-volume. As we shall see, this correction in the source plays an important role in obtaining the logarithmic term in eq.(\ref{phinh}). In fact, in obtaining the correct solution to higher orders, keeping such terms (i.e., those originating from departures from the near-horizon geometry) becomes crucial, as explained in \S\ref{ss-ho}.

Near the horizon, the form of $r^*$ (\ref{defrw}) is given by,
\begin{equation}
\label{rsnh}
r^* = \int \frac{\mathrm{d} r}{F(r)}.
\end{equation}
In terms of this variable, the equation (\ref{phi2}) above can be cast as,
\begin{equation}
\label{phi3}
\frac{\mathrm{d}^2 \phi^{(1)}_{\mathrm{in}} }{\mathrm{d} {r^*}^2} - 2 \mathrm{i} \omega \frac{\mathrm{d} \phi^{(1)}_{\mathrm{in}} }{\mathrm{d} {r^*}} = \frac{2 \mathrm{i} \omega \phi^{(0)} }{r_h} F(r).
\end{equation}
We can write the interior solution in the form,
\begin{equation}
\label{phin1}
\phi^{(1)}_{\mathrm{in}}  = A^{(1)}_{\mathrm{in}} + B^{(1)}_{\mathrm{in}} e^{2\mathrm{i} \omega r^*} + \frac{2 \mathrm{i} \omega \phi^{(0)} }{r_h} e^{2\mathrm{i} \omega r^*} \int_{-\infty}^{r^*} \mathrm{d} {r^*}' e^{-2\mathrm{i} \omega {r^*}' } \int^{r'}_{r_h} \mathrm{d} r''.
\end{equation}
Note that we have chosen the lower limit of both the inner and the outer integral to be on the horizon. We impose ingoing boundary conditions on the horizon. This sets $B^{(1)}_{\mathrm{in}}=0$, since the last term does not contribute to an outgoing mode at the horizon, as we will see shortly.

For the extremal geometry, in the near-horizon region,
\begin{equation}
\label{rse}
r^* = - \frac{1}{6 (r-r_h)},
\end{equation}
and we can therefore write,
\begin{equation}
\label{nph1}
\phi^{(1)}_{\mathrm{in}} =  A^{(1)}_{\mathrm{in}}  - \frac{\mathrm{i} \omega \phi^{(0)} }{3r_h} e^{2\mathrm{i} \omega r^*} \int_{-\infty}^{r^*} \frac{\mathrm{d} {r^*}' }{{r^*}' } e^{-2\mathrm{i} \omega {r^*}' }.
\end{equation}
We note that we can scale $\omega$ out of the integrand above and keep it only in the integration limits,
\begin{equation}
\label{nph2}
\phi^{(1)}_{\mathrm{in}} =  A^{(1)}_{\mathrm{in}}  + \frac{\mathrm{i} \omega \phi^{(0)} }{3r_h} e^{2\mathrm{i} \omega r^*} \int^{\infty}_{-2\zeta} \frac{\mathrm{d} t}{t } e^{\mathrm{i} t },
\end{equation}
Here we have defined the rescaled variable,
\begin{equation}
\label{zeta}
\zeta \equiv \omega r^* = - \frac{\omega}{6(r-r_h)} \, (<0).
\end{equation}
This variable is going to be very useful in the subsequent discussion.

The integral above is known as an exponential integral function \cite{AShandbook},
\begin{equation}
\label{defexp}
\int^{\infty}_{-2\zeta} \frac{\mathrm{d} t }{t } e^{\mathrm{i} t } \equiv E_1 (2 \mi \zeta). 
\end{equation}
The asymptotic behaviour of the function $E_1$ for large $|\omega r^*|$ is given by,
\begin{equation}
\label{eih}
E_1 (2\mathrm{i} \omega r^*) =  \frac{ e^{-2\mathrm{i} \omega {r^*} } }{2\mathrm{i} \omega {r^*}} \pqty{ 1 + \mathcal{O} \pqty{ \frac{1}{\omega r^*} } }.
\end{equation}
This can also be seen directly from eq.(\ref{nph1}). It follows then that the last term in eq.(\ref{phin1}) does not give rise to an outgoing mode, as we had mentioned above. 
From eq.(\ref{eih}), it also follows that   the function  eq.(\ref{nph2}) is well behaved on the horizon and takes the  value  $A^{(1)}_{\mathrm{in}}$.

Let us  examine the behaviour of the response near the boundary of $\mathrm{AdS}_2$, where $|\omega r^*| \ll 1$. The behaviour of the exponential integral in this region is given by,
\begin{equation}
\label{eibn}
E_1 (2\mathrm{i} \omega r^*)  =  -\gamma - \log ( 2 \mathrm{i} \omega r^*) + \mathcal{O} (\omega r^*),
\end{equation}
where $\gamma\approx 0.5772$ is the well-known Euler-Mascheroni constant. We can write the above expression as,
\begin{equation}
\label{eibn2}
E_1 (2\mathrm{i} \omega r^*)  = \pqty{ \frac{\mathrm{i} \pi}{2} - \gamma + \log 3 } + \log \frac{r-r_h}{\omega} + \mathcal{O} \pqty{ \frac{\omega}{r-r_h} }.
\end{equation}

Using this result, we are able to write down $\phi_{in}^{(1)}$  near the boundary region as an expansion in $\omega$, accurate up to the linear order in $\omega$, for the purpose of matching with the outside solution.
\begin{equation}
\label{phnh2}
\phi^{(1)}_{\mathrm{in}}  = A^{(1)}_{\mathrm{in}} + \frac{\mathrm{i} \omega \phi^{(0)}}{3 r_h} \pqty{ \frac{\mathrm{i} \pi}{2} - \gamma + \log 3 } +  \frac{\mathrm{i} \omega \phi^{(0)}}{3 r_h} \log (\frac{r-r_h}{\omega} ) + \cdots.
\end{equation}
Here, the ellipsis denotes terms which are subdominant in $\omega/(r-r_h)$. Comparing with the outside solution in eq.(\ref{phinh}), we see that  $\phi^{(1)}_{\mathrm{in}}$ does not contain any term going like $(r-r_h)^{-1}$ which is linear in $\omega$ (there is such a term but it is $\mathcal{O}(\omega^2/(r-r_h))$ and therefore, of higher order). Therefore, matching the inside and outside solutions around $r=r_B$ gives the condition eq.(\ref{a1out}). Comparing the inside and outside solution with this  choice of $A^{(1)}_{\mathrm{out}}$, shows that the coefficient of the logarithmic term  also matches.

With this choice of  $A^{(1)}_{\mathrm{out}}$, the outer solution  for $\phi^{(1)}_{\mathrm{out}}$ is completely determined to be 
\begin{align}
\phi^{(1)}_{\mathrm{out}} =  -\frac{\mi \omega \phi^{(0)} }{12r_h  } \bqty{ 2 \sqrt{2} \pqty{ \frac{\pi}{2} - \tan^{-1} \frac{r+r_h}{\sqrt{2}r_h}  } - 2 \log\frac{(r-r_h)^2}{r^2 + 2rr_h + 3r_h^2} }\label{phex2}.
\end{align}

Before proceeding, let us note that eq.(\ref{a1out}), which ensures that the coefficient of the $1/(r-r_h)$ vanishes, could be arrived at by using the scaling symmetry of $\mathrm{AdS}_2$ mentioned above, eq.(\ref{scalads2}) ---  this  will also be important in the subsequent discussion at higher orders. Using the scaling symmetry eq.(\ref{scalads2}), we see that 
eq.(\ref{phi3}) can be written as 
\be
\label{phi3a}
\frac{\mathrm{d}^2 \phi^{(1)}_{\mathrm{in}}}{\mathrm{d} \zeta^2} - 2 \mi \frac{\mathrm{d} \phi^{(1)}_{\mathrm{in}}  }{\mathrm{d} \zeta} =  \frac{\mi \omega \phi^{(0)} }{3 r_h} \frac{1}{\zeta^2}.
\ee
We note from its definition,  eq.(\ref{zeta}), that $\zeta$  is invariant under this scaling symmetry. 
It then follows  from eq.(\ref{phi3a}) that $\phi^{(1)}_{\mathrm{in}}$ is $\mathcal{O}(\omega)$ when expressed as a function of $\zeta$. Now, a term going like ${\omega/ (r-r_h)}$ in $\phi^{(1)}_{\mathrm{in}}$ would be of $\mathcal{O}(\omega^0)$ when expressed in terms of $\zeta$, thus such a term cannot arise from the inner solution to order $\omega$ and by matching, must also be absent in $\phi_{\mathrm{out}}^{(1)}$.

Expanding the exterior solution, eq.(\ref{phex2}), near $r=r_B$ to  $\mathcal{O}((r-r_h)/r_h)^0)$ gives,
\begin{equation}
\label{phex3}
\phi^{(1)}_{\mathrm{out}} = \frac{\mathrm{i} \omega \phi^{(0)}}{3 r_h} \log (\frac{r-r_h}{r_h} ) - \frac{\mathrm{i} \omega \phi^{(0)} }{12 r_h} \pqty{ \sqrt{2} \pi + 2 \log 6 -2 \sqrt{2} \tan^{-1} \sqrt{2} } + \mathcal{O} \pqty{ \frac{r-r_h}{r_h}  },
\end{equation}
Comparing with $\phi_{\mathrm{in}}^{(1)}$ in eq.(\ref{phnh2}) gives, 
\begin{equation}
\label{Ain1}
A^{(1)}_{\mathrm{in}} = \frac{\mathrm{i} \omega \phi^{(0)}}{3 r_h} \log \frac{\omega}{r_h}  - \frac{\mathrm{i} \omega \phi^{(0)} }{12 r_h} \pqty{ (\sqrt{2}+2\mathrm{i})  \pi + 6 \log 3 + 2 \log 2 -2 \sqrt{2} \tan^{-1} \sqrt{2} - 4 \gamma }.
\end{equation}
This fixes the inner solution.

Some important comments are now worth making. 
First, note that the logarithmic term near $r_B$ in the interior solution (\ref{phnh2}) is cut off by $\omega$, while in the exterior solution (\ref{phex3}), it is cut-off by $r_h$,  and hence, as we see from eq.(\ref{Ain1})  matching the two gives $A_{\mathrm{in}}^{(1)}$ to be of order $\omega \log(\omega/r_h)$. As a result, the inside solution is  non-analytic in $\omega$,  as might be expected on general grounds from the $\mathrm{AdS}_2\times \mathbb{R}^2$ nature of the near-horizon spacetime. However, interestingly, this non-analyticity is not present in the outside solution which is linear in $\omega$, at this order.

Second, with $A^{(1)}_{\mathrm{out}}$ satisfying the condition eq.(\ref{a1out}), the solution is manifestly independent of  $r_B$, i.e., where the inside and outside solutions are matched, as we can see from eq.(\ref{phex2}) and (\ref{nph2}, \ref{Ain1}). In fact, to this order, we can set the limit of the inner integral in eq.(\ref{sphi}) to be $r_h$ instead of $r_B$, making the outside solution manifestly independent of $r_B$; $A_{\mathrm{out}}^{(1)}$  will then end up vanishing.

Third, while the inner solution is  non-analytic in $\omega$, the correction is still small in $\epsilon$ since it is of order $\epsilon \log (\epsilon) \ll 1$. 

Fourth, the response  to the non-normalisable source turned on eq.(\ref{brass}) can be obtained using the standard AdS/CFT dictionary from the behaviour of the normalisable mode of $\phi$ near the boundary. Since the outside solution is analytic in $\omega$, this is also true of the response. 
We will see below, when we go to higher orders  that this feature is no longer true, and the outside solution, along with the response,  is non-analytic in $\omega$ with logarithmic corrections, e.g. at second order, the corrections we find  go like $\mathcal{O}(\omega^2 \log(\omega))$. However, there continues to be a sensible perturbative expansion, since the $\mathcal{O}(\omega^2 \log(\omega))$ terms are small compared to the leading order terms of $\mathcal{O}(\omega)$. This feature also persists to higher orders. We will see when we consider gravitational and electromagnetic perturbations of the coupled Maxwell-Einstein system that the first order solution in the exterior region is also analytic in $\omega$, like in this prototype system,  and the higher order corrections, while being  non-analytic,  will be smaller than the leading order terms. 

Finally, it is worth contrasting the analysis above with the case of a scalar in the Schwarzschild black brane background at temperature $T$.  The scalar is taken to vary slowly with $\omega/T, \, k_x/T\ll 1$. This is the analogue of the situation for  conventional fluid-mechanics, eq.(\ref{condusual}).  
In this case, the  exterior solution eq.(\ref{sphi}) can be directly  extended all the way to the horizon, $r=r_0$,  by choosing the lower limit in the inner integral to be $r_0$, i.e. setting $r_B=r_0$, so that 
\be
\label{fosc}
\phi^{(1)}= B^{(1)} -\int_{r}^\infty \frac{\mathrm{d} r'}{{r'}^4 f(r')} \pqty{ A^{(1)} +  \int_{r_0}^{r'} \mathrm{d} r'' \, 2\mathrm{i} \omega   r'' \phi^{(0)} } .
\ee
Normalisability at $r\rightarrow \infty$ sets $B^{(1)}=0$.  
Since $f(r)=1-r_0^3/r^3$ has a first order zero at $r=r_0$, by choosing $A^{(1)}=0$ a potential divergence at $r=r_0$ is removed and the integral is well defined all the way to the horizon. The explicit form is given by,
\begin{equation}
\phi^{(1)} =  \frac{\mathrm{i} \omega \phi^{(0)} }{2 r_0} \log \frac{r^2}{r^2 + r r_0 + r_0^2 } -\frac{\sqrt{3} \mathrm{i} \omega \phi^{(0)}}{6 r_0} \pqty{ \pi - 2 \tan^{-1}\pqty{ \frac{2r+r_0}{\sqrt{3}r_0} }  } .
\end{equation}
In contrast, we saw that since in  the extremal case $f(r)$ has a second order zero, by a suitable choice of $A^{(1)}$, only the leading order divergence can be removed, but a sub-leading logarithmic divergence remains and the integral cannot be extended directly to the horizon.  The frequency--dependent terms on LHS of eq.(\ref{radp1}) always get important sufficiently close to the horizon and the analysis has to be carried out by considering the inner and outer regions separately as above.

\subsection{Outline of Higher Order Calculations}\label{ss-ho}
Let us now sketch the procedure for higher order calculations   in the  scalar field model. We continue to work at $T=0$ in this subsection. We will see that the derivative expansion will break down at the next order with corrections going like $\omega^2\log(\omega)$.

The differential equation for $\phi^{(n)}$ in the exterior region would be given by,
\begin{equation}
\label{phino}
\frac{\mathrm{d} }{\mathrm{d} r} \pqty{r^4 f(r) \frac{\mathrm{d} \phi^{(n)}_{\mathrm{out}} }{\mathrm{d} r} } = s^{(n)}_{\mathrm{out}},
\end{equation}
where $s^{(n)}_{\mathrm{out}}$ is a source term of $\mathcal{O}(\epsilon^n)$ determined by the solution up  to the previous orders, $\phi^{(n-1)}_{\mathrm{out}}, \cdots,  \phi^{(1)}_{\mathrm{out}}$, $\phi^{(0)}_{\mathrm{out}}$, with $n$ derivatives in the $v,x$ directions acting on $\phi^{(0)}$. 
This yields the solution,
\begin{equation}
\label{nphio}
\phi^{(n)}_{\mathrm{out}} = B^{(n)}_{\mathrm{out}} - \int_r^\infty \frac{\mathrm{d} r'}{ {r'}^4 f(r')}  \pqty{ A^{(n)}_{\mathrm{out}} + \int_{r_B}^{r'} \mathrm{d} r'' \, s^{(n)}_{\mathrm{out}} (r'') }.
\end{equation}
As before, we set $B^{(n)}_{\mathrm{out}}=0$ by demanding normalisability at asymptotic infinity (it can be shown  that $s^{(n)}_{\mathrm{out}}$ falls off fast enough). 
As before, for the extremal geometry, we have to use $f(r)=f_0(r)$, eq.(\ref{f0r}).

The interior solution is obtained from an equation taking the form

\begin{equation}
\label{phin}
\frac{\mathrm{d}^2 \phi^{(n)}_{\mathrm{in}} }{\mathrm{d} {r^*}^2 } - 2 \mathrm{i}  \omega \frac{\mathrm{d} \phi^{(n)}_{\mathrm{in}} }{\mathrm{d} {r^*} } = s^{(n)}_{\mathrm{in}}.
\end{equation}
Here $r^*$ is given by eq.(\ref{rsnh}) with $F(r)$ given in eq.(\ref{Fdef}), with $T=0$. 
The source term $s^{(n)}_{\mathrm{in}}$ is obtained as follows. We noted above that $\zeta$, eq.(\ref{zeta}), is invariant under the  scaling symmetry, eq.(\ref{scalads2}).  By expressing  the source terms obtained from the lower order solutions  as a function of $\zeta$ and retaining terms of $\mathcal{O}(\epsilon^n)$   we obtain $s^{(n)}_{\mathrm{in}}$. 
Note that, to obtain the source term, one must also take into account departures from the $\mathrm{AdS}_2\times \mathbb{R}^2$ near-horizon geometry to the required order.

The solution to eq.(\ref{phin}) can be written as,
\begin{equation}
\label{inphi}
\phi^{(n)}_{\mathrm{in}} = A^{(n)}_{\mathrm{in}} + B^{(n)}_{\mathrm{in}}  e^{2\mathrm{i} \omega r^*}  + e^{2\mathrm{i} \omega r^*} \int_{-\infty}^{r^*} \mathrm{d} {r^*}' \, e^{-2\mathrm{i} \omega {r^*}' } \int_{-\infty}^{ {r^*}' } \mathrm{d} {r^*}'' \, s^{(n)}_{\mathrm{in}} ({r^*}'').
\end{equation}
One can show iteratively that  the source term $s^{(n)}_{\mathrm{in}}$ decays sufficiently fast towards the horizon, and the  integral on the RHS is well behaved and gives a vanishing contribution on the horizon. Imposing the ingoing boundary conditions on the horizon then sets $ B^{(n)}_{\mathrm{in}} = 0$.

It is worth emphasising that the inside and outside solutions have been obtained
to order $n$ with respect to two {\it different} expansions. In the outer region we simply work to the $n$\textsuperscript{th} order in the derivative expansion with respect to frequency $\omega$, as well as momentum, $k_x$. Whereas in the inner region, we work to the $n$\textsuperscript{th} order with respect to $k_x$, but with respect to $\omega$, 
keep terms  of $\mathcal{O}(\omega^n)$ after rescaling the $r^*$ dependence by $\omega$ and expressing the solution as  a function of $\zeta$ as mentioned above. 
Thus the inner solution at the $n$\textsuperscript{th} order, when expanded near the boundary actually contains higher powers of $\omega$ 
which arise due to its additional  dependence on $\zeta$ and while matching the two solutions, some of the $n$\textsuperscript{th} order terms   on the outside actually 
match against the lower order terms on the inside due to this additional dependence on $\omega$.

In fact, this feature already allows us to conclude from $\phi_{\mathrm{in}}^{(1)}$ that the integration constant $A_{\mathrm{out}}^{(2)}$ which appears in $\phi_{\mathrm{out}}^{(2)}$ must behave like 
$\omega^2 \log(\omega/\mu)$ and thus, is non-analytic in $\omega$. 

From eq.(\ref{nph2}) is follows that near $r=r_B$ 
\begin{align}
\phi_{\mathrm{in}}^{(1)} &= \frac{\mi \omega \phi^{(0)} }{3 r_h} \pqty{ 1-\frac{\mi \omega}{3(r-r_h)}} \bqty{ \pqty{ \frac{\mi \pi}{2} - \gamma +\log 3 } +  \log(\frac{r-r_h}{\omega})  - \frac{\mi \omega}{3(r-r_h)} }  \nonumber \\
&\quad + A_{\mathrm{in}}^{(1)} + \cdots \label{behphaa},
\end{align}
where we have now kept terms up to $\mathcal{O}(\omega^2)$ and the ellipsis indicates higher order corrections. In addition, there are $\mathcal{O}(\omega^2)$ terms  which are present in $\phi_{in}^{(2)}$, these are discussed in appendix \ref{app-so}) and do not contain any term going like $1/(r-r_h)$. 
From appendix \ref{app-so}, we also see that the outside solution to $\mathcal{O}(\omega^2)$ has a term going like (see eq.(\ref{phtoa}))\footnote{We have neglected the $k_x^2 \phi^{(0)}$ term in the source, which arises at this order, see eq.(\ref{radp}). The corresponding analysis is the same as that of $\phi^{(1)}$ in the first part of this section. \label{foot1}}
\be
\label{our}
\phi^{(2)}_{\mathrm{out}}=  \frac{ \omega^2 \phi^{(0)} }{9 r_h } \frac{\log (\frac{r-r_h}{r_h})}{r-r_h} - \frac{A^{(2)}_{\mathrm{out}} +\text{$r_B$--dependent terms} }{6 r_h^2 (r-r_h)} + \cdots.
\ee
Equating the coefficient of the $1/(r-r_h)$ term we  get that 
\be
\label{valaso}
A^{(2)}_{\mathrm{out}}= \frac{2 r_h \phi^{(0)} }{3} \omega^2 \log({\omega \over r_h}) + \cdots.
\ee
The additional terms  in $A^{(2)}_{\mathrm{out}}$, denoted by the ellipses, also   obtained by matching, are  analytic in $\omega$ and $\mathcal{O}(\omega^2)$. Thus, at second order, the exterior solution is not analytic in $\omega$ due to the $\log({\omega/r_h})$ enhancement in $A^{(2)}_{\mathrm{out}}$.

The argument above can be extended to higher orders and one finds that, in fact, 
\be
\label{a2outh}
A_{\mathrm{out}}^{(n)} \sim \epsilon^n \bqty{ \log({\omega\over r_h})}^{n-1}.
\ee

It  follows from these considerations then  that  the  response to the slowly varying source $\phi^{(0)}$, which is obtained from $\phi_{\mathrm{out}}$  is not analytic in the frequency, beyond the first order.
However the first order corrections in $\omega/\mu$ are analytic and the higher order contributions, while being non-analytic, are small for ${\omega /\mu}\ll 1$. Similarly, one can show that  as far as correction in the momentum,  which go   like $(k_x/\mu)^n$ are concerned, in $\phi_{\mathrm{out}}$ these are 
analytic at order $\epsilon^2$  going like $({k_x^2 / \mu^2})$, with a non-analyticity at the next order due to terms going like $\omega ({k_x^2 / \mu^2}) \log ({\omega / \mu})$. Once again, the non-analytic corrections are small.

\subsection{Finite Temperature}\label{ss-ft}
Next, we incorporate the effects of finite temperature in our discussion for the scalar field model. 
We consider a near-extremal black brane, whose temperature satisfies the condition, 
 \begin{equation}
\label{conT}
T  \ll \mu \sim r_h.
\end{equation}
We will consider frequencies $\omega$ and spatial variation characterised by momentum $k_x$ to be {\it bigger} than $T$,
\be
\label{conk}
T \ll \omega,k_x\ll \mu.
\ee
We note that this is a different regime from the conventional one considered in fluid mechanics where condition (\ref{condusual}) holds.

A  procedure similar to the one outlined above in the extremal case can be used to construct the solution, by first finding the solution in the inner and outer regions and  then matching them around $r=r_B$ which is located to satisfy, conditions (\ref{condrba}) and (\ref{condrba2}). From conditions (\ref{conk})  and (\ref{condrba}), it follows that
\begin{equation}
\label{tll}
T \ll (r_B - r_h).
\end{equation}

We will find that the inner solution is most conveniently expanded in the parameters, $\epsilon \sim \omega/\mu, k_x/ \mu$ and in terms of the variable ${\widetilde T}$ defined previously, see (\ref{valttil}) 
\be
\label{deftt}
{\widetilde T} \equiv \frac{T}{\omega},
\ee
 while
the outer solution is obtained in an expansion in ${\omega/ \mu} , {k_x/ \mu}$ and $T/ \mu$. 
In addition, we note that in the inside region perturbation theory in $\epsilon$ is defined, analogous to the zero temperature case, as follows. 
We first rescale $r^* \rightarrow  \zeta = \omega r^*$ and then keep $\zeta, \, {\widetilde T}$ fixed to obtain the required order in $\epsilon$.

In the region near $r=r_B$, we do  the  matching procedure as follows. We carry out a double expansion in both $\epsilon$ and $ {\widetilde T}$. We first expand to a given order in $\epsilon$ and then working to that order, expand in ${\widetilde T}$ to obtain the required perturbative expansion. Note that this procedure is a good approximation
when, $\epsilon$ is sufficiently small compared to ${\widetilde T}$. For example, if we retain terms $\mathcal{O}(\epsilon {\widetilde T})$ while ignoring terms $\mathcal{O}(\epsilon^2)$, this would be valid when 
\be
\label{conddd}
\epsilon \ll {\widetilde T},  
\ee
i.e., when, 
\be
\label{addend}
{\omega^2\over \mu} ,{ \omega k\over \mu}  \ll T \ll \omega, k.
\ee
We will mostly consider the regime where condition (\ref{addend}) is valid in this paper. 

Before proceeding let us mention  that in the opposite limit, when 
\be
\label{condof}
{\widetilde T} \ll \epsilon,
\ee
 one needs to carry out the double expansion in the opposite order, expanding to a given order in ${\widetilde T}$ and then expanding in $\epsilon$. The first term in this expansion to $\mathcal{O}({\widetilde T}^0)$ is the behaviour of the system in the extremal background studied in the previous section. 

We proceed below with an analysis in the limit where eq.(\ref{conddd}), eq.(\ref{addend}) are valid. We regard the inside and outside solutions as functions of $\epsilon= \omega/r_h$, $(r-r_h)/r_h$ and ${\widetilde T }$ and expand first to a given order in $\epsilon$.
Working to this order we keep the most relevant terms in an expansion in $(r-r_h)/r_h$ for purposes of equating $\phi$ and its first derivative and solve the resulting conditions to obtain 
the integration constants as a function of ${\widetilde T}$. We also note that while in general we use $\epsilon$ to denote both $\omega/\mu$ and $k_x/\mu$, in much of the following discussion it will refer to $\omega/\mu$.

Let us begin by first considering the solution to $\mathcal{O} (\epsilon^1)$. We begin with the inside solution. 
The geometry can be taken to be 
\be
\label{geomnext}
\mathrm{d} s^2 =  -6 (r-r_h) (r-r_h - T) \bqty{ 1 + \mathcal{O} \pqty{ \frac{r-r_h}{r_h}, \frac{T}{r_h} } } \dd v^2 + 2\,\dd v\, \dd r+ r^2 (\mathrm{d} x^2+\mathrm{d} y^2) .
\ee
 Note the terms in the $r$-$v$ plane which we have explicitly shown correspond to an $\mathrm{AdS}_2$ black hole; the additional corrections  of $\mathcal{O}( (r-r_h)/r_h, T/r_h)$ arise due to  
departures from the $\mathrm{AdS}_2$ black hole  geometry --- we ignore them for now, since they will lead to higher order corrections, as we will see below.

The near-horizon analysis in eqs.(\ref{phi2}--\ref{phin1}) is completely general, including both extremal and near-extremal black branes. For the near-extremal case, the expression for $r^*$ is different from eq.(\ref{rse}),
\begin{equation}
\label{rsne}
r^* =  \frac{1}{6T} \log \frac{r-r_h -T}{r - r_h}.
\end{equation}
Replacing $r_h$ by $r_+ = r_h +T$ in the inner integral of eq.(\ref{phin1}), we end up with, (instead of eq.(\ref{nph1})),
\begin{equation}
\label{phinT}
\phi^{(1)}_{\mathrm{in}} =  A^{(1)}_{\mathrm{in}}  + \frac{2\mathrm{i} \omega \phi^{(0)} }{r_h} e^{2\mathrm{i} \omega r^*} \int_{-\infty}^{r^*} \mathrm{d} {r^*}' e^{-2\mathrm{i} \omega {r^*}' } \frac{T }{e^{-6 {r^*}' T} -1 }.
\end{equation}
Now, in fact, it is even more clear than the zero-temperature case that the response is well-behaved towards the horizon: the integrand above decays exponentially towards the horizon, ${r^*}' \to -\infty$. 

Let us analyse the integral above in more detail. Let
\begin{equation}
\label{irto}
I^{(1)}_{\mathrm{in}} (r^*; T, \omega) \equiv \int_{-\infty}^{r^*} \mathrm{d} {r^*}' e^{-2\mathrm{i} \omega {r^*}' } \frac{T }{e^{-6 {r^*}' T} -1 }.
\end{equation}
This integral can be expressed in terms of  the Gauss hypergeometric function,
\begin{equation}
\label{ihypgeo}
I^{(1)}_{\mathrm{in}} (r^*; T, \omega) = \frac{1}{6 \pqty{ 1- \frac{\mi \omega}{3T} }} e^{-2\mi \omega r^*} e^{6 T r^*} \, {}_2 F_1 \pqty{ 1,  \,1- \frac{\mi \omega}{3T} , \, 2 -\frac{\mi \omega}{3T}; \, e^{6 T r^*} }.
\end{equation}

We can, of course, analyse the features of this function by examining the properties of hypergeometric functions \cite{AShandbook}. Let us instead adopt an approach that is more transparent. We first take the partial derivative of $I^{(1)}_{\mathrm{in}} (r^*; T,\omega)$ with respect to $\omega$ and split the integral into two parts:
\begin{equation}
\label{delIsp}
\frac{\partial I^{(1)}_{\mathrm{in}} (r^*; T,\omega)}{\partial \omega} = \int_{-\infty}^{0} \mathrm{d} {r^*}'  \frac{e^{-2\mathrm{i} \omega {r^*}' } (-2\mi {r^*}' ) T }{e^{-6 {r^*}' T} -1 } + \int_{0}^{r^*} \mathrm{d} {r^*}'  \frac{e^{-2\mathrm{i} \omega {r^*}' } (-2\mi {r^*}' ) T }{e^{-6 {r^*}' T} -1 } .
\end{equation}
The first term on the RHS above is manifestly independent of $r^*$ and is, in fact, a representation of the polygamma function $\psi^{(1)}$ \cite{AShandbook}. The integrand in the second integral above can be expanded term by term in small $\omega {r^*}'$ and $T {r^*}'$ near $r=r_B$, where the conditions (\ref{condrba}) and (\ref{tll}) hold. We easily obtain, then, around $r=r_B$,
\begin{equation}
\label{delIsp2}
\frac{\partial I^{(1)}_{\mathrm{in}} (r^*; T,\omega)}{\partial \omega} = \frac{\mi}{18 T} \psi^{(1)} \pqty{ 1- \frac{\mi \omega}{3T} } + r^* \bqty{ \frac{\mi}{3}  +  \mathcal{O}\pqty{ { \omega r^*,  T r^*}} }.
\end{equation} 
Integrating this function with $\omega$ yields,
\begin{equation}
\label{Ispi}
I^{(1)}_{\mathrm{in}} (r^*; T,\omega)  = - \frac{1}{6} \psi \pqty{ 1- \frac{\mi \omega}{3T} } + \omega r^* \bqty{ \frac{\mi}{3}  +  \mathcal{O}\pqty{ \omega r^*, T r^*}  } + I_2 (r^*; T).
\end{equation}
where $I_2(r^*,T)$ is the integration ``constant'' and 
 $\psi$ is the digamma function \cite{AShandbook}. We can evaluate $I_2$ by taking $\omega\to 0$ limit of (\ref{Ispi}) above and noting that,
\begin{equation}
\label{omeo}
I^{(1)}_{\mathrm{in}} (r^*; T,\omega = 0) = -\frac{1}{6} \log (1 - e^{6Tr^*} ),
\end{equation}
and from \citep{AShandbook},
\begin{equation}
\label{psi1}
\psi(1) = -\gamma.
\end{equation}
We then obtain from eq.(\ref{rsne}),
\begin{align}
I_2 (r^*; T) &= - \frac{1}{6} \bqty{ \gamma +  \log (1 - e^{6Tr^*} ) } \nonumber \\
&= - \frac{1}{6} \gamma + \frac{1}{6} \log( \frac{r-r_h}{T} ) . \label{I2}
\end{align}

In summary, near $r=r_B$, we have,
\begin{equation}
\label{iine}
I^{(1)}_{\mathrm{in}} (r^*; T,\omega) = \frac{1}{6}  \bqty{\log ({{r-r_h}\over T}) - \psi   \pqty{ 1- \frac{\mi \omega}{3T} } -\gamma   } + \mathcal{O} \pqty{ \frac{\omega}{r-r_h}  } .
\end{equation}
From eq.(\ref{phinT}), we therefore get the solution to be
\be
\label{solina}
\phi_{\mathrm{in}}^{(1)}= A_{\mathrm{in}}^{(1)} + \frac{\mathrm{i} \omega \phi^{(0)} }{3r_h} \bqty{\log ({{r-r_h}\over T}) - \psi   \pqty{ 1- \frac{\mi \omega}{3T} } -\gamma   }+ \mathcal{O} \pqty{ {\omega^2\over r_h(r-r_h)} }.
\ee
There are additional subheading terms on the RHS  which we are not  mentioning here.
Eq.(\ref{solina}) is the inside solution to $\mathcal{O}(\epsilon)$.

 For future reference we note that the imaginary  part of $\psi(1-i \omega/ 3 T)$, for small $T/\omega$, is  given by \cite{AShandbook},
 \begin{align}
  \Im \psi \pqty{ 1- \frac{\mi \omega}{3T} }  &= \frac{3 T}{2\omega}  - \frac{\pi}{2} \coth( \frac{\pi \omega}{3T} ) \nonumber \\
  &= - \frac{\pi}{2} + \frac{3 T}{2\omega} + \mathcal{O} \bqty{ \exp(-2\pi\omega/3T) } \label{psiasim},
 \end{align}
while the real part has an asymptotic expansion in $T/\omega$ given by ,
\begin{equation}
\label{psias}
\Re \psi \pqty{ 1- \frac{\mi \omega}{3T} }  =  \log( \frac{\omega}{3T} ) + \mathcal{O} \pqty{ \frac{T^2}{\omega^2} } .
\end{equation}

As mentioned above, we are constructing the solution in a double expansion in $\epsilon$ and ${\widetilde T}$. We see that the terms we have retained in eq.(\ref{solina}) give the correct solution to order $\epsilon$ and to  all orders in ${\widetilde T}$. In eq.(\ref{solina}), we have not included   departures from the 
$\mathrm{AdS}_2$ black hole geometry; including these will give corrections to $\phi_{\mathrm{in}}$ which are of order ${\omega T/ \mu^2}$. These terms  are of higher order since they go  like $\epsilon^2 {\widetilde T}$  --- we therefore continue to ignore them for now.

In the exterior region, the solution is still given by eq.(\ref{sphi}), where $f(r)$  also now includes the $T$-dependence.   
Near $r=r_B$ we  can obtain  $\phi_{\mathrm{out}}^{(1)}$ by carrying out  an  expansion simultaneously in $(r-r_h)/r_h$, $T/(r-r_h)$ and $T/r_h$. This gives,
\begin{align}
\phi^{(1)}_{\mathrm{out}} &= - \frac{A^{(1)}_{\mathrm{out}} - \mathrm{i} \omega \phi^{(0)} (r_B^2 - r_h^2) }{6r_h^2(r-r_h)}\pqty{ 1 + \frac{T}{2(r-r_h)} }  + \frac{2A^{(1)}_{\mathrm{out}} - \mathrm{i} \omega \phi^{(0)} (2r_B^2 + r_h^2) }{9r_h^2 (r-r_h)} \frac{T}{r_h} \nonumber \\
&\quad -\pqty{ \frac{A^{(1)}_{\mathrm{out}}  - \mathrm{i} \omega \phi^{(0)} (r_B^2 + 2r_h^2 ) }{9 r_h^3} - \frac{3A^{(1)}_{\mathrm{out}}  - \mathrm{i} \omega \phi^{(0)} (3r_B^2 + 2r_h^2 ) }{18 r_h^3} \frac{T}{r_h} }\log \frac{r-r_h}{r_h} + \cdots. \label{phoT}
\end{align}
As explained at the beginning of this subsection, to match the inside and outside solutions, we  work to a given order in  $\epsilon$, and then 
expand the solution in terms of  $(r-r_h)/r_h$, ${\widetilde T}$. We are working here to $\mathcal{O}(\epsilon)$, One can see explicitly that all the  temperature-dependent terms in eq.(\ref{phoT}) 
are of order ${\omega T / r_h^2} \sim  \epsilon^2 {\widetilde T}$
 and thus of higher order in $\epsilon$; setting these to zero 
makes $\phi^{(1)}_{\mathrm{out}}$  agree exactly with its $T=0$ value, eq.(\ref{phex1}). 

As a result, repeating the analysis from the $T=0$ case here, one finds again that since $\phi^{(1)}_{\mathrm{in}}$ has no $1/(r-r_h)$ term,  $A_{\mathrm{out}}^{(1)}$ is given by  its zero temperature value, eq.(\ref{a1out}). 
However,  
\be
\label{defa1ina}
A_{\mathrm{in}}^{(1)} = \frac{\mathrm{i} \omega \phi^{(0)} }{3r_h} \bqty{\log \frac{T}{r_h} + \psi   \pqty{ 1- \frac{\mi \omega}{3T} } + 	\gamma  - \frac{\sqrt{2} \pi + 2 \log 6 -2 \sqrt{2} \tan^{-1} \sqrt{2}}{4}  },
\ee 
and acquires a $T$-dependence. 
We see, noting eq.(\ref{psias}), that this constant has a non-analytic term going like $\omega \log(\omega/r_h)$ --- which was present at $T=0$.
 The $\log(T)$ dependence  is cancelled by a corresponding term appearing in eq.(\ref{psias}) the resulting   $T$ dependence can be obtained in an asymptotic expansion in ${\widetilde T}$ up to exponentially small corrections, eqs.(\ref{psiasim}, \ref{psias}).

 We can now go to $\mathcal{O} (\epsilon^2)$. At this order,  we will see that $A_{\mathrm{out}}^{(2)}$ acquires a $T$-dependence. 
 The analysis is similar to the $\mathcal{O} (\epsilon^2)$ case at $T=0$ and we will be brief. From eq.(\ref{phinT}), we  see that $\phi^{(1)}_{\mathrm{in}}$ is given by 
  \be
 \label{conga}
\phi^{(1)}_{\mathrm{in}}= A^{(1)}_{\mathrm{in}}  + \frac{2\mathrm{i} \omega \phi^{(0)} }{r_h} e^{2\mathrm{i} \omega r^*} I^{(1)}_{\mathrm{in}} (r^*; T, \omega).
 \ee
 Expanding $I^{(1)}_{\mathrm{in}} (r^*; T, \omega)$   in the matching region, eq.(\ref{iine}),  gives a contribution to $\phi_{\mathrm{in}}$ of $ \mathcal{O}(\epsilon^2)$  
 \be
 \label{expo}
 \phi_{\mathrm{in}}= \frac{\omega^2 \phi^{(0)} }{9r_h(r-r_h)} \bqty{\log ({{r-r_h}\over T}) - \psi   \pqty{ 1- \frac{\mi \omega}{3T} } -\gamma   }   + \frac{\omega^2 \phi^{(0)} }{9r_h(r-r_h)}.
 \ee
 In fact, these are the only terms near $r_B$ in $\phi_{in}$  going like $1/(r-r_h)$ to $\mathcal{O}(\epsilon^2)$.

 In the outside region, the corresponding terms are of the form
 \be
 \label{testa} \phi_{\mathrm{out}} \sim -\frac{A_{\mathrm{out}}^{(2)} }{6r_h^2 (r-r_h)} + \frac{\omega^2 \phi^{(0)}}{9r_h (r-r_h)}  \bqty{  \log({r-r_h\over r_h}) + \mathcal{O} \pqty{ \frac{T}{\omega} } + \mathcal{O}(1) }.
 \ee

 Matching the coefficient of the $1/(r-r_h)$ terms between the inside and outside then gives,
\be
\label{a2outxx}
A^{(2)}_{\mathrm{out}}= \frac{2 r_h \phi^{(0)} }{3} \omega^2 \bqty{ \log({\omega \over r_h}) + \mathcal{O} \pqty{\frac{T}{\omega} } + \mathcal{O}(1) }.
\ee
We see that the resulting expression has a term going like $\omega^2 \log (\omega)$ which was present at $T=0$. In addition there are corrections in a power series  in ${T/ \omega}$, arising from $\psi$, eq.(\ref{psias}). These are  small when the condition (\ref{conk})  is met, but completely non-local in terms of a derivative expansion in time derivatives. 
 
 Going to even higher orders, the corrections to $A_{\mathrm{out}}$ will continue to be non-analytic in $\omega$, as was already clear in the $T=0$ case. 
 
 Let us make one comment before ending this section. The expression for $\phi_{\mathrm{in}}^{(1)}$ obtained above, eq.(\ref{solina}) is, in fact, valid for all values of $T/\omega$, including ${\omega / T } \ll 1$, which is the limit in which conventional fluid mechanics is obtained.  We can therefore use the above analysis and also work out the solution in this limit. One finds that in this case $A_{\mathrm{out}}^{(1)}$ receives $\mathcal{O}(T/r_h)$ corrections. If we therefore na{\"i}vely extrapolate the result  obtained in the fluid mechanics limit to $T\rightarrow 0$, it will agree with the correct answer obtained above from the near-extremal analysis. This is true even though this extrapolation is, strictly speaking, not valid,  since the approximation 
 $\omega /T \ll 1$ breaks down when $T\rightarrow 0$, keeping $\omega $ fixed.  At higher orders, of course, the solution in the limit considered here, eq.(\ref{conk}), eq.(\ref{conddd}) and in the conventional fluid mechanics limit, differ, since the non-analytic behaviour  in $\omega $ we find is not present in the fluid mechanics limit.   
 
 We also note that in the limit $\omega /T \ll 1$,
 \be
 \label{exppsi}
   \psi   \pqty{ 1- \frac{\mi \omega}{3T} }  = -\gamma  - \frac{\mi \pi^2 \omega}{18 T} + \mathcal{O} \pqty{ \frac{\omega^2}{T^2} },
 \ee
 and 
 $\phi_{\mathrm{in}}^{(1)}$ eq.(\ref{solina}) therefore  takes the form 
\begin{equation}
\label{phnh7}
\phi^{(1)}_{\mathrm{in}}  = A^{(1)}_{\mathrm{in}}  +  \frac{\mathrm{i} \omega \phi^{(0)}}{3 r_h} \log (\frac{r-r_h}{T} ) + \cdots.
\end{equation}
Thus, the logarithm in this limit is cut off by $T$ instead of $\omega$, when\footnote{More generally the logarithm  is cut  off by whichever scale, $\omega$ or $T$, is bigger.} $\omega /T \ll 1$ and additional corrections are in powers of ${\omega /T}$. This shows that the  non-analyticity in $\omega$  disappears 
 in the fluid mechanics limit, as mentioned above. This is to be expected  since a derivative expansion in $\omega, \, k_x$ should be  valid in the fluid mechanics limit.

 \subsection{Background Geometry with Slowly Varying Chemical Potential and Temperature}\label{ss-varbg}
 
We have so far discussed the prototype scalar field model in a background geometry with a fixed chemical potential $\mu = \sqrt{3} r_h$ and temperature $T$. Our real interest is in gravity duals of systems with chemical potential, temperature and boost parameters varying with the boundary coordinates $x^\mu$. With this in mind,  we now turn to a brief discussion of the scalar field in a gravitational background with  $\mu$ and $T$ varying with $x^\mu$. In such cases, the metric and gauge field still assume the forms (\ref{metv}, \ref{gfv}), with the chemical potential and temperature being functions of the boundary coordinates,
\begin{equation}
\mu = \mu (x^\nu), \quad T= T(x^\nu). 
\end{equation}
Such a metric and gauge field configuration obviously does not solve the Maxwell and Einstein equations, but that is not relevant for the current discussion.

Since the chemical potential and the temperature are slowly varying on the length and time scales of our interest with,
\be
\label{narrate}
{\partial \mu \over \mu^2 }, \, \, {\partial T \over T \mu } \sim \epsilon \ll 1,
\ee
(where $\partial \mu, \partial T$ denote a  derivative generically along $x^\mu$) we can use the adiabatic approximation to construct the solution to a non-normalisable deformation of $\phi$, order by order in our approximations. In particular, it turns out that we can easily obtain the first order solution by simply substituting the parameters $r_h$ and $T$ in the solutions obtained above, with the local values for $r_h, T$ in the slowly varying case. For instance, for the extremal case, the first order solution on the inside and outside would look respectively like,
\begin{equation}
\label{phinvar}
\phi^{(1)}_{\mathrm{in}}= A^{(1)}_{\mathrm{in}} (x^\nu) + \frac{\mi \omega \phi^{(0)} }{3 r_h (x^\nu) } e^{2 \mi \omega r^* (x^\nu)} E_1 (2\mi \omega r^* (x^\nu)) ,
\end{equation} 
and,
\begin{equation}
\label{phiovar}
\phi^{(1)}_{\mathrm{out}} = -\frac{\mi \omega \phi^{(0)} }{12r_h (x^\nu) } \bqty{ 2 \sqrt{2} \pqty{ \frac{\pi}{2} - \tan^{-1} \frac{r+r_h(x^\nu)}{\sqrt{2}r_h(x^\nu)}  } - 2 \log\frac{(r-r_h(x^\nu))^2}{r^2 + 2rr_h(x^\nu) + 3r_h(x^\nu)^2} },
\end{equation}
see eqs.(\ref{nph2}) and (\ref{phex2}). Note that $r^*$ and $A^{(1)}_{\mathrm{in}}$ are functions of $x^\nu$ through their dependence on $r_h$, see eqs.(\ref{rse}) and (\ref{Ain1}). 

There is, however, a change in the second order calculations. There would now be additional source terms involving derivatives of $r_h$. The matching procedure goes through analogously and we can construct the required solution to the second order. In \S\ref{appsub-hocp} of appendix \ref{app-so}, we have discussed the matching procedure for this case.   Similar considerations would apply for a varying temperature $T$ as well, where we can replace the various $T$-dependent terms in the first order solution in \S\ref{ss-ft} with $T(x^\nu)$.

\section{Perturbative Expansion for the Metric and Gauge Field Components }\label{sec-deo}

We now extend our analysis for the metric and gauge field system. Our starting point is the near-extremal black brane solution, eqs.(\ref{metvb}) and (\ref{gfvb}) with a slowly varying temperature $T(x^\mu)$, chemical potential $\mu(x^\mu) = \sqrt{3} r_h(x^\mu)$ and velocity $u^\mu(x^\nu)$. We will be interested in situations meeting the condition (\ref{condtwo}) which is different from the condition (\ref{condusual}) met in conventional fluid mechanics. As mentioned in \S\ref{ss-rfg}, we gauge fix some components of the metric to take the form eq.(\ref{gauge}) and then  calculate  the metric and gauge field corrections to the starting configuration, eqs.(\ref{met0}) and (\ref{gauge0}).

Our basic strategy, like for the scalar field, is  to calculate the solution in a double expansion in $\epsilon$ and ${\widetilde T}$, these variables are defined in  (\ref{valeps}) and (\ref{valttil}). Note that we treat a term involving one boundary derivative as $\mathcal{O}(\epsilon)$, including those in which the derivative acts on the local temperature.
We first work to a given order in $\epsilon$ and then, at this order, carry out an expansion in ${\widetilde T}$. Retaining terms for example at $\mathcal{O}(\epsilon {\widetilde T})$ while ignoring those at $\mathcal{O}(\epsilon^2)$ would be justified when the condition (\ref{conddd}) is met. 
The resulting analysis is  quite similar to the scalar field case. We divide the geometry into the near $\mathrm{AdS}_2$ region and the far region and  carry out the appropriate perturbative expansions in both these regions. We impose ingoing boundary conditions at the horizon in the near region solution and normalisability at the $\mathrm{AdS}_4$ boundary in the far region. The full solution is then obtained by matching in the vicinity of the boundary of the near region, $r=r_B$ which meets the conditions (\ref{condrba}), (\ref{condrba2}) and  (\ref{tll}). 
In this way, we find  the full corrected solution to the Einstein-Maxwell equations in the double expansion mentioned above. 

We will find that  at  $\mathcal{O}(\epsilon^1)$, the resulting constitutive relations for the boundary stress tensor and charge current are local in the boundary variables,
and include corrections of order $\mathcal{O}(\epsilon {\widetilde T})$. At higher orders in $\epsilon$, the resulting constitutive relations are no longer local --- this is analogous to what we saw in the scalar field case. At $\mathcal{O}(\epsilon^2)$,  for example, the lack of non-locality is both due to corrections going $\omega^2 \log(\omega)$, as well as terms of the type $\omega^2 {\widetilde T}^n$ with arbitrarily high powers $n$.

It is worth noting, as we learned in the scalar case,  that quite generally, since we are working in the regime given by (\ref{conk}), as far as the outside solution is concerned, the temperature-dependent corrections at $\mathcal{O}(\epsilon)$ should  actually  be thought of as corrections at  $\mathcal{O}(\epsilon^2)$, since $\epsilon {T / r_h}=\epsilon (\omega / r_h) (T /\omega)=\epsilon^2 (T/ \omega)$. Thus, at order $\epsilon$, the $T$-dependent corrections can be dropped and the matching to fix the outside solution  can be carried out in the extremal limit itself. 

We will not go into further details here, since they are analogous to the scalar case, and only focus on the key points below.

\subsection{Dynamical Equations}
We will mostly focus on the outside region below. The resulting solutions for various modes will be obtained up to some integration constants which will need to be fixed from the inner solution by matching. We will illustrate some of these details in appendix \ref{app-nhsm}. 
In the outer region the radial dependence of the perturbations at any order can be obtained by solving  ordinary  differential equations in the radial variable alone,  locally for any value of $x^\mu$. In carrying out the analysis below it will be  convenient to go to the local rest frame at this value of $x^\mu$, i.e., where 
$u^\mu(x^\nu)= (-1, \mathbf{0} )$. 
The metric and gauge field components can then be classified into different irreducible representations of the $\mathrm{SO}(2)$ rotation group acting on the $xy$-plane: symmetric traceless tensor, vector and scalar. In the inner region we need to solve partial differential equations in the radial and time variables;
however, even here, since $u^\mu$ is slowly varying with time, we can go to the instantaneous rest frame and then carry out the analysis in the adiabatic 
approximation analogous to the scalar case described in \S\ref{ss-varbg}. 

Working mostly in the outside region we  write down the explicit form of the differential operators,  analogous to $\mathbb{D}$  (\ref{diffop}) in conventional fluid mechanics, in the different  sectors below. More details are given in appendix \ref{app-leq}. We show that several integration constants can be fixed by recourse to normalisability of the perturbations and the Landau frame conditions, eq.(\ref{landau}), and discuss how the remaining integration constants are fixed by matching with  the interior to obtain the full solution in the outside region.

We will refer to perturbations in the metric and gauge field at  $\mathcal{O}(\epsilon^n)$ by $g^{(n)}_{MN}, a^{(n)}_M$.  We also display only the radial dependence of the functions explicitly.

 \subsubsection{Tensor Sector}\label{sss-tensor}

This sector involves traceless symmetric tensors $g^{(n)}_{ij}$ of $\mathrm{SO}(2)$: $(g^{(n)}_{xx}-g^{(n)}_{yy})/2$ and $g^{(n)}_{xy}$. The dynamical equations for these modes is of the same form as that for the massless scalar considered in \S\ref{sec-toy}. In particular, the dynamical equation on the outside is given by,
\begin{equation}\
\label{ten1}
\frac{\del}{\del r} \pqty{r^4 f(r) \frac{\del}{\del r}  \pqty{ r^{-2} g^{(n)}_{ij} }  } = s^{(n)}_{ij},
\end{equation}
see eq. (\ref{phino}). This admits the solution, like eq.(\ref{nphio})
\begin{equation}
\label{sol}
r^{-2} g^{(n)}_{ij}  = c^{(n)}_{ij,2} - \int_{r}^\infty \frac{\mathrm{d} r'}{ {r'}^4 f(r') } \pqty{c^{(n)}_{ij,1} + \int^{r'}_{r_B} \mathrm{d} r'' \, s^{(n)}_{ij}(r'')  }.
\end{equation}
Assuming that $s^{(n)}_{ij}$ falls off sufficiently rapidly, demanding normalisability at the asymptotic infinity sets $c^{(n)}_{ij,2} =0$.
The dynamical equation on the inside needs to be obtained more carefully, as discussed in \S\ref{sec-toy}, since the frequency--dependent terms need to be kept, eq.(\ref{phi2}). And one then needs to match the inner and outer solutions  in the matching region, near  $r=r_B$, which satisfies the conditions (\ref{condrba}), (\ref{condrba2}) and  (\ref{tll}).
We will not go into these details here, since we can borrow directly the analysis of the scalar field system and only focus on obtaining the solution of the outside region from which the boundary stress tensor and charge current can be calculated. 

\subsubsection*{$n=1$ Case}

It was mentioned above that on general grounds  the temperature-dependent corrections at $\mathcal{O}(\epsilon)$ are, in fact, $\mathcal{O}(\epsilon^2 {\widetilde T})$ and thus can be neglected  if we are interested in the  $\mathcal{O}(\epsilon)$ solution. We will therefore work by setting $T=0$ here. 

Let us explicitly illustrate how we can fix $c^{(n)}_{ij,1}$ using the interior  analysis, when $n=1$. In this case, as we shall see later, the source $s^{(1)}_{ij}$ remains well-behaved towards the horizon. Therefore, the lower limit of the inner integral in (\ref{sol}) above can be pushed to  $r''=r_h$, see discussion in the second paragraph after eq.(\ref{Ain1}). We then  have, near $r=r_B$,
\begin{equation}
\label{gijsim}
r^{-2} g^{(1)}_{ij} \sim \frac{c^{(1)}_{ij,1}}{r-r_h} + \cdots.
\end{equation}
Note that for this mode, the interior equation (\ref{nhxy}) is same  in form as the scalar equation (\ref{phi2}).

We also saw in   \S\ref{sec-toy} that at first order in the derivatives, there are no terms going like $(r-r_h)^{-1}$ in the inner solution near $r=r_B$. Matching the inner and outer solutions then forces us  to set $c^{(1)}_{ij,1}=0$. (See also appendix \ref{app-nhsm}).

In this way,  both the constants of integration are fixed in the outer region and the full solution is obtained to be,
\begin{equation}
\label{solij}
r^{-2} g^{(1)}_{ij}  = - \int_{r}^\infty \frac{\mathrm{d} r'}{ {r'}^4 f_0(r') } \int^{r'}_{r_h} \mathrm{d} r'' \, s^{(1)}_{ij}(r'') ,
\end{equation}
where $f_0(r)$ is the $f(r)$ corresponding to extremality, defined in eq.(\ref{f0r}).

\subsubsection{Vector Sector}\label{sss-vector}

The vectors of $\mathrm{SO}(2)$ are the metric and gauge field components $g^{(n)}_{vi}$ and $A^{(n)}_i$. These components satisfy coupled second order differential equations. We find it convenient to study the $\mathsf{M}^i=0$ and $\mathsf{E}_{ri}=0$ equations. (Note that $\mathsf{E}_{ri}=0$ is  \emph{not} the constraint equation $\mathsf{E}^{r} {}_i=0$,) These equations lead to (see appendix \ref{app-leq}) 
\begin{align}
\frac{\del}{\del r} \pqty{r^2 f(r) \frac{\del}{\del r} A^{(n)}_i } + Q \frac{\del}{\del r} \pqty{ r^{-2} g^{(n)}_{iv} } &= s^{(n)}_{i,1}, \label{vec1} \\
\frac{\del}{\del r}\pqty{ r^4 \frac{\del}{\del r} \pqty{ r^{-2} g^{(n)}_{iv} } } + 4 Q \frac{\del}{\del r} A^{(n)}_i &= s^{(n)}_{i,2}. \label{vec2}
\end{align}

Both these equations can be integrated once to give,
\begin{align}
r^2 f(r) \frac{\del}{\del r} A^{(n)}_i + Q  r^{-2} g^{(n)}_{iv}  &= c^{(n)}_{i,1}+\int_{r_B}^r \mathrm{d} r' \, s^{(n)}_{i,1} (r')  \label{vec3}, \\
 r^4 \frac{\del}{\del r} \pqty{ r^{-2} g^{(n)}_{iv} } + 4 Q A^{(n)}_i &=c^{(n)}_{i,2}+  \int_{r_B}^r \mathrm{d} r' \, s^{(n)}_{i,2} (r'). \label{vec4}
\end{align}

We can use the expression (\ref{vec3}) in the equation (\ref{vec2}) to obtain a decoupled second order equation for  $g^{(n)}_{iv}$. We obtain,
\begin{equation}
\frac{\del}{\del r}\pqty{ r^4 \frac{\del}{\del r} \pqty{ r^{-2} g^{(n)}_{iv} } }  - \frac{4Q^2}{r^2 f(r)}  \pqty{ r^{-2} g^{(n)}_{iv} } =  s^{(n)}_{i,2} - \frac{4Q}{r^2 f(r)} \pqty{ c^{(n)}_{i,1}+\int^r_{r_B} \mathrm{d} r' \, s^{(n)}_{i,1} (r')}. \label{vec5}
\end{equation}

It is useful to make the transformation of variables,
\begin{equation}
\label{trvec}
 r^{-2} g^{(n)}_{iv} = f(r) H^{(n)}_i. 
\end{equation}
This transformation results in the simple differential equation,
\begin{equation}
\label{vec6}
 \frac{\del}{\del r} \pqty{r^4 f(r)^2 \frac{\del H^{(n)}_i}{\del r} } =  s^{(n)}_i,
\end{equation}
where we have defined the source to be the (known) function,
\begin{equation}
\label{srcv}
s^{(n)}_i (r) \equiv f(r) s^{(n)}_{i,2} - \frac{4Q}{r^2} \pqty{ c^{(n)}_{i,1}+\int_{r_B}^r \mathrm{d} r' \, s^{(n)}_{i,1} (r') }.
\end{equation}
The solution to eq.(\ref{vec6}) is given by,
\begin{equation}
\label{vec7}
 H^{(n)}_i =  -\int_{r}^\infty \frac{\mathrm{d} r'}{ {r'}^4 f(r')^2} \pqty{ c^{(n)}_{i,3} + \int^{r'}_{r_B} \mathrm{d} r'' \, s^{(n)}_i (r'') }  + c^{(n)}_{i,4}.
\end{equation}

The metric perturbation $g^{(n)}_{iv}$ then follows from eq.(\ref{trvec}) and the gauge field perturbation $A^{(n)}_{i}$ can then be obtained from $g^{(n)}_{iv}$
using eq.(\ref{vec4}).
We see that for each value of index $i$, there are four integration constants: $c_{i,1}^{(n)}, c_{i,2}^{(n)},c_{i,3}^{(n)}, c_{i,4}^{(n)}$.  Two of these integration constants can be fixed by demanding normalisibility at asymptotic infinity for the metric and gauge field components. It is clear that  $g^{(n)}_{iv}$ will not have any non-normalisable mode when $c^{(n)}_{i,4}=0$ (when the source falls off fast enough). Demanding normalisability of the gauge field component $A^{(n)}_i$ fixes the constant $c^{(n)}_{i,2}$  in terms of $c^{(n)}_{i,1}$ and $c^{(n)}_{i,3}$.  To fix the two remaining constants, $c_{i,1}^{(n)}$ and $c_{i,3}^{(n)}$, we have to compare with the interior analysis (in the near-extremal case) and impose additional conditions on the boundary fluid (Landau frame), as we discuss below for the $n=1$ case.

\subsubsection*{$n=1$ Case}

In this case, too, the  source terms for eqs.(\ref{vec1}--\ref{vec2}) are well-behaved towards the horizon and therefore, the lower limit of the integrals in eqs.(\ref{vec3}), (\ref{vec4}) and (\ref{srcv}) and the lower limit of the inner integral in eq.(\ref{vec7}) can be pushed to $r'$ (or $r''$) $=r_h$.

As noted at the beginning of  this section, for the $\mathcal{O}(\epsilon)$ corrections in the outside region, we can work in the extremal limit itself. 
We  find that 
\begin{equation}
\label{Hi1}
 H^{(1)}_i \sim \frac{c^{(1)}_{i,3} }{(r-r_h)^3} + \cdots.
\end{equation}
Since $g^{(1)}_{iv}$ is related to $H^{(1)}_i$ through eq.(\ref{trvec}), we  have, near $r=r_B$,
\begin{equation}
\label{giv1}
g^{(1)}_{iv} \sim \frac{c^{(1)}_{i,3} }{(r-r_h)} + \cdots.
\end{equation}
Since $A^{(1)}_i$ is determined by acting with a radial derivative on $g^{(1)}_{iv}$, eq.(\ref{vec4}), we have,
\begin{equation}
\label{Ai1}
A^{(1)}_{i} \sim \frac{c^{(1)}_{i,3} }{(r-r_h)^2} + \cdots.
\end{equation}
The ellipses in these equations denote terms which are subdominant in $(r-r_h)/r_h$. 

We thus see that the leading behaviour in both eq.(\ref{giv1}) and eq.(\ref{Ai1}) is associated with a non-zero $c^{(1)}_{i,3} $. In appendix \ref{app-nhsm}, we have analysed the near-horizon differential equations for these modes. It turns out that such a leading term does not arise when we impose ingoing boundary conditions. Therefore, ingoing boundary conditions set $c^{(1)}_{i,3} =0$. As mentioned  before, requiring normalisability of $g^{(1)}_{iv}$  sets  $c^{(1)}_{i,4} =0$.

So the solution for the metric component is given by,
\begin{equation}
\label{solgiv1}
r^{-2} g^{(1)}_{iv} (r) =  -f_0 (r) \int_{r}^\infty \frac{\mathrm{d} r'}{ {r'}^4 f_0(r')^2}  \int^{r'}_{r_h} \mathrm{d} r'' \, s^{(1)}_i (r'') .
\end{equation}
Here $s^{(1)}_i$ is given in eq.(\ref{srcv}) for $n=1$, with the appropriate replacements for the extremal background, $f(r) = f_0 (r)$ and $Q=\sqrt{3}r_h^2$.  On the other hand, the gauge field component is given by,
\begin{equation}
\label{solAi1}
 A^{(1)}_i = \frac{1}{4\sqrt{3}r_h^2} \pqty{ c^{(1)}_{i,2}+  \int_{r_h}^r \mathrm{d} r' \, s^{(1)}_{i,2} (r') -  r^4 \frac{\del}{\del r} \pqty{ r^{-2} g^{(1)}_{iv} (r) } }.
\end{equation}

Now, as mentioned before, we can use normalisability of the gauge field component to fix $c^{(1)}_{i,2}$, in terms of  $c^{(1)}_{i,1}$. This leaves one undetermined constant, $c^{(1)}_{i,1}$ which we fix by going to the Landau frame, eq.(\ref{landau}), as will be illustrated explicitly in the linearised analysis in \S\ref{sss1-vector}. All the constants are then fixed.

\subsubsection{Scalar Sector}\label{sss-scalar}

The metric and gauge field components relevant in this sector are $g^{(n)}_{vv}$, $g^{(n)}_{vr}$ and $A^{(n)}_{v}$. The trace part of the metric $g^{(n)}_{xx} + g^{(n)}_{yy}$ is determined in terms of $g^{(n)}_{vr}$ through the last of the gauge condition (\ref{gauge}). The Einstein equation $\mathsf{E}_{rr} = 0$ (see eq.(\ref{Err})) leads to a simple dynamical equation,

\begin{equation}
\label{gnvr}
 \frac{\del}{\del r} \pqty{ r^4 \frac{\del}{\del r} g^{(n)}_{vr}  } = s^{(n)}_{vr}.
\end{equation}
The solution to this equation is easily seen to be,
\begin{equation}
\label{gvrs}
g^{(n)}_{vr} =  -\int^\infty_{r} \frac{\mathrm{d} r' }{{r'}^4} \int^{r'}_{r_B} \mathrm{d} r'' \, s^{(n)}_{vr} (r'') + \frac{c^{(n)}_{vr,1}}{r^3} + c^{(n)}_{vr,2}.
\end{equation}

The next equation to solve is the Maxwell equation $\mathsf{M}^v=0$, which has the form (see eq.(\ref{Mv})),
\begin{equation}
\label{av1}
- \frac{1}{r^2}  \frac{\del}{\del r} \pqty{r^2  \frac{\del}{\del r} A^{(n)}_{v} } + \frac{2Q}{r^2} \frac{\del}{\del r} g^{(n)}_{vr} =  s^{(n)}_{v} .
\end{equation}

Since we have already solved for $g^{(n)}_{vr}$, we can absorb the second term on the left hand side above into the definition of the source term, giving us an equation,
\begin{equation}
\label{av2}
 \frac{\del}{\del r} \pqty{r^2  \frac{\del}{\del r} A^{(n)}_{v} } = {\widehat s}^{(n)}_{v} ,
\end{equation}
where ${\widehat s}^{(n)}_{v}$ includes the contribution from the second term,
\begin{equation}
\label{svhat}
{\widehat s}^{(n)}_{v} \equiv -r^2 s^{(n)}_{v} +2 Q \frac{\del}{\del r} g^{(n)}_{vr}
\end{equation}
Eq.(\ref{av2}) has  the straightforward solution,
\begin{equation}
\label{avsol}
A^{(n)}_{v} =  -\int_r^\infty \frac{\mathrm{d} r' }{ {r'}^2 } \int^{r'}_{r_B} \mathrm{d} r'' \, \widehat{s}^{(n)}_{v} (r'') + \frac{c^{(n)}_{v,1}}{r} + c^{(n)}_{v,2}
\end{equation}

Finally, from the $\mathsf{E}_{xx} + \mathsf{E}_{yy} =0$ equation, we can determine $g^{(n)}_{vv}$. Absorbing the known solutions $ A^{(n)}_{v}$ and $g^{(n)}_{vr}$ into the source term, we have an equation of the form (see eq.(\ref{Expy})),
\begin{equation}
\label{gnvv}
\frac{\del}{\del r} \pqty{ r^2 \frac{\del}{\del r}  g^{(n)}_{vv} } = s^{(n)}_{vv}.
\end{equation}
This again has the solution\footnote{Note that here we have not used $r'=\infty$ as the upper limit of the outer integral, because at first order, the source $s^{(1)}_{vv} (r)$ goes like $r$, and so $g^{(1)}_{vv} \sim r$. This behaviour of $g_{vv}^{(1)}$ is consistent with it being a normalisable deformation.}
\begin{equation}
\label{gvvs}
 g^{(n)}_{vv} = \int_{r_B}^r \frac{\mathrm{d} r'}{{r'}^2} \int_{r_B}^{r'} \mathrm{d} r'' \, s^{(n)}_{vv} (r'') + \frac{c^{(n)}_{vv,1}}{r} + c^{(n)}_{vv,2}.
\end{equation}

We see that there are  a total of six ``constants of integration'': $c^{(n)}_{vr,1}$, $c^{(n)}_{vr,2}$, $c^{(n)}_{v,1}$, $c^{(n)}_{v,2}$ and $c^{(n)}_{vv,1}$ and $c^{(n)}_{vv,2}$.

\subsubsection*{$n=1$ Case}
We have not discussed the form of the equations in the scalar sector in the inside region here. 
In section \ref{sec-hdm}, we will consider linearised perturbations and consider both the inside and outside regions at first order, i.e., $n=1$. 
It will turn out that for the scalar sector,  the analysis is quite simple. By using a residual gauge freedom, one can set  $g_{vr}^{(1)}=0$ and 
as a result, no frequency-dependent terms then appear in the dynamical equations we considered above in this sector. This means the outside solutions continue to be valid  all the way  till the horizon.

It then turns out that of the $6$ constants of integration needed to 
obtain the outside solution, $2$ can be fixed by demanding normalisibility at the boundary, setting $c^{(1)}_{vr,2} = 0 = c^{(1)}_{v,2}$. Furthermore, to ensure that the stress tensor at the asymptotic boundary is finite, we must set the $r$-independent term in $g_{vv}^{(1)}$ to zero, thus fixing $c^{(1)}_{vv,2}$ in eq.(\ref{gvvs}).  The forms of $g^{(1)}_{vr}$  (\ref{gvrs}) and $A^{(1)}_{v}$ (\ref{avsol}) are thus given respectively by (after setting $r_B=r_h$),
\begin{equation}
\label{gvrs1}
g^{(1)}_{vr} =  -\int^\infty_{r} \frac{\mathrm{d} r' }{{r'}^4} \int^{r'}_{r_h} \mathrm{d} r'' \, s^{(1)}_{vr} (r'') + \frac{c^{(1)}_{vr,1}}{r^3} ,
\end{equation}
and,
\begin{equation}
\label{avsol1}
A^{(1)}_{v} =  -\int_r^\infty \frac{\mathrm{d} r' }{ {r'}^2 } \int^{r'}_{r_h} \mathrm{d} r'' \, \widehat{s}^{(1)}_{v} (r'') + \frac{c^{(1)}_{v,1}}{r} .
\end{equation}

The remaining three constants can be fixed by imposing the Landau frame conditions, eq.(\ref{landau}), 
   and by  using a  residual gauge symmetry present at $\mathcal{O}(\epsilon)$, as discussed in \S\ref{sss1-scalar}.

\section{Linearised Perturbations  and Constitutive Relations at First Order}\label{sec-hdm}

In this section, we obtain the constitutive relations for $T_{\mu\nu}, J_\mu$  to the first order in the $\epsilon$-expansion. 
We will do this by carrying out the double expansion mentioned above, working to first order in $\epsilon$ and  at both $\mathcal{O}(\epsilon^0)$ and $\mathcal{O}(\epsilon^1)$ obtaining corrections to all orders  in ${\widetilde T}$. We had argued above that up to $\mathcal{O}(\epsilon^1)$, the corrections in the constitutive relations are local in spacetime.  Thus, for  obtaining the  $\mathcal{O}(\epsilon^1)$ corrections to the constitutive relations, we can set $T=0$ in the source terms, since any temperature dependence would be higher order in $\epsilon$. For example, a term linear in $T$ in the   first order corrections would actually be $\mathcal{O}(\epsilon (T/ \mu))$ i.e., $\mathcal{O}(\epsilon^2 {\widetilde T})$. 

To obtain the constitutive relations, we will first work out the linearised perturbations in the system.  For this, there is one important subtlety which we need to be careful about. 
It will turn out that  for obtaining all  the linearised perturbations, we need to allow  the temperature fluctuation, $\delta T$ and chemical potential fluctuation 
$\delta \mu$ to be  of the same order. 
 Noting that $\mu$ is related to $r_h$, as  given in eq.(\ref{extmu}), this means, 
\be
\label{jcp}
\delta T \sim \delta r_h.  
\ee
Note that  when eq.(\ref{jcp})  is met, the  fractional change in $T$  is much bigger than in $r_h$, since 
\be
\label{comfluc}
{\delta T\over T}={\delta T \over r_h}{r_h\over T} \sim {r_h\over T} {\delta r_h\over r_h} \gg {\delta r_h \over r_h},
\ee
and  condition (\ref{condtwo}) is valid. 
As a result, for carrying out the linearised perturbation analysis, we will need to obtain the zeroth order constitutive relations to $\mathcal{O}(T^2/\mu^2)$, as we will see below. It will turn out  that eq.(\ref{jcp}) is true for the charge diffusion mode. 
The linearised  equations  we obtain will also allow us to obtain information about other modes where, instead of eq.(\ref{jcp}), the fluctuation $\delta T$ is parametrically smaller than $\delta r_h$,
as will happen for the sound modes, see \S\ref{ss-consr} for a more complete discussion.  

From the linearised analysis, we will be able to deduce the general non-linear constitutive relations up to $\mathcal{O}(\epsilon)$ --- this will include terms of $\mathcal{O}(\epsilon {\widetilde T})$.

To carry out the linearised analysis, the zeroth order metric and gauge field that we start with is taken to be of the form 
eqs.(\ref{met0}) and (\ref{gauge0}), with 
\begin{align}
r_h(x^\sigma) &= r_h + \delta r_h \, e^{-\mathrm{i} \omega v + \mathrm{i} k_x x}, \label{drh} \\ 
T( x^\sigma) &= T + \delta T \, e^{-\mathrm{i} \omega v + \mathrm{i} k_x x}, \label{dte} \\
u_\mu (x^\sigma) &= \pqty{ -1,\, \delta \beta_x \, e^{-\mathrm{i} \omega v + \mathrm{i} k_x x} ,\, \delta \beta_y \, e^{-\mathrm{i} \omega v + \mathrm{i} k_x x} }.  \label{dumu}
\end{align}
Note that  the chemical potential $\mu(x^\sigma)$ is  given in terms of $r_h(x^\sigma)$ by eq.(\ref{extmu}). 
We see from eqs.(\ref{drh}--\ref{dumu}) that the $x^\mu$-dependent terms in  $g_{MN}^{(0)}$,  $A_M^{(0)}$ are both slowly varying and small in amplitude. 
We also note that $r_h$, $T$ and $\delta \beta_i$ appearing on the RHS of eq.(\ref{drh}--\ref{dumu}) are constants, independent of $x^\mu$. 

We will  now study the Einstein-Maxwell equations to first order in 
$\epsilon$, in the double expansion mentioned above, and also work to  first order in the amplitude, $\delta T$,  $\delta r_h$ and $\delta \beta_i$.

The resulting linearised Maxwell and Einstein equations are written down in appendix \ref{app-leq} (the notation is also explained in the appendix).
The source terms in these equations arise from the zeroth order terms varying in the $x^\mu$ directions.

We can write the zeroth order metric as 
\be
\label{zero met}
\mathrm{d} s^2=g_{MN}^{(0,0)} \, \mathrm{d} x^M \mathrm{d} x^N + h_{MN}^{(0)} (r) e^{-\mathrm{i} \omega v + \mathrm{i} k_x x} \, \mathrm{d} x^M  \mathrm{d} x^N,
\ee
and the gauge field as
\be
\label{zero a}
A_M=A_M^{(0,0)} + a_M^{(0)} (r) e^{-\mathrm{i} \omega v + \mathrm{i} k_x x},
\ee
where $g_{MN}^{(0,0)}, \, A_M^{(0,0)}$ are  given by eqs.(\ref{metvb}, \ref{gfvb}). The perturbations $h_{MN}^{(0)}, $ are chosen to satisfy the gauge conditions, eq.(\ref{gauge}).
From eqs.(\ref{drh}),  (\ref{dte})  and (\ref{dumu}), we see that the non-zero components of $h^{(0)}_{MN}$ and $a^{(0)}_{M}$ are 
\begin{align}
 h_{ir}^{(0)} &= -\delta \beta_i, \label{hir0} \\
h_{iv}^{(0)} &= r^2 \pqty{f(r)-1} \delta \beta_i, \label{hiv0} \\
a_i^{(0)} &= -g(r) \delta \beta_i,  \label{ai0}\\
h_{vv}^{(0)} &= \frac{6}{r^2} \pqty{ 2r_h^2 (r-r_h) + T r_h (2r-3r_h)} \delta r_h   +\frac{6}{r^2} \pqty{ r_h + 2 T  }  r_h (r-r_h) \delta T, \label{hvv0} \\
a_v^{(0)} &= -\frac{\sqrt{3}}{r} \bqty{ \pqty{2r_h + T} \delta r_h + (r_h+T) \delta T }. \label{av0}
\end{align}

This corresponds to the boundary stress tensor and charge current of a perfect fluid,
\begin{align}
T^{(0)}_{\mu \nu} &= \frac{1}{2} \mathcal{E} \pqty{ \eta_{\mu \nu} + 3 u_\mu u_\nu } , \label{tmn0} \\
J^{(0)}_{\mu} &= \rho u_\mu . \label{jm0}
\end{align}
Here, up to the required order,  the energy density  $\mathcal{E}$ is given by, 
\begin{align}
\mathcal{E} &= \frac{1}{8\pi G} \Big[ 4 r_h^3 + 6 r_h^2 T + 6 r_h T^2  + 6   \pqty{ (2r_h^2 + 2r_h T ) \, \delta r_h  + r_h(r_h +2 T)\, \delta T }  e^{-\mathrm{i} \omega v + \mathrm{i} k_x x} \Big],  \label{enden}
\end{align}
and the charge density $\rho$  is,
\begin{equation}
\rho = \frac{\sqrt{3}  }{8 \pi G} \bqty{ 2r_h^2 + 2r_h T  + T^2+ 2 \pqty{ \pqty{2r_h + T} \, \delta r_h + (r_h +T)\, \delta T }  e^{-\mathrm{i} \omega v + \mathrm{i} k_x x}  } . \label{charden}
\end{equation}

Let us evaluate the conservation equations (\ref{nsft}--\ref{nsfj}) inputting the zeroth order stress tensor and  current (\ref{tmn0}--\ref{jm0}).

The energy conservation equation  $\del_\mu T^{\mu 0}=0$ equation gives,
\begin{equation}
\label{enercon}
 k_x r_h (2r_h +3  T )\, \delta \beta_x - 2 \omega \bqty{  (2r_h + 2 T) \delta r_h +( r_h + 2 T) \, \delta T } = 0.
\end{equation}
On the other hand, the $\del_\mu T^{\mu \nu}=0$ equations give, for $\nu = x, y$  respectively,
\begin{equation}
\label{momconx}
k_x \bqty{  (2r_h + 2 T) \delta r_h +( r_h + 2 T) \, \delta T } - \omega r_h \pqty{ 2 r_h + 3  T  } \delta \beta_x  =0,
\end{equation}
and
\begin{equation}
\label{momcony}
\omega \, \delta \beta_y = 0,
\end{equation}
which are the  conservation equations for momentum in the $x,y$ directions respectively.
The equation for conservation of the charge current (\ref{nsfj}) gives,
\begin{equation}
\label{chargecon}
2\omega [\pqty{2r_h + T} \, \delta r_h + (r_h +T)\, \delta T ]- k_x (2r_h^2 + 2r_h T) \delta \beta_x = 0.
\end{equation}
We see that these relate $\delta T$, $\delta r_h$ and $\delta \beta_i$ to each other. 

We now  study the dynamical  Einstein and Maxwell equations in the different irreps of $\mathrm{SO}(2)$. These will involve the $\mathcal{O} (\epsilon)$ corrections to the metric and gauge fields. The first order corrections
were denoted as $g_{MN}^{(1)}, A_M^{(1)}$ above. These are related to the variables we use below by,
\be
\label{rela}
g^{(1)}_{MN}= e^{-\mi \omega v + \mi k_x x}  h^{(1)}_{MN} , \quad A^{(1)}_{M}= e^{-\mi \omega v + \mi k_x x} a^{(1)}_{M} .
\ee

\subsection{Explicit Solutions to the Dynamical Equations in the Outer Region}\label{ss-explicit}

In this subsection, drawing from the discussion in section \ref{sec-deo}, we provide the explicit solutions for the first order metric and gauge field components. Towards this purpose, we need to find out the sources appearing in various dynamical equations, arising from the zeroth order metric and gauge field perturbations,  eqs.(\ref{hir0}--\ref{av0}). 

Note that we had kept both $T$- and $\delta T$- dependent terms in these perturbations. However, as we argued above, to calculate the corrections to the metric and gauge field at $\mathcal{O}(\epsilon)$ in the perturbation expansion we are carrying out, we can drop     all the $T$-dependent terms,  in the perturbations eqs.(\ref{hir0}--\ref{av0}), while  retaining the terms involving $\delta T$. For example, in eq.(\ref{hvv0}), we drop the second term, $T r_h(2r-3r_h) \delta r_h,$ while retaining the third term $r_h^2 (r-r_h) \delta T$. The reason for this is that we are interested in obtaining the behaviour of the system, including its constitutive relations, up to $\mathcal{O}(\epsilon {\widetilde T})$, as mentioned above. 
Each perturbation, $\delta r_h, \, \delta T, \,
\delta \beta_i$ gives rise to a source with one derivative due to its spacetime variation and therefore is $\mathcal{O}(\epsilon)$. The terms which involve an additional factor of $T$, 
like  $T r_h(2r-3r_h) \delta r_h$ in eq.(\ref{hvv0}), result in a source going like   $ \sim \omega \epsilon {\widetilde T} $ which is of order $\epsilon^2 {\widetilde T}$ and therefore,  at $\mathcal{O}(\epsilon^1)$, can be dropped.

\subsubsection{Tensor Sector}\label{sss1-tensor}
This sector was discussed in \S\ref{sss-tensor}. The relevant metric components in this sector are ${h^y {}_x}^{(1)}$ and $({h^x {}_x}^{(1)} - {h^y {}_y}^{(1)} )/2 \equiv  \alpha^{(1)}$. 

These components satisfy the same equation as eq.(\ref{ten1}), with the source term
\begin{equation}
\label{s1xy}
s^{(1)}_{ij} = - 4 r \sigma_{ij} . 
\end{equation}

Here, 
\begin{equation}
\sigma_{xx} = -\sigma_{yy} = \frac{1}{2} \mi k_x \, \delta \beta_x e^{-\mathrm{i} \omega v + \mathrm{i} k_x x} , \quad \sigma_{xy}= \sigma_{yx} = \frac{1}{2} \mi k_x \, \delta \beta_y e^{-\mathrm{i} \omega v + \mathrm{i} k_x x} \label{sigmac}.
\end{equation}

We see that the source (\ref{s1xy}) is well-behaved towards the horizon. Let us define $\widetilde{\sigma}$ after factoring out the exponential dependence,
\begin{equation}
\label{sigtil}
\sigma_{ij} \equiv \widetilde{\sigma}_{ij} e^{-\mathrm{i} \omega v + \mathrm{i} k_x x}.
\end{equation}

From the discussion for $n=1$ caee in \S\ref{sss-tensor}, we thus have, (see eq.(\ref{solij}))
\begin{equation}
\label{solhxy}
 r^{-2} h_{ij}^{(1)} = 2 \widetilde{\sigma}_{ij}  \int_r^\infty \frac{\mathrm{d} r'}{ {r'}^4 f_0(r')} ({r'}^2 - r_h^2) .
\end{equation}
The metric components can explicitly be written as,
\begin{equation}
\label{hxy1full}
{h^y {}_x}^{(1)}  = \frac{\mi k_x \, \delta \beta_y }{2} \mathcal{T}(r),
\end{equation}
and
\begin{equation}
\label{alp1full}
{\alpha}^{(1)}  = \frac{\mi k_x \, \delta \beta_x }{2} \mathcal{T}(r),
\end{equation}
where,
\begin{equation}
\label{tenfun}
\mathcal{T}(r) \equiv  \frac{1}{6r_h} \bqty{ 2 \sqrt{2} \pqty{ \frac{\pi}{2} - \tan^{-1} \frac{r+r_h}{\sqrt{2}r_h}  } - 2 \log\frac{(r-r_h)^2}{r^2 + 2rr_h + 3r_h^2} }.
\end{equation}
Note the similarity with the scalar field solution (\ref{phex2}), which is expected because of the similarity of the dynamical equations (\ref{radp2}) and (\ref{ten1}).

\subsubsection{Vector Sector}\label{sss1-vector}

We now turn to the vector sector, discussed in \S\ref{sss-vector}, involving the metric and gauge field components $h_{iv}$ and $a_i$. We can write the source terms  which arise from  the perturbations, eq.(\ref{hir0})--eq.(\ref{av0}).

Let us look at the vector sector with $i=y$ first. The components $h_{yv}^{(1)}$ and $a_y^{(1)}$ satisfy eqs.(\ref{vec1}--\ref{vec2}) with the source terms,
\begin{align}
s^{(1)}_{y,1} &=  - \frac{\mathrm{i} \sqrt{3} r_h^2 \, \omega\, \delta \beta_y }{r^2}  e^{-\mathrm{i} \omega v + \mathrm{i} k_x x} \label{s1y1}, \\
s^{(1)}_{y,2} &= 2 \mathrm{i} r \, \omega \, \delta\beta_y e^{-\mathrm{i} \omega v + \mathrm{i} k_x x}. \label{s1y2}
\end{align}
Remembering the discussion for the $n=1$ case in \S\ref{sss-vector}, we thus obtain the solution from (\ref{solgiv1}),
\begin{align}
\label{hyv1}
{h^y {}_v}^{(1)} &=- f_0(r)  \int_r^\infty \frac{\mathrm{d} r'}{{r'}^4 f_0(r')^2}  \int_{r_h}^{r'} \mathrm{d} r'' \, \widetilde{s}^{(1)}_y (r'')  ,
\end{align}
where, from eq.(\ref{srcv}),
\begin{equation}
\label{sty1}
\widetilde{s}^{(1)}_y (r) = -\frac{4\sqrt{3} r_h^2}{r^2} \widetilde{c}^{(1)}_{y,1} +  2 \mathrm{i} \omega \delta \beta_y \pqty{ rf_0(r) + \frac{6r_h^3 (r-r_h)}{r^3} } .
\end{equation}
The solution for the gauge field perturbation is also obtained from eq.(\ref{solAi1}),
\begin{equation}
\label{ay1}
a^{(1)}_y  = \frac{1}{4\sqrt{3}r_h^2} \pqty{ \widetilde{c}^{(1)}_{y,2} + \mathrm{i} (r^2 - r_h^2) \omega \, \delta \beta_y - r^4 \frac{\mathrm{d}}{\mathrm{d} r} { h^y {}_v}^{(1)} }.
\end{equation}

Note the $c$'s defined with a tilde differ from those in section \ref{sec-deo} by an exponential factor,
\begin{equation}
\label{tildec}
c= e^{-\mi \omega v + \mi k_x x}  \widetilde{c} .
\end{equation}
We will continue to use this notation through the rest of this section.

One of the  constants, $\widetilde{c}^{(1)}_{y,1}$,  can be fixed  by going to the Landau frame: the $\nu=y$ component of the first equation (\ref{landau}). 
The relevant components of $T^{(1)}_{\mu \nu}$ to linear order in the perturbations are,
\begin{align}
T_{yv}^{(1)} = T_{vy}^{(1)} &= \frac{2 \sqrt{3} r_h   \widetilde{c}^{(1)}_{y,1}}{8\pi G} e^{-\mi \omega v + \mi k_x x}.  \label{Tyv1} 
\end{align}
We thus find, using (\ref{dumu}) that, to the required order,
\begin{equation}
u^\mu T^{(1)}_{\mu y} =  \frac{2 \sqrt{3} r_h   \widetilde{c}^{(1)}_{y,1}}{8\pi G} e^{-\mi \omega v + \mi k_x x}. \label{landauy}
\end{equation} 
The Landau frame condition therefore forces us to set,
\begin{equation}
\label{c1y1}
\widetilde{c}^{(1)}_{y,1} = 0.
\end{equation}

The normalisability of the gauge field component at the asymptotic boundary then sets,
\begin{equation}
\widetilde{c}^{(1)}_{y,2} = \mathrm{i} \omega \, \delta \beta_y r_h^2. \label{c1y2}
\end{equation}

The leads to  explicit forms for ${h^y {}_v}^{(1)}$ and $a_y^{(1)}$,
\begin{align}
{h^y {}_v}^{(1)} &= -2\sqrt{3}\mi \omega r_h^2 \delta \beta_y \mathcal{V}_1 (r) - \frac{\mi \omega \delta \beta_y}{r} , \label{hyv1full}\\ 
a^{(1)}_y &= -2 \sqrt{3} \mi \omega r_h^2 \delta \beta_y \mathcal{V}_2 (r). \label{ay1full}
\end{align}

where
\begin{align}
\mathcal{V}_1 (r) &\equiv \frac{f_0(r)}{72 \sqrt{3} r_h^3} \bqty{ 7 \sqrt{2} \pqty{ \frac{\pi}{2} - \tan^{-1} \frac{r+r_h}{\sqrt{2}r_h}  } - 4 \log\frac{(r-r_h)^2}{r^2 + 2rr_h + 3r_h^2} } \nonumber \\
&\quad - \frac{1}{12 \sqrt{3} r^4 r_h^2} (5r^3 -r^2 r_h -r r_h^2 + 3r_h^3), \label{vecfun1}\\ 
\mathcal{V}_2 (r) &\equiv -\frac{(r-r_h) }{72r_h^2 r} \bqty{ 7 \sqrt{2} \pqty{ \frac{\pi}{2} - \tan^{-1} \frac{r+r_h}{\sqrt{2}r_h}  } - 4 \log\frac{(r-r_h)^2}{r^2 + 2rr_h + 3r_h^2} }  - \frac{1 }{12 r_h r}. \label{vecfun2}
\end{align}

Next we deal with the vector index involving $x$. 
The source in eq.(\ref{vec2}), is simply obtained by the substitution $\delta \beta_y \to \delta \beta_x$ in eq.(\ref{s1y2}),
\begin{equation}
\label{s1x2}
s^{(1)}_{x,2} (r) = 2 \mathrm{i} r \, \omega \, \delta\beta_x e^{-\mathrm{i} \omega v + \mathrm{i} k_x x}.
\end{equation}
The source for the other equation eq.(\ref{vec1}) is, however, quite different:
\begin{equation}
\label{s1x1}
s^{(1)}_{x,1} (r) = \frac{\sqrt{3} \mathrm{i} }{r^2} \pqty{  r_h k_x \, (\delta T + 2 \delta r_h) - r_h^2 \, \omega\,\delta \beta_x } e^{-\mathrm{i} \omega v + \mathrm{i} k_x x}.
\end{equation}
The solution is then given by eq.(\ref{solgiv1}),
\begin{align}
\label{hxv11}
{h^x {}_v}^{(1)} &= -f_0(r)  \int_r^\infty \frac{\mathrm{d} r'}{{r'}^4 f_0 (r')^2}   \int_{r_h}^{r'} \mathrm{d} r'' \, \widetilde{s}^{(1)}_x (r'') ,
\end{align}
where, using (\ref{srcv}),
\begin{align}
\widetilde{s}^{(1)}_x (r) &= - \frac{12 \mathrm{i} r_h^2}{r^2}  \pqty{ \frac{1}{r_h} - \frac{1}{r} }\pqty{  r_h k_x ( \delta T + 2\delta r_h)- r_h^2 \, \omega\,\delta \beta_x } - \frac{4\sqrt{3}r_h^2}{r^2} \widetilde{c}^{(1)}_{x,1}  \nonumber  \\  &\quad  + 2 \mathrm{i} r \omega f_0(r)\, \delta \beta_x  \label{s1x}.
\end{align}
The gauge field component is given by,
\begin{equation}
a^{(1)}_x = \frac{1}{4\sqrt{3}r_h^2} \pqty{ \widetilde{c}^{(1)}_{x,2} + \mathrm{i} (r^2 - r_h^2) \omega \, \delta \beta_x - r^4 \frac{\mathrm{d}}{\mathrm{d} r} { h^x {}_v}^{(1)} }. \label{ax1}
\end{equation}

The two constants here are fixed by the Landau frame condition (\ref{landau}) and normalisability of the gauge field, as before, leading to,
\begin{align}
\widetilde{c}^{(1)}_{x,1} &= -\frac{\sqrt{3} \mathrm{i}}{2}  k_x (\delta T + 2 \delta r_h) , \label{c1x1} \\
\widetilde{c}^{(1)}_{x,2} &= \mathrm{i} \omega \, \delta \beta_x r_h^2. \label{c1x2}
\end{align}

Explicitly, the solutions read,
\begin{align}
{h^x {}_v}^{(1)} &= \sqrt{3} \mi \pqty{  r_h k_x \, (\delta T + 2 \delta r_h) - 2 r_h^2 \, \omega\,\delta \beta_x } \mathcal{V}_1 (r) - \frac{\mi \omega\,\delta \beta_x  }{r} , \label{hxv1full} \\
a_x^{(1)} &=  \sqrt{3} \mi \pqty{  r_h k_x \, (\delta T + 2 \delta r_h) - 2 r_h^2 \, \omega\,\delta \beta_x } \mathcal{V}_2 (r), \label{ax1full}
\end{align}
where $\mathcal{V}_1(r)$ and $\mathcal{V}_2 (r)$ are as defined in eqs.(\ref{vecfun1}) and (\ref{vecfun2}).

\subsubsection{Scalar Sector}\label{sss1-scalar}

This sector was discussed in \S\ref{sss-scalar} above. 
Note that we have defined 
$h_{xx} (r) + h_{yy} (r) = 2r^2 \sigma(r)$ (see eqs.(\ref{hxx}--\ref{hyy})). By the choice of the gauge (\ref{gauge}), we have,
\begin{equation}
h_{vr}^{(1)} = - \sigma^{(1)} (r) \label{hvrs}.
\end{equation}

Note  that $h_{xr}=r^2 h^{x} {}_r= -\delta \beta_x$ is a constant, so that there is no source term in eq. (\ref{gnvr}). This equation thus becomes,
\be
\label{errata2}
\pqty{ r^4 {h_{vr}^{(1)} }' (r) }'=0.
\ee

 The  solution is given by eq.(\ref{gvrs1}),
\begin{equation}
\label{s1r}
h_{vr}^{(1)}  = \frac{\widetilde{c}^{(1)}_{vr,1}}{r^3}.
\end{equation}
As discussed  in \cite{Bhattacharyya:2008jc}, the constant $\widetilde{c}^{(1)}_{vr,1}$ can be set to zero by using a  residual gauge freedom in the gauge, eq.(\ref{gauge}) which allows for the transformation
\begin{equation}
\label{resg}
r \to r \pqty{1 +  \frac{\widetilde{c}^{(1)}_{vr,1} e^{-\mathrm{i} \omega v + \mathrm{i} k_x x}}{r^3}}.
\end{equation}
In this way, we  see that $h_{vr}^{(1)}$, and hence by eq.(\ref{hvrs}), $\sigma^{(1)}(r)$ can be set to vanish.

The equation (\ref{av2}) for $a_v^{(1)}$ also turns out to be a source-free one. The solution is therefore, from eq.(\ref{avsol1}),
\begin{equation}
\label{av12}
a_v^{(1)} = \frac{\widetilde{c}_{v,1}^{(1)}}{r}.
\end{equation}
The  equation (\ref{gnvv}) then has as the source for $h_{vv}^{(1)}$,
\begin{equation}
\label{s1vv}
s^{(1)}_{vv} (r) = \pqty{ 2 \mathrm{i} r \, k_x \, \delta \beta_x + \frac{4 \sqrt{3} \widetilde{c}_{v,1}^{(1)}  r_h^2}{r^2} } e^{-\mathrm{i} \omega v + \mathrm{i} k_x x} ,
\end{equation}
with  the solution eq.(\ref{gvvs}) (with $r_B=r_h$ in the integrals),
\begin{equation}
\label{h1vv}
h^{(1)}_{vv} = \frac{\mathrm{i} (r - r_h)^2 \, k_x \, \delta \beta_x }{r} + 2\sqrt{3} r_h^2 \widetilde{c}_{v,1}^{(1)}  \pqty{\frac{1}{r} - \frac{1}{r_h}}^2 + \frac{\widetilde{c}^{(1)}_{vv,1}}{r} + \widetilde{c}^{(1)}_{vv,2}. 
\end{equation}
Finiteness of the stress tensor at the boundary requires,
\begin{equation}
\label{cvv2}
\widetilde{c}^{(1)}_{vv,2} = 2 \mi  k_x r_h \, \delta \beta_x.
\end{equation}
The constants $\widetilde{c}^{(1)}_{vv}$ and $\widetilde{c}_{v,1}^{(1)}$ are fixed by the Landau frame conditions (\ref{landau}),
\begin{align}
\widetilde{c}_{v,1}^{(1)} &= 0, \label{cv1} \\
\widetilde{c}^{(1)}_{vv,1} &= -\mathrm{i}  k_x \, \delta \beta_x  r_h^2. \label{cvv1} 
\end{align}

 Therefore, in summary, in the scalar sector,
\begin{align}
h_{vr}^{(1)} = - \sigma^{(1)} (r)  &= 0, \nonumber   \\
a_v^{(1)} &= 0 \label{scal1} , \\ 
h^{(1)}_{vv} &= \mathrm{i} r \, k_x \, \delta \beta_x. \nonumber
\end{align}

\subsection{Constitutive Relations and Dispersion Relations}\label{ss-consr}

From the explicit results of \S\ref{ss-explicit}, we have all the metric and gauge field components to construct the stress tensor and charge current in the linearised approximation,  up to 
$\mathcal{O}(\epsilon {\widetilde T })$.  
The zeroth order $T^{(0)}_{\mu\nu}$, $J^{(0)}_\mu$  were written down in eqs.(\ref{tmn0}) and (\ref{jm0}) (see also eqs.(\ref{enden}) and (\ref{charden})).  

From the analysis of linearised perturbations  above it follows that up to the order we are working, 
\begin{align}
T^{(1)}_{\mu \nu} &= -  \frac{r_h^2}{16 \pi G} e^{-\mi \omega v + \mi k_x x}   \begin{pmatrix}
0 & 0 & 0 \\
0 & \mi k_x \, \delta \beta_x & \mi k_x \, \delta \beta_y \\
0 & \mi k_x \, \delta \beta_y & -\mi k_x \, \delta \beta_x
\end{pmatrix}, \label{T1mn} \\
J^{(1)}_{\nu} &= \frac{\sqrt{3}}{4\pi G}e^{-\mi \omega v + \mi k_x x}  \pqty{ 0, \,\, \frac{1}{2} (2 \mi \omega \, \delta \beta_x r_h - \mi k_x (2\delta r_h + \delta T) ) , \, \,  \mi \omega \, \delta \beta_y \, r_h  } . \label{J1m1}
\end{align}
In eq.(\ref{T1mn}) and eq.(\ref{J1m1}), the order of the components refers to   $v$, $x$ and $y$ respectively. 

As an aside, we note that we will \emph{not} substitute the zeroth order conservation relations (\ref{enercon}--\ref{chargecon}) to simplify the constitutive relations above.
Such an additional approximation is not always valid --- e.g..  in the study of linearised approximations which follows the dispersion relation for some modes 
 goes like $\omega \sim k^2$; this  makes the zeroth and first order terms comparable.

Now, using the stress tensor given by (\ref{tmn0}) and (\ref{T1mn}) and the charge current given by (\ref{jm0}) and (\ref{J1m1}), and using the conservation equations (\ref{nsft}) and (\ref{nsfj}), we obtain the four homogeneous equations, 
\begin{align}
\mi k_x r_h  (2r_h +3T)\, \delta \beta_x - 2\mi \omega \bqty{ r_h (2 \delta r_h + \delta T) + 2 T (\delta r_h + \delta T) } &= 0 \label{cont0} , \\
6 \mi k_x [ r_h (2 \delta r_h + \delta T)+ 2 T (\delta r_h+ \delta T) ] + k_x^2 r_h \delta \beta_x -6 \mi \omega r_h (2r_h + 3 T) \delta \beta_x &= 0,  \label{contx} \\
 (k_x^2 r_h -6 \mi \omega r_h (2r_h + 3 T) ) \delta \beta_y &= 0,  \label{conty}   \\
2\mi k_x r_h (r_h +T + \mi \omega)\, \delta \beta_x - 2\mi \omega [ r_h(2 \delta r_h + \delta T ) + T (\delta r_h+ \delta T) ) + k_x^2 (2\delta r_h + \delta T) &= 0 \label{conj}. 
\end{align}

The first three equations are  obtained from the stress tensor conservation equations (\ref{nsft}) for $\nu = v, \, x, \, y$ respectively and the last equation is obtained from the current conservation equation (\ref{nsfj}). See eqs.(\ref{enercon}--\ref{chargecon}) for comparison.

These give rise to four modes which  map to the four modes --- a shear mode, two sound modes and one charge diffusion mode ---  that  arise in the hydrodynamic limit of conventional fluid mechanics, where condition (\ref{condusual}) is met. Below, we obtain the dispersion relations correct up to $\mathcal{O} (\epsilon^2 \widetilde{T} )$ and relations between the perturbations correct up to $\mathcal{O} (\epsilon \widetilde{T} )$.

The shear mode,  with only $\delta \beta_y$ non-zero, has the dispersion relation, 
\begin{equation}
\label{dishea}
\omega = - \mi \frac{k_x^2}{12 r_h}.
\end{equation}
We have another pair of modes, which are related to  sound modes in the hydrodynamic limit, having the dispersion relation,
\begin{equation}
\label{disso}
\omega = \pm \frac{k_x}{\sqrt{2}} - \mi \frac{k_x^2}{24 r_h}.
\end{equation} 
In the  pair of sound modes, the perturbations are related by,
\begin{align}
\delta \beta_y &= 0,  \nonumber \\
\delta \beta_x &=   \frac{\omega}{k_x r_h } \bqty{ (2\delta r_h + \delta T) + \frac{T}{2r_h} (\delta T - 2 \delta r_h)  },
  \label{sonre}\\
 \delta T &=  \mathcal{ O }\pqty{ {k_x^2 \over T r_h} } \delta r_h \nonumber.
 \end{align}
Note that in the regime where the double expansion as described above is valid, ${\widetilde T}\sim T/k_x \gg \epsilon \sim {k_x/r_h}$.
As a result, $\delta T \sim ({\epsilon/ {\widetilde T}} ) \delta r_h $  is parametrically smaller than $\delta r_h$. To obtain $\delta T$ in terms of $\delta r_h$ in  more detail,
it turns out that  one needs to go to $\mathcal{O}(\epsilon^2)$ in the constitutive relations, due to some near-cancellations.  We will not pursue this further here. 

Finally, there is a mode, which is related to the charge diffusion mode in conventional fluid mechanics, having the dispersion relation,
\begin{equation}
\label{cddis}
\omega = - \mi \frac{k_x^2}{r_h}.
\end{equation}
In this case, the perturbations are related as,
\begin{equation}
\label{cdpert}
\delta \beta_i = 0 = 2 (r_h + T) \delta r_h + (r_h + 2T)\delta T.
\end{equation}
Note in particular that from  eq.(\ref{cdpert})
\be
\label{realaa}
\delta T =  -  2 \pqty{ 1- {T \over r_h}} \delta r_h,
\ee
thereby meeting  the condition eq.(\ref{jcp}), which we had mentioned in the beginning of the section  the charge diffusion mode would satisfy. Actually, this condition implies that the variation of the energy density is zero,
\begin{equation}
\delta \mathcal{E} =0, \label{delE}
\end{equation}
see eq.(\ref{enden}).

In conventional hydrodynamics, the linearised perturbations also consist of the shear mode, two sound modes and a charge diffusion mode, as discussed below in \S\ref{sec-amp}. Comparing with eqs.(\ref{dishea3}--\ref{discdm3}) in the limit $T\to 0$, we see that the dispersion relations for the shear and sound modes we have obtained in the limit being considered here match with those in conventional hydrodynamics, but the dispersion relation for the charge diffusion mode does not, and is in fact off by a factor of $2$. 
Actually,  it turns out that our analysis above is not valid for obtaining the dispersion relation for the charge diffusion mode  and one needs to go to the next higher order in the $\epsilon$-expansion for obtaining it correctly. Once we do so, the result we obtain for the limit we are considering  does agree with that in conventional hydrodynamics, as we explain now.

Before proceeding, let us mention that the need to 
 go to one higher order can actually be seen from an elegant argument based on thermodynamics. 
 For the charge diffusion mode the pressure and energy density do not vary, it then follows  from thermodynamic relation $T \, \mathrm{d} s=-\mu \, \mathrm{d}\rho$ that the change in the charge density is at least of $\mathcal{O}(T)$. Since the mode we are dealing with is a diffusive mode, this leads to the conclusion that one must go  to the next higher order{\footnote{We are deeply grateful to Richard Davidson for explaining the  elegant thermodynamic argument given here and also pointing out the additional terms which need to be kept to obtain the correct dispersion relation. }.

 Let us now turn to a more detailed analysis. 
To obtain eq.(\ref{cddis}), after setting $\delta \beta_i=0$, we obtain from eq.(\ref{cont0}) the relation eq.(\ref{realaa}). Substituting this relation  next in eq.(\ref{conj}) however   results in a cancellation so that eq.(\ref{conj}) takes the form, 
\begin{equation}
-2 \mi T \omega \delta r_h +  \frac{2T}{r_h} k_x^2 = 0. \label{nearcanc}
\end{equation}
We see that due to the cancellation the term containing the  $k_x^2$ factor  above is of order $\epsilon^2 T \sim \epsilon^3 {\widetilde T}$.
The constitutive relations we started with  were however obtained only by retaining terms up to order $\epsilon$. There are corrections to them of order $\epsilon^2$ and  these corrections will also yield extra terms in the conservation equations of order $\epsilon^3$. One therefore needs to include these corrections in the constitutive relations to obtain the dispersion relation for the charge diffusion mode correctly. 
As  noted already, in general, the $\mathcal{O}(\epsilon^2)$ terms in the constitutive relations will be non-local and it would seem that obtaining them would be rather involved. However, it turns out to be possible to do so for the limited purpose of studying the charge diffusion mode correctly. 
We start with the relation between the perturbations, given by eq.(\ref{cdpert}). Inputting this into the zeroth order' terms (\ref{hir0}--\ref{av0}), we find that leading non-vanishing source terms are of order $ \epsilon^2 \widetilde{T}$.

The zeroth order metric perturbations are now given by, up to order $T$,
\begin{align}
h_{vv}^{(0)} &= - \frac{6r_h^2}{r^2} T \, \delta r_h , \label{thvv0} \\
a_v^{(0)} &= - \frac{\sqrt{3}}{r} T \, \delta r_h \label{tav0}.
\end{align}
The only non-vanishing source term is one in the vector sector, for which,
\begin{equation}
\label{ts1x1}
s^{(1)}_{x,1} = \frac{ \mi \sqrt{3} k_x }{r^2} T \, \delta r_h .
\end{equation}
We can now repeat the analysis of \S\ref{sss1-vector}, with the only non-zero source term above.  Note that since we have included the leading $T$-dependent term above, we can carry out the rest of the analysis in the extremal background, imposing similar boundary conditions (normalisability of the gauge field and metric, ingoing boundary condition on the horizon and the Landau frame condition). With the source (\ref{ts1x1}), we obtain the relevant metric and gauge field perturbation to be,
\begin{align}
{h^x {}_v}^{(1)} (r) &= \sqrt{3} \mi k_x T \, \delta r_h \, \mathcal{V}_1 (r) , \label{thxv1} \\
a_x^{(1)} (r) &= \sqrt{3} \mi k_x T \, \delta r_h \, \mathcal{V}_2 (r) . \label{tax1}
\end{align}
These give rise to the corrections in the  constitutive relations (beyond zeroth order),
\begin{align}
T_{\mu \nu}^{(1)} &= 0,  \label{ttmn1} \\ 
J_{\nu}^{(1)} &=   \frac{\sqrt{3}}{4\pi G}e^{-\mi \omega v + \mi k_x x}  \pqty{ 0, \,\, -\mi k_x \frac{T}{2r_h} \delta r_h , \, \,  0 }. \label{jjmn1}
\end{align}
We now only need to look at the current conservation equation. Using the zeroth order current given by eqs.(\ref{jm0}) and (\ref{charden}) and using eq.(\ref{realaa}) to substitute $\delta T$ in terms of $\delta r_h$, we end up with now  the correct  dispersion relation 
\begin{equation}
\label{cddisc1}
\omega = - \mi  \frac{k_x^2}{2r_h}.
\end{equation}
This agrees with the result in conventional hydrodynamics limit below, eq.(\ref{discdm2}), and differs from that obtained in eq.(\ref{cddis}) by a factor of $1/2$. 
Note that since this relation is obtained by keeping all possible terms which are linear in the temperature, the further corrections to the dispersion relation will be small.

We now list the dispersion relations in coordinate-free forms in terms of the chemical potential.  We have, from eqs.(\ref{dishea}), (\ref{disso}) and (\ref{cddisc1}),
\begin{align}
\label{dishea2}
\omega &= - \mi \frac{\mathbf{k}^2}{4\sqrt{3} \mu} \quad & \text{(shear mode)}, \\
\label{disso2}
\omega &= \pm \frac{|\mathbf{k}|}{\sqrt{2}} - \mi \frac{\mathbf{k}^2}{8\sqrt{3} \mu} \quad & \text{(sound modes)}, \\
\label{discdm2}
\omega &= - \mi \frac{\sqrt{3} \mathbf{k}^2}{2\mu} \quad & \text{(charge diffusion mode)}.
\end{align}

\subsection{Summary: The Stress Tensor and the Charge Current}\label{ss-linss}

The linearised analysis allows us to surmise the  full  covariant expressions for the stress tensor and charge current up to  $\mathcal{O}(\epsilon {\widetilde T})$. These expressions are   also consistent with the Landau frame condition eq.(\ref{landau}).

The covariant stress tensor  consistent with the linearised expression in eq. (\ref{T1mn}) can be written as,
\begin{equation}
\label{t1mn}
T_{\mu \nu} = \frac{1}{2} \mathcal{E} \pqty{ \eta_{\mu \nu } + 3u_\mu u_\nu} - 2 \eta \sigma_{\mu \nu},
\end{equation}
where $\mathcal{E}$, (\ref{cm}), is given to $\mathcal{O}(\epsilon \widetilde{T})$ by,
\begin{equation}
\label{elint}
\mathcal{E} = \frac{1}{4\pi G} \pqty{ \frac{2\mu^3}{3\sqrt{3}} + T \mu^2 }.
\end{equation}
Here $\sigma_{\mu \nu}$ is the shear tensor given
by 
\begin{equation}
\sigma_{\mu \nu} \equiv P_{\mu} {}^\kappa P_{\nu} {}^{\lambda} \pqty{ \frac{\del_\lambda u_\kappa  + \del_\kappa u_\lambda }{2} - \frac{1}{2} \theta \eta_{\lambda \kappa}   }, \label{sigma}
\end{equation}
and $\theta$ is the expansion defined by,
\begin{equation}
\label{expansion}
\theta \equiv \del_\nu u^\nu.
\end{equation}

 In the linearised approximation  for the velocity field (\ref{dumu}), the non-zero components $\sigma_{\mu \nu}$ are given in eq.(\ref{sigmac}). The value of $\theta$ is given by,
 \begin{equation}
 \label{thetacomp}
 \theta = \mi k_x \, \delta \beta_x e^{-\mi \omega v + \mi k_x x}  .
\end{equation}  
 
 Also, $\eta$ is the shear viscosity, whose value, on comparison with eq.(\ref{T1mn}),  is found to be,
\begin{equation}
\label{etav}
\eta = \frac{r_h^2}{16 \pi G} = \frac{\mu^2}{48 \pi G}.
\end{equation}
Given the location of the horizon, eq.(\ref{crp}), we can find the entropy density from the Bekenstein-Hawking formula. At extremality,
\begin{equation}
\label{entd}
s =  \frac{r_h^2}{4G}.
\end{equation}
At extremality, we thus recover the famous result for the ratio of the shear viscosity to entropy density as another consistency check,
\begin{equation}
\label{etas}
\frac{\eta}{s} = \frac{1}{4\pi}.
\end{equation}

The covariant  charge current  is found to be,
\begin{equation}
\label{jm1}
J_\nu = \rho u_\nu - \chi_1 \mathfrak{a}_\nu -\chi_2 P_{\nu} {}^\lambda \del_\lambda \pqty{ \mu + \frac{\sqrt{3} T}{2}  }    ,
\end{equation}
where $\rho$ is given to $\mathcal{O}(\epsilon \widetilde{T})$ by,
\begin{equation}
\rho = \frac{\mu}{4\pi G} \pqty{ \frac{\mu}{\sqrt{3}} + T } \label{rholint}
\end{equation} 

Here, $\mathfrak{a}_\nu$ is the acceleration field defined by,
\begin{equation}
\mathfrak{a}_{\nu} \equiv u^\lambda \del_\lambda u_\nu. \label{acc}
\end{equation} 
For the velocity field (\ref{dumu}), it is given by,
\begin{equation}
\label{compacc}
\mathfrak{a}_\nu = - \mi \omega e^{-\mi \omega v + \mi k_x x} \pqty{ 0, \, \delta \beta_x , \, \delta \beta_y  }.
\end{equation}
The transport coefficients in eq.(\ref{jm1}), $\chi_1$ and $\chi_2$ are given respectively by,
\begin{align}
\chi_1 &=  \frac{\mu}{4\pi G} \label{chi1} , \\
\chi_2 &=  \frac{1}{4\pi G} \label{chi2} ,
\end{align}

Note that  in   the constitutive relations above, for $T_{\mu \nu}(x^\lambda), J_\nu (x^\lambda)$ eqs.(\ref{t1mn}) and (\ref{jm1}), $\mu, T, u^\mu$  which appear are the local values at $x^\lambda$. 
We have also verified directly by going to the local rest frame for the velocity $u^\mu$ and working beyond the linearised approximation, with   a   chemical potential  and temperature near-extremality, that eq.(\ref{t1mn}), eq.(\ref{jm1}) are correct. 

Note that the Newton constant $G$ appears in the transport coefficients above. If we reinstate the $\mathrm{AdS}_4$ length scale $L_{\mathrm{AdS}}$ in appropriate places, then $G$ would appear in the dimensionless combination $L_{\mathrm{AdS}}^2/G$, which by the holographic dictionary, measures the number of degrees of freedom of the boundary field theory. 

Let us also mention that the constitutive relations above are accurate up to $\mathcal{O}(\epsilon {\widetilde T})$; this is not sufficient to describe  the charge diffusion mode and obtain its dispersion relation. For describing this mode, one would need to include  terms of $\mathcal{O}(T^2) \sim \mathcal{O}(\epsilon^2 {\widetilde T}^2)$ in $\mathcal{E}$ and $\rho$, eqs.(\ref{enden}) and (\ref{charden}), while in eq.(\ref{elint}) we have only kept terms up to $\mathcal{O}(T)$. 
 The extra $\mathcal{O}(T^2)$ terms   were needed   in the linearised analysis since in   the charge diffusion mode  $\delta T$  which satisfies the relation, eq.(\ref{jcp}),  is  fractionally anomalously large meeting eq.(\ref{comfluc}). In addition, to describe the charge diffusion mode,  one needs to retain a  term in $J_\nu$ which is $\mathcal{O}(\epsilon T)$ as discussed above. This term, obtained in eq.(\ref{jjmn1})  for the linearised case  (when eq.(\ref{cdpert}) is met)  can be written more generally as 
  \be
  \label{extJ}
 J_\nu^{\mathrm{cd}} = -  \frac{\sqrt{3} T}{8\pi G \mu} P_\nu {}^{\lambda} \del_\lambda \mu.  
  \ee

\subsection{More Comments on First Order Fluid Dynamics Beyond Small Amplitudes}\label{sec-amp}

 The metric and gauge field up to $\mathcal{O}(\epsilon {\widetilde T})$, which give rise to the constitutive relations eqs.(\ref{t1mn}) and (\ref{jm1}) can also be written down in a covariant form and are given by,

\begin{align}
\mathrm{d} s^2 &= - r^2 f(r; \mu, T) \, u_\mu u_\nu \, \mathrm{d}  x^\mu \, \mathrm{d} x^\nu - 2 \, u_\mu \, \mathrm{d}  x^\mu\, \mathrm{d} r + r^2 P_{\mu \nu} \, \mathrm{d} x^\mu \, \mathrm{d} x^\nu  \nonumber \\
&\quad + r \theta \, u^\mu u^\nu \, \mathrm{d} x^\mu \, \mathrm{d} x^\nu + r^2 \mathcal{T}(r) \sigma_{\mu \nu} \,  \mathrm{d} x^\mu \, \mathrm{d} x^\nu \label{metglo}     \\
&\quad - 2r^2 u_\mu P^{\lambda} { }_{\nu} \bqty{ \sqrt{3}r_h \Big( \del_\lambda (2 r_h + T) + 2 r_h \mathfrak{a}_\lambda \Big) \mathcal{V}_1 (r) +   \frac{\mathfrak{a}_\lambda }{r}}   \mathrm{d} x^\mu \, \mathrm{d} x^\nu  \nonumber,
\end{align}
and, 
\begin{equation}
A = -g(r; \mu, T) u_\mu \, \mathrm{d}  x^\mu +  \sqrt{3}r_h \Big( \del_\lambda (2 r_h + T) + 2 r_h \mathfrak{a}_\lambda \Big)  \mathcal{V}_2 (r)  P^\lambda {}_\mu \mathrm{d} x^\mu . \label{gaugeg}
\end{equation}

The functions $\mathcal{T}(r)$, $\mathcal{V}_1 (r)$ and $\mathcal{V}_2(r)$ are defined in eqs.(\ref{tenfun}), (\ref{vecfun1}) and (\ref{vecfun2}) respectively and the enrgy density $\mathcal{E}$ and charge density $\rho$ are to be kept to the order $\mathcal{O}(\epsilon \widetilde{T})$, eqs.(\ref{elint}) and (\ref{rholint}). The quantities $\theta$ and $\sigma_{\mu \nu}$ were defined in eqs.(\ref{expansion}) and (\ref{sigma}).

It is interesting to compare the results with conventional fluid dynamics. We do not describe the corresponding calculations in detail, but simply state the results. The stress tensor is of the same form as eq.(\ref{t1mn}), 
\begin{equation}
\label{t1mncon}
T_{\mu \nu} = \frac{1}{2} \mathcal{E} \pqty{ \eta_{\mu \nu } + 3u_\mu u_\nu} - 2 \eta \sigma_{\mu \nu},
\end{equation}
with the value of $\eta$ given by
\begin{equation}
\label{eta2}
\eta = \frac{r_+^2}{16\pi G}.
\end{equation}
The charge current is given by,
\begin{equation}
\label{Jmucon}
J_\nu = \rho u_\nu  -\widetilde{\chi}_1 \mathfrak{a}_\nu - \widetilde{\chi}_2  P_{\nu} {}^\lambda \del_\lambda \rho ,
\end{equation}
where now, the transport coefficients are given by,
\begin{align}
\widetilde{\chi}_1 &= \frac{1}{6\pi G} \frac{Q(Q^2 + 3r_+^4)}{r_+ (Q^2 + r_+^4)} , \label{chi21} \\
\widetilde{\chi}_2 &= \frac{(Q^2 + 3r_+^4)}{3r_+ (Q^2 + r_+^4)}, \label{chi22}
\end{align}
Note that the full expressions for $\mathcal{E}$ eq.(\ref{cm}) and $\rho$ (\ref{cq}) enter eqs. (\ref{t1mncon}) and (\ref{Jmucon}) respectively. Also, we have to use the full expressions for $r_+$ and $Q$ in (\ref{eta2}) and (\ref{Jmucon}).

In conventional fluid mechanics, the dispersion relations in the rest frame of the fluid are given by, (see the definition of $\widetilde{\chi}_2$ above),
\begin{align}
\label{dishea3}
\omega &= - \mi \frac{r_+^3 \mathbf{k}^2}{3(Q^2 +r_+^4) } \quad & \text{(shear mode)}, \\
\label{disso3}
\omega &= \pm \frac{|\mathbf{k}|}{\sqrt{2}} - \mi \frac{r_+^3 \mathbf{k}^2}{6(Q^2 +r_+^4) } \quad & \text{(sound modes)} ,\\
\label{discdm3}
\omega &= - \mi  \widetilde{\chi}_2 \mathbf{k}^2 \quad & \text{(charge diffusion mode)}.
\end{align}
For a near-extremal system conventional fluid mechanics arises when 
\be
\label{condusuala}
\omega, k \ll T\ll \mu.
\ee
In this limit, the transport coefficients $\eta$, $\widetilde{\chi}_1$, $\widetilde{\chi}_2$ are given by,
\begin{align}
\eta &=  \frac{\mu^2}{48\pi G} \pqty{ 1 + \frac{2\sqrt{3}T}{\mu} +  \frac{6T^2}{\mu^2} + \cdots }, \label{etacon} \\
\widetilde{\chi}_1 &=  \frac{\mu}{4\pi G} \pqty{ 1 + \frac{\sqrt{3}T}{2\mu} +  \frac{3T^2}{4\mu^2 } + \cdots  }, \label{chi1con} \\
\widetilde{\chi}_2 &= \frac{\sqrt{3}}{2\mu} \pqty{ 1 - \frac{\sqrt{3}T}{2\mu} +  \frac{3T^2}{4\mu^2 } + \cdots  } \label{chi2con}.
\end{align}
We also have to keep $\mathcal{E}$ and $\rho$ to an appropriate order in $T/\mu$.
Interestingly, the dispersion relations eq.(\ref{dishea3})--eq.(\ref{discdm3}) at near-extremality for conventional fluid mechanics,   
eq.(\ref{condusuala}) are also given by eqs.(\ref{dishea2}--\ref{discdm2}). 
These dispersion relations have corrections at $\mathcal{O}({T/ \mu})$. 

The charge current, eq.(\ref{Jmucon}) takes the form 
\begin{equation}
\label{Jmucon2}
J_\nu = \rho u_\nu - \frac{2\mu + \sqrt{3} T }{8\pi G} \mathfrak{a}_\nu - \frac{P_\nu {}^\lambda}{4\pi G} \bqty{ \del_\lambda \mu + \frac{\sqrt{3}}{2} \pqty{1+\frac{\sqrt{3} T}{2 \mu} } \del_\lambda T }.
\end{equation}
This is to be compared with eqs.(\ref{jm1}, \ref{chi1}, \ref{chi2})  above.

For the sound  modes  discussed in \S\ref{ss-consr} above,  in  conventional fluid mechanics the  relation between the different fluctuations is given by 
\begin{align}
\delta \beta_y &= 0  \nonumber \\
\delta \beta_x &= \pm \sqrt{2}  \frac{\delta r_h}{r_h}  - \mi \frac{ k_x}{12 r_h^2} \delta r_h, \label{sonre2}\\
 \delta T &= T \frac{\delta r_h}{r_h}   \nonumber.
\end{align}
The signs in (\ref{disso3}) and (\ref{sonre2}) are correlated. This should be contrasted with eq.(\ref{sonre}) above.

It is worth commenting on the constitutive relations eqs.(\ref{t1mn}) and (\ref{jm1}) we have obtained  here in some more detail. 
The conservation equation obtained  from the constitutive relations can be thought of as dynamical equations which determine the time development of the fluid. More specifically, there are $4$ parameters, $T$, $\mu$, $\beta^i$, $(i=1,2)$ which determine the constitutive relations,  and also  four equations of stress energy and charge current conservation. These four equations   can be thought of as determining the time evolution of $T$, $\mu$, $\beta^i$,  starting with their initial values, say at  $t=0$, $T(t=0,{\mathbf{x}})$, 
 $\mu(t=0, {\mathbf{x}})$, $\beta^i(t=0, {\mathbf{x}})$. 
Since the constitutive relations were obtained only up to $\mathcal{O}(\epsilon)$ (more accurately $\mathcal{O}(\epsilon {\widetilde T})$)  above, with  corrections  at $\mathcal{O}(\epsilon^2)$ or higher orders being generically non-local in time and therefore complicated, it is natural to wonder what limitations this imposes on obtaining the time evolution of the  system.

At first sight, one might think that since the resulting conservation equations, which involve one additional derivative compared to the constitutive relations,  would be accurate up to $\mathcal{O}(\epsilon^2)$,
one should be able to  obtain the time derivatives of the four parameters mentioned above, and thus the time development of the system, accurately up to $\mathcal{O}(\epsilon^2)$. 
However, this is not true as we will see shortly, and in fact generically,  the time derivatives, ${\dot T}, {\dot \mu}$ can only be obtained up to $\mathcal{O}(\epsilon)$, more precisely $\mathcal{O}(\epsilon/{\widetilde T})$ from the constitutive relations at $\mathcal{O}(\epsilon)$.  The reasons for this are tied to our discussion of the charge diffusion mode where, as the reader will recall, we had to go to $\mathcal{O}(\epsilon^3)$ to obtain the dispersion relation.

We carry out the discussion below in the local rest frame of the fluid at $t=0$. Expanding the conservation equations  obtained from the constitutive relations eqs.(\ref{t1mn}) and (\ref{jm1}) up to leading order in the velocities $\beta^i$, we get $4$  equations from the $\mu=0, i,=1,2$ components of stress energy conservation and charge conservation respectively which take the form,  see appendix \ref{app-constime},
\be
\label{tevo}
\begin{pmatrix}
2 \pqty{ \frac{\mu^2}{\sqrt{3}} + \mu T}  &0 &0  & \pqty{ \mu^2 + 2 \sqrt{3} \mu T } \\
0 & 1 & 0 & 0 \\
0 & 0 & 1 & 0 \\
\pqty{ \frac{2 \mu}{\sqrt{3}} + T +  2\theta_{(1)}   } & 0 &0 &  \pqty{ \mu + \sqrt{3} T +  \sqrt{3} \theta_{(1)}  }
\end{pmatrix}  \begin{pmatrix} \dot{\mu} \\ \dot{\beta}^1 \\ \dot{\beta}^2 \\ \dot{T} \end{pmatrix} = \begin{pmatrix} B_1 \\ B_2\\ B_3 \\ B_4 \end{pmatrix},
\ee
where $\theta_{(1)} \equiv \del_i \beta^i$, see appendix \ref{app-constime}. In obtaining eq.(\ref{tevo}) we have actually kept terms in the perfect fluid stress tensor eq.(\ref{tmn0}) accurately up to $\mathcal{O}(T^2)\sim \epsilon^2$. This might not seem consistent at first sight since the first order corrections to the constitutive relations have been obtained only up to $\mathcal{O}(\epsilon)$. 
As we will see shortly though,  the time derivative of the temperature ${\dot T}$ will turn out to be  $\mathcal{O}(\epsilon)$ and therefore comparable to ${\dot \mu}$ and ${\dot{\beta}^i}$, even though the initial value $T \ll \mu$. Thus retaining the $\mathcal{O}(T^2)$ terms from the perfect fluid allow us to obtain the left hand side of eq.(\ref{tevo}) accurately up to $\mathcal{O}(\epsilon^2)$.
We also note that $B_1, \cdots B_4$ can be obtained accurately up to $\mathcal{O}(\epsilon^2)$ and their detailed form is given in appendix \ref{app-constime}. 

The second and third rows  in eq.(\ref{tevo}) immediately yield the values of ${\dot \beta^1}, {\dot \beta^2}$ and we see in particular that they are can be obtained reliably up to $\mathcal{O}(\epsilon^2)$. It is convenient to carry out the remaining analysis by considering the first row and a suitable  linear combination of the first and fourth rows, this gives, (see appendix \ref{app-constime} for an explanation of the notation)
\begin{align}
 \frac{\mu^2}{\sqrt{3}} \pqty{ 2\dot{\mu} + \sqrt{3} \dot{T} }+ 2\mu T \pqty{ \dot{\mu} + \sqrt{3} \dot{T} } &= B_1^{(1)} \epsilon + B_1^{(2)} \epsilon^2,  \nonumber \\
\mu T \pqty{ \dot{\mu} + \sqrt{3}  \dot{T} } - \mu \theta_{(1)} \pqty{ 2\dot{\mu} + \sqrt{3} \dot{T} } &= \hat{B}^{(2)} \epsilon^2. \label{secede}
\end{align}
Noting that  $T$ and  $\theta_{(1)}$ are $\mathcal{O}(\epsilon)$  we see  the coefficient on the LHS of the second equation in eq.(\ref{secede}) is $\mathcal{O}(\epsilon)$ and therefore small. As described in appendix \ref{app-constime},l it then follows that 
\be
\label{valah}
{\dot \mu}+\sqrt{3}  {\dot T} =  \frac{\hat{B}^{(2)} \epsilon^2}{\mu T} + \frac{\sqrt{3}}{\mu^2 T} \theta_{(1)} B_1^{(1)} \epsilon. 
\ee
and therefore the LHS of eq.(\ref{valah}) is $\mathcal{O}(\epsilon^2/T) \sim \mathcal{O}(\epsilon/{\widetilde T})$.  In contrast,  generically (for $\theta\ne 0$) $(2{\dot \mu}+\sqrt{3} {\dot T}) \sim \mathcal{O}(\epsilon)$. 
Thus, we find that generically both ${\dot \mu}, {\dot T} \sim {\epsilon / {\widetilde T}}$. We also learn that these time derivatives cannot be obtained to any higher order, in particular to $\mathcal{O}(\epsilon^2)$, reliably from the corrections to the perfect fluid terms in the constitutive relations only up to $\mathcal{O}(\epsilon)$. The $\mathcal{O}(\epsilon^2)$ uncertainties in the time derivatives this results in however cancel  out, in the linear combination
$2{\dot \mu} + \sqrt{3} {\dot T}$, which is $\mathcal{O}(\epsilon)$ as noted above,  and can actually be obtained accurately up to $\mathcal{O}(\epsilon^2)$. 

It is worth emphasising that the $\mathcal{O}(\epsilon^2)$ corrections we are missing in ${\dot T}$, ${\dot \mu}$ are not insignificant, since dissipation arises at $\mathcal{O}(\epsilon^2)$ in the system. 
The bottom line therefore is that the constitutive relations we obtained above, eqs.(\ref{t1mn}) and (\ref{jm1}) have limited utility in obtaining the full dynamics of the system. Going beyond is however complicated since it involves keeping track of non-local terms. Sometimes,  as in the linearised analysis for the charge diffusion mode, these  corrections can be tractably obtained; more generally,   we have sketched out the procedure which would needed to be followed, along the lines of the analysis in section \ref{sec-toy} to obtain them. 
We also note that since ${\dot T}\sim \mathcal{O}(\epsilon/{\widetilde T})$, starting with a very cold fluid  where condition (\ref{condtwo}) is met,  on a time scale $\delta t \sim {\widetilde T}/\mu$, in the  local rest frame of the fluid,  the temperature  will become $\sim \mathcal{O}(\epsilon)$ so that condition (\ref{condtwo})  will no longer be met.

Let us also note that the small value of the  coefficient on the LHS of second equation in eq.(\ref{secede}) is tied to the issues which arose in our discussion above for the charge diffusion mode, and our conclusion 
here that we cannot obtain the time derivatives ${\dot T}, {\dot \mu}$ more accurately than $\mathcal{O}(\epsilon/{\widetilde T})$ is tied to why we had to go to one higher order in the constitutive relation in obtaining the dispersion relation for that mode.

We end this section with one more comment. The viscosity and other transport coefficients can be calculated using  the standard Kubo formulae in linear response theory. 
However these calculations are somewhat involved in the charged case, since the different perturbations need to be decoupled by using master fields etc., see e.g., \cite{Edalati:2010hk}. 
The procedure  of solving the Einstein equations systematically in the derivative expansion provides an easier way of obtaining these coefficients, this is true both in the conventional limit (\ref{condusual}) and the unconventional one, (\ref{condtwo}) considered here. 
In fact, strictly speaking, if we want to obtain the linear response exactly at extremity, we need to first set $T=0$, turn on a small $\omega$ and then take the $\omega \rightarrow 0$ limit in the Kubo formulae. This means we would be working in the unconventional limit considered here since with $T=0$, condition (\ref{condtwo}) would be met.

\section{Time-Independent Solutions in Extremal Background}\label{sec-hs}

We considered general time-dependent solutions in the limit eq.(\ref{condtwo}) above. 
One might also be interested in the analogue of hydrostatic solutions, where there is no $\omega$ dependence but the temperature still varies slowly compared to the momentum, meeting  the condition,
\be
\label{condkka}
T\ll k \ll  \mu.
\ee
We will consider a few such solutions here, in the linearised approximation  which correspond to extremal, zero temperature, solutions that   arise when  external sources  are turned on in the boundary theory. The solutions will be of an attractor type and the perturbations due to the external sources will die away at the horizon, restoring the $\mathrm{AdS}_2\times \mathbb{R}^2$ nature of the geometry very close to the horizon. We note that there is a vast literature on the attractor mechanism by now, see, for example, \cite{Ferrara:1995ih, Ferrara:1996dd, Sen:2005wa,  Goldstein:2005hq, Astefanesei:2006dd} and \cite{Andrianopoli:2006ub, Larsen:2006xm, Sen:2007qy} for reviews.

Our starting point is the extremal RN black brane solution, eqs.(\ref{metv}, \ref{gfvb}) with the functions $f(r)=f_0(r)$, eq.(\ref{f0r}) and $g(r)=g_0(r)$, eq.(\ref{g0r}). Starting with this solution, we turn on suitable 
external sources at the boundary of $\mathrm{AdS}_4$ and find the resulting solution of the Einstein-Maxwell system in the linearised approximation. 
We  take the spatial variation of the extremal sources to be of the form $e^{\mi k_x x}$, this is then also the form of  spatial dependence of the 
resulting metric and gauge field perturbations.  The solutions we consider  when evaluated near the horizon, have a  simple power-law scaling form,
\begin{equation}
\label{had}
h_{MN}, a_M \sim \pqty{ \frac{r-r_h}{r_h} }^\Delta,
\end{equation}
where $\Delta$ the exponent is a function of the momentum $k_x$.
Here we only consider cases where $\Delta>0$ so that these solutions are of an attractor type as mentioned above. 
In addition, we will consider solutions where the momentum is much smaller than $\mu$ meeting the condition, eq.(\ref{condkka}), $k_x \ll \mu$. 

The actual strategy we adopt is to start near the  $\mathrm{AdS}_2$ horizon, and consider perturbations of a scaling type, eq.(\ref{had}) which satisfy the linearised Einstein Maxwell equations.  We then match this  to the solution in the far region, meeting condition, 
\begin{equation}
\label{farcond}
\frac{r-r_h}{r_h} \gg 1,
\end{equation}
as was done in our analysis of time-dependent situations above. This allows us then to construct the solution in the bulk in a derivative expansion; near the boundary of $\mathrm{AdS}_4$; the solution,  in general,   has non-normalisable components turned on. From these components, we  read off the resulting external sources in the boundary theory.   From the  point of view of the renormalisation  group, these perturbations in the UV CFT become irrelevant in the deep IR.

In this section, we choose to work with the gauge,
\begin{equation}
\label{hrm0}
h_{r M}  = 0 = a_r.
\end{equation} 

We find that the three perturbations, $( h^y {}_v, h^y {}_x, a_y)$ decouple from the others. We therefore consider two kinds of solutions.
The first kind,  discussed in \S\ref{ss-stn}, in which only these three perturbations are turned on - these solutions are stationary but not static. And a  second kind discussed in \S\ref{ss-stat}, in which some of the other perturbations  are turned on --- these turn out to be static. 
In the solutions below, the full perturbations would be organised as,
\begin{align}
h_{\mu \nu} &= h_{\mu \nu}^{(0)} +  h_{\mu \nu}^{(1)} +  h_{\mu \nu}^{(2)}, \label{hmnexp}\\
a_{\mu } &= a_{\mu }^{(0)} +  a_{\mu }^{(1)} +  a_{\mu }^{(2)} \label{amexp},
\end{align}
where the superscript index refers to the number of powers of $k_x$ associated with the perturbation.

\subsection{Stationary Sector}\label{ss-stn}

Let us analyse the near-horizon form of the equations of motion involving $( h^y {}_v, h^y {}_x, a_y)$. Specifically, we analyse the Einstein equations $\mathsf{E}_{xy} =0  = \mathsf{E}_{ry}$ (\ref{Exy}) and (\ref{Eyr}) and the Maxwell equation $\mathsf{M}^y = 0$ (\ref{My}). These equations  read, respectively.
\begin{align}
- \frac{r_h^2}{2} \frac{\mathrm{d}}{\mathrm{d} r} \pqty{ F(r) \frac{\mathrm{d} h^y {}_x }{\mathrm{d} r}  }  + \frac{\mi k_x}{2} r_h^2  \frac{\mathrm{d} h^y {}_v }{\mathrm{d} r} &= 0, \label{hsExy} \\
\frac{r_h^2}{2} \frac{\mathrm{d}^2 h^y {}_v}{\mathrm{d} r^2} + 2 \sqrt{3} \frac{\mathrm{d} a_y }{\mathrm{d} r} + \frac{\mi k_x}{2} \frac{\mathrm{d} h^y {}_x }{\mathrm{d} r} &= 0 ,\label{hsEyr} \\
\frac{1}{r_h^2} \frac{\mathrm{d}}{\mathrm{d} r} \pqty{ F(r) \frac{\mathrm{d} a_y }{\mathrm{d} r}  } + \sqrt{3}  \frac{\mathrm{d} h^y {}_v}{\mathrm{d} r} - \frac{k_x^2 a_y}{r_h^4} &= 0. \label{hsMy}
\end{align}
Here, $F(r) = 6 (r-r_h)^2$, the $T=0$ case of eq.(\ref{Fdef}).

It is easy to see  in  a solution of the power-law type, eq.(\ref{had}) the  exponents  which appear for the three perturbations are related as follows,
\begin{align}
a_y (r) &= c_{s,1} \pqty{ \frac{r-r_h}{r_h} }^\Delta, \nonumber \\
h^y {}_x (r) &= c_{s,2} \pqty{ \frac{r-r_h}{r_h} }^\Delta, \label{hssh} \\
h^y {}_v (r) &= c_{s,3} \pqty{ \frac{r-r_h}{r_h} }^{\Delta+1}. \nonumber
\end{align} 
The possible solutions for $\Delta$ are,
\begin{equation}
\label{dsf}
\Delta = -1, \quad 0, \quad -\frac{1}{2} \pm \frac16 \sqrt{45 + \frac{6k_x^2}{r_h^2} \pm 36 \sqrt{1 + \frac{k_x^2}{3 r_h^2}}} .
\end{equation}
Here we consider the case where 
\be
\label{valdelta}
\Delta = -\frac{1}{2} + \frac{1}{6} \sqrt{45 + \frac{6k_x^2}{r_h^2} + 36 \sqrt{1 + \frac{k_x^2}{3 r_h^2}}}
\ee
Note that of the remaining solutions above for $\Delta$, three have negative exponents,  one, with $\Delta=0$, is a gauge transformation, while  another  at small $k_x$ has an exponent $\mathcal{O}(k_x^4)$.

We expand the  value of  $\Delta$, eq.(\ref{valdelta}) 
at  small $k_x$  to get,
\begin{equation}
\label{dsf1}
\Delta \approx  1 + \frac{k_x^2}{9r_h^2}. 
\end{equation}
The constants $c_{s,1}$, $c_{s,2}$ and $c_{s,3}$, are also related with only one of them being independent. Calling $c_{s,1}\equiv c_s $ as the independent constant, we have the following form of the perturbations near the $\mathrm{AdS}_2$ boundary, written in a derivative expansion:
\begin{align}
a_y (r) &= c_s \frac{(r-r_h)}{r_h} \bqty{ 1 +  \frac{k_x^2}{9r_h^2} \log\frac{(r-r_h)}{r_h}    }, \label{hsay} \\
h^y {}_x (r) &= -c_s  \frac{\mathrm{i} k_x}{\sqrt{3} r_h^2} \frac{(r-r_h)}{r_h} ,\label{hshxy}\\
h^y {}_v (r) &= -\frac{2 \sqrt{3} c_s}{r_h}  \pqty{\frac{r-r_h}{r_h}}^2 \bqty{ 1+ \frac{k_x^2}{36r_h^2} \pqty{ 1+ 4 \log\frac{(r-r_h)}{r_h} } }. \label{hshyv}
\end{align}

We can now solve for the solution in the outer region. We will do so in a derivative expansion in $k_x$.   The zeroth order solutions on the outside, which matches with the corresponding leading terms on the boundary are given by,
\begin{align}
a^{(0)}_y (r) &= c_s \pqty{1 - \frac{r_h}{r} }, \label{hsay0} \\
{h^y {}_x}^{(0)} (r) &= 0 ,  \label{hshxy0} \\
{h^y {}_v}^{(0)} (r) &= -\frac{c_s}{\sqrt{3} r_h } f_0(r). \label{hshyv0}
\end{align}

At the first order in the derivative expansion, it turns out the equations for $a_y$ and $h^y {}_v$ are sourceless and so,
\begin{equation}
a_y^{(1)} = 0 = {h^y {}_v}^{(1)}. \label{hsay1}
\end{equation}
The equation for $h^y {}_x$ is, however, sourced at the first order by ${h^y {}_v}^{(0)}$ and we are able to write the solution explicitly by matching with the near-horizon behaviour (\ref{hshxy}) above,
\begin{equation}
\label{hshxy1}
{h^y {}_x}^{(1)} (r) =  c_s  \frac{\mathrm{i} k_x}{\sqrt{3} r_h} \pqty{ \frac{1}{r} - \frac{1}{r_h}}.
\end{equation}
To the second order in derivative expansion, there is no source for $h^y {}_x$ and so,
\begin{equation}
\label{hsxy2}
{h^y {}_x}^{(2)} (r) = 0.
\end{equation}

The sources for the Maxwell equation (\ref{vec1}) and the Einstein equation (\ref{vec2}) are given by,
\begin{align}
s^{(2)}_{y,1} &= c_s k_x^2 \frac{(r-r_h)}{r^3} e^{\mathrm{i} k_x x}, \label{s2y1} \\
s^{(2)}_{y,2} &= -\frac{c_s k_x^2}{\sqrt{3} r_h} e^{\mathrm{i} k_x x} . \label{s2y2}
\end{align}
This gives, after matching near $r=r_B$, eq.(\ref{hshyv}),
\begin{equation}
{h^y {}_v}^{(2)} =\frac{f_0(r) c_s k_x^2}{\sqrt{3} r_h} \bqty{ \frac{12 -\xi }{144 r_h^2} + \int_r^\infty \frac{(r'-r_h)^3(r'+3r_h)}{{r'}^7 f_0(r')^2} \mathrm{d} r' }, \label{hyv2}
\end{equation}
and,
\begin{equation}
a_y^{(2)} = \frac{1}{12 r_h^2} \bqty{ c_s k_x^2 \pqty{ 1-\frac{r}{r_h} } -\sqrt{3} r^4 \frac{\mathrm{d} {h^y {}_v}^{(2)} }{\mathrm{d} r}  } \label{ay2}.
\end{equation}
Here $\xi$ is a constant,
\begin{equation}
\xi  \equiv 7\sqrt{2} \pi -14\sqrt{2} \tan^{-1} \sqrt{2} + 8 \log 6 \approx 26.52 \label{xi}.
\end{equation}

We can write down closed form expressions for eqs.(\ref{hyv2}) and (\ref{ay2}) above:
\begin{align}
{h^y {}_v}^{(2)} (r) &= \frac{f_0(r) c_s k_x^2 }{72\sqrt{3} r_h^3} \Bigg[  7\sqrt{2} \pqty{ \frac{\pi}{2} - \tan^{-1} \frac{r+r_h}{\sqrt{2}r_h} } - 4\log  \frac{(r-r_h)^2}{r^2 + 2rr_h + 3r_h^2}   - \frac{\xi}{2} \nonumber \\
&\quad + \frac{6r(r-3r_h)}{r^2 + 2rr_h + 3r_h^2} \Bigg], \label{hyv2hs} \\
a_y^{(2)} (r) &= - \frac{c_s k_x^2 (r-r_h) }{72 r_h^2 r} \bqty{  7\sqrt{2} \pqty{ \frac{\pi}{2} - \tan^{-1} \frac{r+r_h}{\sqrt{2}r_h} } - 4\log  \frac{(r-r_h)^2}{r^2 + 2rr_h + 3r_h^2}   - \frac{\xi}{2} }. \label{ay2hs}
\end{align}

Note that the resulting  spacetime  is a stationary spacetime, but not static. More precisely, $\del / \del v$ is a timelike Killing vector, but is not hypersurface orthogonal, see e.g., \cite{hawking_ellis_1973, Wald:1984rg}. 
If we work in the $(t,r)$ coordinate system, eq.(\ref{metrt}),  instead of the ingoing Eddington-Finkselstein coordinates, we find that the metric  is $t$-independent but does not possess a $t\to - t$ symmetry. 

We can find the forms of the metric and gauge field on the boundary manifold from the leading asymptotic behaviour of the bulk metric and gauge field. The behaviour is given by,

\begin{equation}
\widetilde{\gamma}_{\mu \nu} \, \mathrm{d} x^\mu\, \mathrm{d} x^\nu = \eta_{\mu \nu} \, \mathrm{d} x^\mu\, \mathrm{d} x^\nu -\frac{2 c_s e^{\mathrm{i} k_x x} }{\sqrt{3} r_h} \bqty{ \frac{\mathrm{i} k_x}{r_h} \, \mathrm{d} x \,\mathrm{d} y + \pqty{ 1 + \frac{(\xi-12) k_x^2}{144 r_h^2} } \mathrm{d} v \, \mathrm{d} y  } \label{hsmsh}.
\end{equation}

The form of the boundary gauge field is given by,
\begin{equation}
\widetilde{A}_\mu \, \mathrm{d} x^\mu = c_s e^{\mathrm{i} k_x x}  \pqty{ 1 + \frac{\xi k_x^2}{144 r_h^2} }\, \mathrm{d} y . \label{hsgsh}
\end{equation}

The stress tensor is given by
\begin{equation}
8\pi G T_{\mu \nu} = 2r_h^3 \mathrm{diag} (2, \, 1, \, 1) + \tau_{\mu \nu}, \label{hstsh}
\end{equation} 
where the non-zero components of $\tau_{\mu \nu}$ are:
\begin{align}
\tau_{vy} &= \frac{4 c_s r_h^2 e^{\mathrm{i} k_x x} }{\sqrt{3} }  \pqty{ 1 + \frac{(\xi-12) k_x^2}{144 r_h^2} }, \label{tavy} \\
\tau_{xy} &= -\frac{2 \mathrm{i} k_x c_s r_h e^{\mathrm{i} k_x x} }{\sqrt{3} } . \label{taxy}
\end{align}

There is a magnetic field turned on,
\begin{equation}
\label{hsm}
\widetilde{F}_{xy} = \mathrm{i} k_x c_s e^{\mathrm{i} k_x x}  \pqty{ 1 + \frac{\xi k_x^2}{144 r_h^2} }.
\end{equation}
The charge current is given by,
\begin{equation}
\label{hsshj}
4\pi G J^\mu = \pqty{\sqrt{3} r_h^2 ,\, 0, \, -\frac{c_s k_x^2}{2 r_h} e^{\mathrm{i} k_x x}  }.
\end{equation}

It is easy to verify that the conservation equations (\ref{nsct}--\ref{nscj}) are met. In fact, the stress tensor is conserved in this sector.

\subsection{Static Sector}\label{ss-stat}

In this sector, the relevant modes, which couple between themselves are $a_x$, $a_v$, $h^x {}_v$, $h_{vv}$, $\sigma \equiv (h^x {}_x + h^y {}_y)/2$ and $\alpha \equiv (h^x {}_x - h^y {}_y)/2$. As before, it is easy to see from the near-horizon form of the equations (\ref{Mv}, \ref{Mx}, \ref{Evr}, \ref{Exr}, \ref{Expy}, \ref{Exmy}) that these components have the following power-law behaviour in the near-horizon region:
\begin{align}
a_x , \, \sigma , \, \alpha &\sim  \pqty{ \frac{r-r_h}{r_h} }^\Delta , \nonumber \\
a_v , h^x {}_v &\sim \pqty{ \frac{r-r_h}{r_h} }^{\Delta+1} , \label{anstat} \\
h_{vv} &\sim \pqty{ \frac{r-r_h}{r_h} }^{\Delta+2}. \nonumber
\end{align}
It turns out that the only non-trivial $\Delta$ for our interest, for which the perturbations decay at the horizon is given by the same $\Delta$ as in the previous subsection, (see also the discussion below eq.(\ref{valdelta}))
\begin{equation}
\label{delso}
\Delta = -\frac{1}{2} + \frac{1}{6} \sqrt{45 + \frac{6k_x^2}{r_h^2} + 36 \sqrt{1 + \frac{k_x^2}{3 r_h^2}}}  \approx 1 + \frac{k_x^2}{9r_h^2}.
\end{equation}
Furthermore, the near-horizon behaviour of the vectors and the tensor is  given  up to $\mathcal{O}(k_x^2)$ by
\begin{align}
a_x (r) &= c_a \frac{(r-r_h)}{r_h} \bqty{ 1 +  \frac{k_x^2}{9r_h^2} \log\frac{(r-r_h)}{r_h} }, \label{hsax} \\
\alpha (r) &= -c_a  \frac{\mathrm{i} k_x}{\sqrt{3} r_h^2} \frac{(r-r_h)}{r_h} , \label{hsal} \\
h^x {}_v (r) &= -\frac{2 \sqrt{3} c_a}{r_h}  \pqty{\frac{r-r_h}{r_h}}^2 \bqty{ 1+ \frac{k_x^2}{36r_h^2} \pqty{ 1+ 4 \log\frac{(r-r_h)}{r_h} } } .\label{hshxv}
\end{align}
Here, $c_a$ is a  constant.

Among the scalars, $\sigma, h_{vv}, a_v$, the perturbation $\sigma(r) = 0$ near the horizon and therefore $\sigma(r)$, when continued into the far region, remains zero to all orders in $k_x$. There is however, a non-trivial near-horizon form of the scalar components $a_v$ and $h_{vv}$,
\begin{align}
a_v (r) &= - c_a \frac{\mathrm{i} k_x}{2r_h } \pqty{ \frac{r-r_h}{r_h} }^2 , \label{hsav} \\
h_{vv} (r) &= c_a \frac{4\mathrm{i} k_x}{\sqrt{3} } \pqty{ \frac{r-r_h}{r_h} }^3 . \label{hshvv}
\end{align}

When we solve in the outside region, the zeroth order solution for the vector and tensor sector turns out to be, 
\begin{equation}
{h^x {}_v}^{(0)} (r) =  -\frac{c_a}{\sqrt{3} r_h } f_0(r), \quad a_x^{(0)} (r) = c_a \pqty{ 1 - \frac{r_h}{r}}, \quad \alpha^{(0)} (r) = 0. \label{hsv0}
\end{equation}
For the scalars, we have
\begin{equation}
\label{hss0}
a_v^{(0)} (r) = 0, \quad h_{vv}^{(0)} = 0.
\end{equation}

To the first order in the derivative expansion, only the components $\alpha, a_v, h_{vv}$ receive non-zero corrections. For $\alpha (r)$, the correction is 
\begin{equation}
\label{hsa1}
{\alpha}^{(1)} (r) =  c_a  \frac{\mathrm{i} k_x}{\sqrt{3} r_h} \pqty{ \frac{1}{r} - \frac{1}{r_h}}.
\end{equation}

The corrections to the scalars at the first order is obtained to be,
\begin{align}
a_v^{(1)} (r) &= - c_a \frac{\mathrm{i} k_x}{2r_h }  \frac{(r-r_h)^2}{r^2} ,\label{hsav1}\\
h_{vv}^{(1)} (r) &= c_a \frac{\mathrm{i} k_x (r+3r_h) (r-r_h)^3}{\sqrt{3} r_h r^3}  . \label{hshvv1}
\end{align}

The only components to receive correction to the second order are $a_x$ and $h^x {}_v$. the solutions are,
\begin{align}
{h^x {}_v}^{(2)} (r) &= \frac{f_0(r) c_a k_x^2 }{72\sqrt{3} r_h^3} \Bigg[  7\sqrt{2} \pqty{ \frac{\pi}{2} - \tan^{-1} \frac{r+r_h}{\sqrt{2}r_h} } - 4\log  \frac{(r-r_h)^2}{r^2 + 2rr_h + 3r_h^2}   - \frac{\xi}{2} \nonumber \\
&\quad + \frac{6r(r-3r_h)}{r^2 + 2rr_h + 3r_h^2} \Bigg], \label{hxv2hs} \\
a_y^{(2)} (r) &= - \frac{c_a k_x^2 (r-r_h) }{72 r_h^2 r} \bqty{  7\sqrt{2} \pqty{ \frac{\pi}{2} - \tan^{-1} \frac{r+r_h}{\sqrt{2}r_h} } - 4\log  \frac{(r-r_h)^2}{r^2 + 2rr_h + 3r_h^2}   - \frac{\xi}{2} }. \label{ax2hs}
\end{align}
Note that in this sector, the vector and tensor solutions are the same in form as those in \S\ref{ss-stn}, with the simple replacement $c_s \to c_a$.

From the full solution, we can actually deduce that the spacetime is actually static: the timelike vector $\del /\del v$, in addition to being a Killing vector, is hypersurface orthogonal. In the $(t, r)$ coordinate system, eq.(\ref{metrt}),  this solution is manifestly static. 

In this case, the boundary metric is given by,
\begin{equation}
\widetilde{\gamma}_{\mu \nu} \, \mathrm{d} x^\mu\, \mathrm{d} x^\nu = \eta_{\mu \nu} \, \mathrm{d} x^\mu\, \mathrm{d} x^\nu -\frac{c_a e^{\mathrm{i} k_x x} }{\sqrt{3} r_h} \bqty{ \frac{\mathrm{i} k_x}{r_h} \, (\mathrm{d} x^2 -\mathrm{d} y^2) + 2\pqty{ 1 + \frac{(\xi-12) k_x^2}{144 r_h^2} } \mathrm{d} v \, \mathrm{d} x  } \label{hsmso},
\end{equation} 
where $\xi$ is given by the eq.(\ref{xi}).
The form of the boundary gauge field is given by,
\begin{equation}
\widetilde{A}_\mu \, \mathrm{d} x^\mu = - c_a e^{\mathrm{i} k_x x} \frac{\mathrm{i} k_x}{2r_h} \, \mathrm{d} v +  c_a e^{\mathrm{i} k_x x}  \pqty{ 1 + \frac{\xi k_x^2}{144 r_h^2} }\, \mathrm{d} x . \label{hsgo}
\end{equation}

The stress tensor is given by
\begin{equation}
8\pi G T_{\mu \nu}  = 2r_h^3 \mathrm{diag} (2, \, 1, \, 1) + \tau_{\mu \nu}, \label{hstso}
\end{equation} 
where the nonzero components of $\tau_{\mu \nu}$ are given by,
\begin{align}
\tau_{vv} &= -2 \sqrt{3} \mathrm{i} k_x c_a r_h e^{\mathrm{i} k_x x}, \label{tavv} \\
\tau_{vx} &= \frac{4 c_a r_h^2 e^{\mathrm{i} k_x x} }{\sqrt{3} }  \pqty{ 1 + \frac{(\xi-12) k_x^2}{144 r_h^2} }, \label{tavx} \\
\tau_{xx} &= -\frac{5 \mathrm{i} k_x c_a r_h e^{\mathrm{i} k_x x} }{\sqrt{3} }  , \label{taxx} \\
\tau_{yy} &= -\frac{\mathrm{i} k_x c_a r_h e^{\mathrm{i} k_x x} }{\sqrt{3} } \label{tayy} .
\end{align}

There is a non-zero electric field turned on,
\begin{equation}
\label{hsef}
\widetilde{F}_{xv} = c_a e^{\mathrm{i} k_x x} \frac{k_x^2}{2r_h}.
\end{equation}
The charge current is given by,
\begin{equation}
\label{hsoj}
4\pi G J^\mu = \pqty{ \sqrt{3} r_h^2 - \mathrm{i} c_a k_x e^{\mathrm{i} k_x x}  ,\, 0 , 0 }.
\end{equation}

In this case, too, it can be easily verified that the conservation equations (\ref{nsct}--\ref{nscj}) are met.

\section{Discussion}\label{sec-co}

In this paper, we considered the behaviour of a near-extremal system  with  varying temperature $T$, chemical potential $\mu$, and three-velocity $u^\nu$, where  condition (\ref{condtwo}) is   satisfied so that the rate of variation is bigger than $T$ but\footnote{We remind the reader that in our notation the physical temperature $\widehat{T}$ is related to $T$ by eq.(\ref{deft}).} smaller than $\mu$. This is to be contrasted with the usual case studied in fluid mechanics where condition (\ref{condusual}) is true and the rate of variation is the smallest scale. 
We found that  a  near-extremal black brane configuration of the type, eq.(\ref{met0}), eq.(\ref{gauge0}), 
with 
\be
\label{condtu}
T(x^\sigma) \ll \mu(x^\sigma),
\ee
 is  a good starting point, even when condition (\ref{condtwo}) is satisfied,  for finding  solutions to Einstein-Maxwell  equations. Corrections  to this starting configuration can be computed systematically in a double expansion in the  parameter $\epsilon$, eq.(\ref{valeps}), and ${\widetilde T}$, eq.(\ref{valttil}), which are  small when condition (\ref{condtwo}) is met. 

At first order in $\epsilon$, the corrections can be incorporated in the boundary theory by adding additional terms in the stress tensor and charge current which  are local in spacetime and involve one spacetime derivative with a coefficient determined  by the viscosity and charge diffusivity  respectively.

At higher orders, the corrections are no longer local in spacetime --- e.g., at  second  order, there are corrections going like $\epsilon^2 \log(\epsilon)$, and  also corrections of $\mathcal{O}(\epsilon^2)$ in an asymptotic series in  ${\widetilde T}$. Despite the logarithmic enhancements, the procedure for calculating the corrections systematically continues to be valid since terms at the $n$\textsuperscript{th} order  are still smaller than those at lower orders when $\epsilon \ll 1$. 
We discuss in the paper how, in principle at least, these corrections can be computed order by order in $n$, although in practice this gets quite complicated beyond $n=1$. 

The fact that such a systematic approximation exists, even in principle, though,  is enough to justify  that the   truncation to first order is a good one for describing the constitutive relations. However, as we argue in the paper, the resulting conservations equations of stress energy and the charge current, obtained from the first order corrections are quite restricted in scope.  In particular. they only allow for the time derivatives ${\dot T}, {\dot \mu}$, in the local  rest frame of the fluid, to be calculated  up to $\mathcal{O}(\epsilon/{\widetilde T})$ --- 
this is not enough to incorporate the leading effects of dissipation in the system accurately since these effects arise from $\mathcal{O}(\epsilon^2)$ terms in the constitutive relations.

It will be interesting to try and calculate these $\mathcal{O}(\epsilon^2)$ corrections to the constitutive relations of the fluid --- while they will be non-local, they are  also constrained by the conformal symmetry of the near $\mathrm{AdS}_2$ region and should not be too difficult to obtain. With these corrections in hand, one can then hope to incorporate all effects of dissipation to  leading order  in a consistent manner and also  possibly extend the analysis to  the full region of parameter space,  $T,k, \omega \sim \epsilon \ll \mu$ without restricting to very low temperatures with    $T/\omega\ll 1$. We leave such an analysis for the future. 

An important feature about near-extremal systems that our analysis highlights is that  in determining the behaviour of the system time derivatives are on a somewhat different footing compared to spatial derivatives. This is tied to the $\mathrm{AdS}_2$ near-horizon geometry since the scaling symmetry of $\mathrm{AdS}_2$ involves the time direction but not the spatial ones. 
This feature suggests that a different regime, where frequency is the smallest scale in the problem instead of the temperature, and condition (\ref{condtwo}) is replaced by 
\begin{equation}
\label{hsfuture}
\omega \ll T \ll k \ll \mu,
\end{equation}
would also be worth studying\footnote{We thank Shiraz Minwalla for this very insightful suggestion.}. One can hope to obtain a systematic understanding of the system in an expansion in ${\omega / T}$, ${k / \mu} $ and ${T / \mu}$, which are all small,  in this limit. 
 
 We are also led to consider this regime by our result that  a very cold system meeting eq.(\ref{condtwo}), heats us generically quite  quickly acquiring a temperature $T \sim \epsilon \mu$ in a time scale of order ${{\widetilde T}/ \mu}$. 
 As the system heats up further, one would then expect to enter the regime where condition (\ref{hsfuture}) is valid and a perturbative expansion in ${\omega / T}$ becomes useful. 
 An interesting special case  to consider is hydrostatic solutions which are time independent ($\omega=0$).  Some zero-temperature static and stationary solutions  were  considered  in section \ref{sec-hs} and exhibited attractor behaviour, the more general solutions we have in mind  would be at $T\ne 0$. 
We hope to report on this work in the future \cite{MMST}.

It will also be interesting to explore whether the behaviour we have uncovered for a near-extremal black hole is exhibited in some strongly coupled field theories as well. In particular, one might expect interesting parallels with systems which in the infrared flow to a fixed point with scaling properties analogous to the $\mathrm{AdS}_2\times \mathbb{R}^2$ near-horizon geometry found here. {Our results suggest that the conclusions regarding the non-locality of the constitutive relations would be generically true for UV systems in a model-independent way, as long as there is a near-horizon $\mathrm{AdS}_2$ geometry in the IR. On the other hand, we expect the results to be different if the near-horizon geometries are different from $\mathrm{AdS}_2$.}

 As was mentioned in the introduction, we  hope to extend this analysis in subsequent work  to  the study of near-extremal Kerr black holes in asymptotically flat space, because of its  obvious observational interest. This would one of the main physical pay-offs one can hope for from this direction of research in the future.

 It should be straightforward to extend these results also to more complicated near-extremal black branes/ holes, including those which have extra gauge fields and neutral scalars. Many of these systems are well known to exhibit  attractor behaviour at extremality \cite{Ferrara:1995ih, Ferrara:1996dd, Sen:2005wa,  Goldstein:2005hq, Astefanesei:2006dd, Andrianopoli:2006ub, Larsen:2006xm, Sen:2007qy} and the interplay between attractor  behaviour and the near-extremal fluid mechanics will be worth investigating in more detail. As was mentioned above, in section \ref{sec-hs} of this paper, we briefly explored this idea for the Einstein-Maxwell system, by studying slowly spatially varying perturbations subject to external forces. We found that in some cases, the attractor behaviour continues to hold, with the perturbations dying away at the horizon resulting in an $\mathrm{AdS}_2$  geometry to good approximation close to the horizon. Another interesting direction would be to incorporate  charged scalars --- this is known to give rise to a rich new set of phenomenon, since the charged scalars can exhibit the phenomenon of superradiance \cite{Brito:2015oca}, and can also condense in the bulk \cite{Gubser:2008px,  Hartnoll:2008vx, Herzog:2009xv, Horowitz:2010gk}. {However, for such systems which do not have a near-horizon $\mathrm{AdS}_2$ geometry, the analysis must be carried out afresh.}
 
 It is worth pointing out that our results  generalise to higher dimensions in quite straightforward fashion. The essential new feature of non-locality involves  only the time direction and is similar in higher dimensions as well. In fact, it is easy to see from the study of the scalar field in section \ref{sec-toy} that 
 terms in $A_{\mathrm{out}}^{(2)}$ with  logarithmic violations  going like $\epsilon^2 \log(\epsilon)$  and  also terms  going like $\epsilon^2 {\widetilde T}^n$, in an asymptotic series in ${\widetilde T}$, arise in general    independent of the number of spatial directions. These will therefore be present   in higher dimensions as well and  should then be  reflected in the shear channel for fluid mechanics as well  leading to  non-local effects  of this type   in higher  dimensions too.

  It is also worth mentioning that the logarithmic effects we have seen might have been expected on the basis of the ``long throat'' which is present in the geometry of near-extremal branes,  as described in the introduction. It is well known from the study of the quasi-normal mode spectrum of extremal and near-extremal branes, e.g., \cite{Denef:2009yy, Edalati:2010hk},  that the poles corresponding to these modes become more and more dense  along the line $\mathrm{Re} (\omega)=0$ as one approaches extremality, eventually coalescing in the extremal limit to a branch cut. This ties in with the logarithmic violations mentioned above, since the logarithmic terms results in a  branch cut in the resulting response  properties of the  system.

As was mentioned in the introduction, we expect the dynamics of near-horizon $\mathrm{AdS}_2$ region  to be reproduced by a 1-dimensional theory involving the time reparametrisation modes and a phase mode. Starting with the near-extremal black brane with varying $T$, $\mu$, $u^\nu$ we showed how corrections can be found by solving Einstein-Maxwell equations in the near-horizon $\mathrm{AdS}_2$ region and the region away from it and then matching the two solutions together at the boundary of the $\mathrm{AdS}_2$ region.  One would expect to be able to replace the $\mathrm{AdS}_2$ region by the $1$-dimensional theory at its boundary and to be able to obtain the corrections by coupling this theory to the far region. 
  We leave a more complete understanding of this for the future as well.

\acknowledgments
We thank Sayantani Bhattacharyya,  Sachin Jain, Gautam Mandal,  Subir Sachdev and Ashoke Sen for useful discussions. We thank Richard Davidson for an important discussion on the charge diffusion mode, especially for pointing out why one needs to go to higher orders to obtain its dispersion relation correctly. We are especially grateful to ``the Professor'',   Shiraz Minwalla,  for suggesting this problem in the first place and for several important conversations and key insights  subsequently which 
were crucial in completing this project. 
We acknowledge the support of the Government of India, Department of Atomic Energy, under Project No. 12-R{\&}D-TFR-5.02-0200 and support from the Infosys Foundation in form of the Endowment for the Study of the Quantum Structure of Spacetime. S. P. T. acknowledges support from a J. C. Bose Fellowship, Department of Science and Technology, Government of India. Most of all, we are grateful to the people of India for generously supporting research in String Theory. 

\appendix

\section{Linearised Einstein-Maxwell Equations}\label{app-leq}

In this appendix, we write down the system of linearised Einstein-Maxwell equations relevant to the main text. 

Our starting point is the background metric $\bar{g}_{MN}$ and gauge field $\bar{A}_M$ given in eqs.(\ref{metv}--\ref{gfv}). We consider linearised perturbations to this background:
\begin{align}
\mathrm{d} s^2 &= \bar{g}_{MN} \, \mathrm{d} x^M \, \mathrm{d} x^N + e^{-\mathrm{i} \omega v + \mathrm{i} k_x x} h_{MN} (r) \, \mathrm{d} x^M  \, \mathrm{d} x^N, \label{ds2} \\
A_M &= \bar{A}_M +   e^{-\mathrm{i} \omega v + \mathrm{i} k_x x} a_M (r). \label{gfp}
\end{align}

For the purpose of this appendix, we work with the ``minimal'' gauge-fixing conditions,
\begin{equation}
\label{mgf}
h_{rr} = 0 = a_r.
\end{equation}
This condition is common to all the gauge-fixing conditions we have used in the text. Depending on the specific case at hand, we have gauge-fixed other components $h_{\mu r}$ of the metric in the main text. For some components of the perturbation, it is useful to raise one index with the background metric,
\begin{equation}
\label{hrais}
h^M {}_N \equiv \bar{g}^{MS} h_{SN}.
\end{equation}
Furthermore, it is convenient to decompose the components $h^x {}_x$ and $h^y {}_y$ into symmetric and anti-symmetric parts:
\begin{align}
h^x {}_x (r) &=  \sigma (r) + \alpha(r),   \label{hxx} \\
h^y {}_y (r) &=  \sigma (r) - \alpha(r) .  \label{hyy}
\end{align}

In the equations below, we have grouped the terms according to the number of boundary spacetime $(x^\mu)$ derivatives. A prime ($'$) denotes a derivative with respect to $r$ below. Let us write down the Maxwell equations.

$\mathsf{M}^v:$
\begin{align}
\bqty{- \frac{1}{r^2} \pqty{ r^2 a'_v (r) }' + g'(r) h'_{vr} (r) -  g'(r) \sigma'(r)} + \mathrm{i} k_x \bqty{ g'(r) h^x {}_r(r) - \frac{a'_x (r)}{r^2} } = 0. \label{Mv}
\end{align}

$\mathsf{M}^r:$
\begin{align}
{}& \mathrm{i} \bqty{ g'(r) \pqty{ \omega h_{vr} (r) - k_x h^x {}_v (r) -  \omega \sigma(r) } - \omega a'_v (r)-k_x f(r) a'_x (r)  } \nonumber \\
&\quad - \frac{1}{r^2} \bqty{ k_x^2 a_v (r) + k_x \omega a_x (r) } = 0.  \label{Mr}
\end{align}

$\mathsf{M}^x:$
\begin{align}
\bqty{ \frac{1}{r^2} \pqty{r^2 f(r) a'_x(r) }' + g'(r) {h^x {}_v}'(r) } - \frac{\mathrm{i}}{r^2} \bqty{k_x a'_v (r) + 2 \omega a'_x (r) - \omega r^2 g'(r) h^x {}_r (r)   } = 0. \label{Mx}
\end{align}

$\mathsf{M}^y:$
\begin{align}
\bqty{ \frac{1}{r^2} \pqty{r^2 f(r) a'_y(r) }' + g'(r) {h^y {}_v}'(r) } - \frac{\mathrm{i} \omega}{r^2} \bqty{ 2 a'_y (r) -  r^2 g'(r) h^y {}_r (r) } - \frac{k_x^2 a_y(r)}{r^4} =0. \label{My}
\end{align}

Let us now write down the Einstein equations.

$\mathsf{E}_{vr}:$
\begin{align}
&\bqty{ r f(r) \pqty{ r \sigma''(r) + 4 \sigma'  } + \frac{r^2}{2} f'(r) \sigma'- 6 h_{vr} (r) -2rf h'_{vr}  - \frac{\pqty{r h_{vv} (r) }' }{r^2} + 2 g'(r) a'_v(r) } \nonumber \\
&\quad - \mathrm{i} \bqty{  k_x r f(r) \pqty{ r {h^x {}_r}'(r) + 4 h^x {}_r  } + \frac{k_x r^2}{2} f' h^x {}_r + \frac{k_x}{2r^4} \pqty{r^4 h^x {}_v (r) }' +  \frac{\omega}{r^2} \pqty{r^2 \sigma(r) }'} \nonumber \\
&\quad  + \frac{1}{2r^2} \bqty{ k_x^2 (\alpha (r) - \sigma(r) -  h_{vr} (r) ) - k_x \omega r^2 h^x {}_r (r) } = 0 . \label{Evr}
\end{align}

$\mathsf{E}_{rr}:$
\begin{align}
\frac{1}{r} \bqty{2 h'_{vr} (r) - 2 \sigma'(r) - r \sigma''(r)  } + \frac{\mathrm{i} k_x}{r^2} \bqty{ r^2 h^x {}_r (r) }' =0. \label{Err}
\end{align}

$\mathsf{E}_{xr}:$
\begin{align}
\bqty{ \frac{1}{2r^2} \pqty{ r^4 {h^x {}_v}' (r) }' + 2 g'(r) a'_x (r) } + \frac{\mathrm{i}}{2} \bqty{\frac{\omega}{r^2} \pqty{r^4 h^x {}_r }' -k_x r^2 \pqty{ \frac{h_{vr}}{r^2} }'  + k_x (\alpha' - \sigma') } =0. \label{Exr}
\end{align}

$\mathsf{E}_{yr}:$
\begin{align}
\bqty{ \frac{1}{2r^2} \pqty{ r^4 {h^y {}_v}' (r) }' + 2 g'(r) a'_y (r) } + \frac{\mathrm{i}}{2} \bqty{\frac{\omega}{r^2} \pqty{r^4 h^y {}_r }' + k_x {h^y {}_x }'(r)  } + \frac{k_x^2}{2} h^y {}_r (r)=0. \label{Eyr}
\end{align}

$\mathsf{E}_{xy}:$
\begin{align}
- \frac{1}{2} \bqty{ r^4 f(r) { h^y {}_x }'(r) }' &+ \frac{\mathrm{i}}{2} \bqty{ 2\omega r \pqty{r h^y {}_x (r) }' +  k_x \pqty{r^4 f(r) h^y {}_r (r) + r^2 h^y {}_v (r)}' }  \nonumber \\
& + \frac{1}{2} k_x \omega r^2 h^y {}_r (r) = 0. \label{Exy}
\end{align}

$\mathsf{E}_{xx}+\mathsf{E}_{yy}:$
\begin{align}
&{} \bqty{ \pqty{r^4 f(r) \sigma'(r) }'  - \pqty{r^4 f(r) }' h'_{vr} (r) - 12 r^2 h_{vr} - \pqty{ r^2 h'_{vv} (r) }' -4r^2 g'(r) a'_v (r) } \nonumber \\
&\quad - \mathrm{i} \bqty{ k_x \pqty{r^4 f(r) h^x {}_r  + r^2 h^x {}_v  }' + 2\omega r^2 h'_{vr}  + 2\omega r \pqty{r \sigma }' } - \bqty{ k_x^2 h_{vr} + k_x \omega r^2 h^x {}_r } = 0 .\label{Expy}
\end{align}

$(\mathsf{E}_{xx}-\mathsf{E}_{yy})/2:$
\begin{align}
- \frac{1}{2} \bqty{ r^4 f(r) \alpha'(r) }' &+ \frac{\mathrm{i}}{2} \bqty{ 2\omega r \pqty{r \alpha (r) }' +  k_x \pqty{r^4 f(r) h^x {}_r (r) + r^2 h^x {}_v (r)}' }  \nonumber \\
& + \frac{1}{2} \bqty{k_x^2 h_{vr} (r) +  k_x \omega r^2 h^x {}_r (r)} = 0 . \label{Exmy}
\end{align}

$\mathsf{E}^r {}_v:$
\begin{align}
 &\mathrm{i} \bqty{  \omega r \pqty{r f(r) \sigma'(r) - \frac{r}{2}  f'(r)\sigma - 2f(r) h_{vr}  - \frac{h_{vv} }{r^2} } + \frac{1}{2} k_x r^2 \pqty{f(r) {h^x {}_v}' -f'(r)  {h^x {}_v} }  } \nonumber \\
 &\quad + \frac{1}{2} \bqty{ k_x^2 \pqty{f(r) h_{vr} + \frac{h_{vv}}{r^2} } + \omega k_x \pqty{2 h^x {}_v + r^2 f(r) h^x {}_r } + 2 \omega^2 \sigma } = 0. \label{Er_v}
\end{align}

$\mathsf{E}^r {}_x:$
\begin{align}
&\frac{\mathrm{i}}{2} \bqty{\frac{k_x}{r^2} (r^4 f(r))' h_{vr} + 4 g'(r) (k_x a_v + \omega a_x) +k_x h'_{vv} + \omega r^2 {h^x {}_v}' + k_x r^2 f(r) (\alpha' -\sigma') }  \nonumber \\
&\quad + \frac{1}{2} \bqty{ k_x \omega (\alpha - \sigma - h_{vr}) - \omega^2 r^2 h^x {}_r } =0 . \label{Er_x}
\end{align}

$\mathsf{E}^r {}_y:$
\begin{align}
& \frac{\mathrm{i}}{2} \bqty{4 \omega g'(r) a_y (r) + \omega r^2 {h^y {}_v}'(r)  + k_x r^2 f(r) {h^y {}_x}' (r) } \nonumber \\
&\quad + \frac{1}{2} \bqty{k_x^2  \pqty{ h^y {}_v + r^2 f(r) h^y {}_r } + \omega k_x h^y {}_x - \omega^2 r^2 h^y {}_r } =0. \label{Er_y}
\end{align}

\section{Higher Order Calculations in the Prototype Scalar Field Model}\label{app-so}

In section \ref{sec-toy}, we had considered a massless minimally coupled scalar field as a prototype for our near-extremal fluid-gravity correspondence. We had formulated a general procedure for all orders in the perturbative expansion in \S\ref{ss-ho}. In this appendix, we work in the extremal case $T=0$, and illustrate the corrections up to 
  second order in derivative expansion, i.e. up to $\mathcal{O}(\epsilon^2)$ and also show how the leading non-analytic behaviour emerges in the outer solution up to  $\mathcal{O}(\epsilon^3)$. 
  Let us also note that since $T=0$,  $f(r)$ below will  actually equal $f_0(r)$, eq.(\ref{f0r}).

Let us first look at the second order solution in the exterior region. We have the differential equation,
\begin{equation}
\label{phio2}
\frac{\mathrm{d} }{\mathrm{d} r} \pqty{r^4 f(r) \frac{\mathrm{d} \phi^{(2)}_{\mathrm{out}} }{\mathrm{d} r} } = s^{(2)}_{\mathrm{out}},
\end{equation}
where the source term $s^{(2)}_{\mathrm{out}}$ is given by, (neglecting the $k_x^2 \phi^{(0)}$ term in the source, see footnote \ref{foot1}), 
\begin{equation}
\label{so2}
s^{(2)}_{\mathrm{out}} =  2 \mi \omega r \frac{\mathrm{d} ( r \phi^{(1)}_{\mathrm{out}} )}{\mathrm{d} r}, 
\end{equation}
with $ \phi^{(1)}_{\mathrm{out}} $ being given by eq.(\ref{phex2}). Near $r=r_B$, this source term can then be expanded in small $(r-r_h)/r_h$,
\begin{equation}
\label{so2e}
s^{(2)}_{\mathrm{out}} = - \frac{2 \omega^2 \phi^{(0)} }{3} \frac{r_h}{r-r_h}-\frac{2 \omega^2 \phi^{(0)} }{3} \log( \frac{r-r_h}{r_h} ) + \cdots.
\end{equation}
In contrast with the first order calculation, where the source is given by,
\be
\label{s1a}
s^{(1)}_{\mathrm{out}}= 2\mi \omega r \phi^{(0)},
\ee
 see eq.(\ref{radp2}), we see that $s^{(1)}_{\mathrm{out}}$ diverges as $r\rightarrow r_h$.

 Following the general principle outlined in \S\ref{ss-ho}, we can  write the solution  of eq.(\ref{phio2}) to be,
\begin{equation}
\label{phio22}
\phi^{(2)}_{\mathrm{out}} = B^{(2)}_{\mathrm{out}} - \int_r^\infty \frac{\mathrm{d} r'}{ {r'}^4 f(r')}  \pqty{ A^{(2)}_{\mathrm{out}} + \int_{r_B}^{r'} \mathrm{d} r'' \, s^{(2)}_{\mathrm{out}} (r'') }.
\end{equation}
We set $B^{(2)}_{\mathrm{out}} =0$ by imposing normalisability. Near $r=r_B$, the solution $\phi^{(2)}_{\mathrm{out}}$ has the form,
\begin{equation}
\label{phtoa}
\phi^{(2)}_{\mathrm{out}} = \frac{ \omega^2 \phi^{(0)} }{9 r_h } \frac{\log (\frac{r-r_h}{r_h})}{r-r_h} - \frac{A^{(2)}_{\mathrm{out}} + \omega^2 \phi^{(0)} f_1 (r_B, r_h) }{6 r_h^2 (r-r_h)} - \frac{ \omega^2 \phi^{(0)} }{54 r_h^2 } \log^2 \pqty{ \frac{r-r_h}{r_h} } + \cdots.
\end{equation}
Here, the form of the function $f_1 (r_B, r_h)$ is explicitly known and rather long, which we do not write  because it is not important for the discussion at hand. Note that the coefficient of the $\log(r-r_h)/(r-r_h)$ term in $\phi^{(2)}_{\mathrm{out}} $ is completely fixed even before fixing $A^{(2)}_{\mathrm{out}}$. This term should then automatically match between the interior and exterior solutions. It is indeed seen from eq.(\ref{behphaa}) that near $r=r_B$, the interior solution $\phi^{(1)}_{\mathrm{in}}$ has the form, including only the $\mathcal{O}(\omega^2)$ terms,
\begin{equation}
\label{ph1in2}
\phi_{\mathrm{in}}^{(1)} = \frac{ \omega^2 \phi^{(0)} }{9 r_h } \frac{\log (\frac{r-r_h}{\omega})}{r-r_h} + \frac{ \omega^2 \phi^{(0)} }{9 r_h (r-r_h) } \pqty{ \frac{\mi \pi}{2} - \gamma +\log 3 + 1}    + \cdots.
\end{equation}
From the scaling symmetry eq.(\ref{scalads2}), referred to in section \ref{sec-toy}, it is easy to see that terms of the first two type, which are shown explicitly on the RHS in eq.(\ref{ph1in2}), 
would   not arise for $\phi^{(n)}_{\mathrm{in}}$ for $n\geq 2$. 
We also note that the terms involving the logarithms in eq.(\ref{phtoa}) and (\ref{ph1in2}) do have the same coefficients. There is one difference though in these terms --- the log term in eq.(\ref{phtoa})   is cut off by $r_h$  while it is cut-off   by $\omega$ in eq.(\ref{ph1in2}). Equating the first two terms in eqs.(\ref{phtoa}) and (\ref{ph1in2}) then leads to, 
\begin{equation}
\label{a2out}
 A^{(2)}_{\mathrm{out}}=   \frac{2  \omega^2 r_h \phi^{(0)} }{3 } \pqty{ \log \frac{\omega}{r_h} -\frac{\mi \pi}{2} + \gamma -\log 3 - 1} - \omega^2 \phi^{(0)} f_1 (r_B, r_h).
\end{equation}
This agrees with eq.(\ref{valaso}).
With the  choice of $A^{(2)}_{\mathrm{out}}$ in eq.(\ref{a2out}), the exterior solution $\phi^{(2)}_{\mathrm{out}}$ turns out to be independent of $r_B$, as expected.

Let us now look at the second order interior solution.  We will argue below from its behaviour that in $\phi_{\mathrm{out}}^{(3)}$, the integration constant 
$A^{(3)}_{\mathrm{out}}$,  eq.(\ref{nphio}) for $n=3$, must have a term of the form 

\be
\label{forma3}
A^{(3)}_{\mathrm{out}} \sim \omega^3 \log^2\pqty{\frac{\omega}{r_h}}. 
\ee
The relevant differential equation for the second order solution  is given by,
\begin{equation}
\label{odphin2}
\frac{\mathrm{d}^2 \phi^{(2)}_{\mathrm{in}}}{\mathrm{d} {r^*}^2} - 2 \mi \omega \frac{\mathrm{d} \phi^{(2)}_{\mathrm{in}} }{\mathrm{d} {r^*} } = \frac{ 2 \mi \omega }{r_h} F(r) \phi^{(1)}_{\mathrm{in}} + \cdots.
\end{equation}
Note that the source term on the right hand side above is consistent with the scaling symmetry eq.(\ref{scalads2}) mentioned in  \S\ref{ss-ho}. The ellipsis above includes terms involving $ \phi^{(0)}$  which arise because of keeping $k_x^2$ terms and departures from the near-horizon geometry, these will be ignored below because, near $r=r_B$, they do not give rise to a term of the form $\log^2((r-r_h)/\omega)$ in eq.(\ref{ph2irb}) below. This is  the kind of term which could give rise to a term in $A^{(3)}_{\mathrm{out}}$  of the form, eq.(\ref{forma3}) that  is our main focus here. The contribution to $\phi_{\mathrm{in}}^{(2)}$  with the  source term  shown on the RHS  of eq.(\ref{odphin2}) is then given by,
\begin{equation}
\label{odphin22}
 \phi^{(2)}_{\mathrm{in}}= A^{(2)}_{\mathrm{in}}+ \frac{  \mi \omega  e^{2\mi \omega {r^*} } }{3r_h} \int_{-\infty}^{r^*} \mathrm{d} {r^*}' e^{-2\mi \omega {r^*}' } \int_{-\infty}^{ {r^*}' } \frac{\mathrm{d} {r^*}'' }{ { {r^*}''}^2} \pqty{ A^{(1)}_{\mathrm{in}} + \frac{\mi \omega \phi^{(0)}}{3r_h} e^{2\mi \omega {r^*}'' } E_1 (2\mi \omega {r^*}'' ) }.
\end{equation}
Here we have used the form of $\phi^{(1)}_{\mathrm{in}}$ given in eq.(\ref{nph2}). It is easy to see that  $\phi^{(2)}_{\mathrm{in}}$ remains well-behaved towards the horizon. The behaviour near $r=r_B$ is also easy to evaluate. We obtain,
\begin{equation}
\label{ph2irb}
 \phi^{(2)}_{\mathrm{in}} =-\frac{ \omega^2 \phi^{(0)} }{18 r_h^2}\log^2 \pqty{ \frac{r-r_h}{\omega} } + \cdots.
\end{equation}
The log-squared term is the leading term near $r=r_B$ at $\mathcal{O}(\omega^2)$. Note that we got a similar term in $\phi^{(2)}_{\mathrm{out}}$, albeit with a different coefficient. The difference in the coefficient of $\phi^{(2)}_{\mathrm{out}}$ is because there is also a contribution of the form in eq.(\ref{ph2irb}) which arises to this order from 
$\phi^{(1)}_{\mathrm{in}}$ which we will not explicitly calculate. Accounting for it, the coefficient of the log-squared term in (\ref{phtoa}) should match. 

A term like eq.(\ref{ph2irb}) gives rise, at $\mathcal{O}(\omega^3)$, to a term going like $\log^2({r-r_h\over \omega})/(r-r_h)$ near $r=r_B$. 
An explicit calculation, which we do not include here, shows that one does get a term of the same  form in  the exterior at $\mathcal{O}(\omega^3)$, with one important difference.
The logarithm term is   cut-off, once again,  by $r_h$ and not $\omega$. 
 Matching the $1/(r-r_h)$ coefficient  then necessarily yields an exterior coefficient of the form eq.(\ref{forma3}).

\subsection{Matching Procedure for Varying Chemical Potential}\label{appsub-hocp}

We now explore the case considered in \S\ref{ss-varbg} when the background geometry is extremal, but the location of the horizon  $r_h$ is varying with $x^\mu$, in particular $v$. In such a case, the differential equation in the exterior now reads,
\begin{equation}
\label{phiov2}
\frac{\del }{\del r} \pqty{r^4 f(r) \frac{\del \phi^{(2)}_{\mathrm{out}} }{\del r} } = s^{(2)}_{\mathrm{out}},
\end{equation}
where the source term $s^{(2)}_{\mathrm{out}}$ is now given by,
\begin{equation}
\label{so2v2}
s^{(2)}_{\mathrm{out}} =  -2 r \bqty{ - \mi \omega  \frac{\del( r \phi^{(1)}_{\mathrm{out}} )}{\del r}  + \dot{r}_h  \frac{\del}{\del r_h}  \frac{\del( r \phi^{(1)}_{\mathrm{out}} )}{\del r}  }, 
\end{equation}
where $\dot{r}_h$ indicates the $v$-derivative of the varying function $r_h(x^\nu)$ and $\phi^{(1)}_{\mathrm{out}}$ is given by eq.(\ref{phiovar}).
We have the same form of the solution as eq.(\ref{phio22}), with $B^{(2)}_{\mathrm{out}} = 0$,
\begin{equation}
\label{phio22v}
\phi^{(2)}_{\mathrm{out}} =  - \int_r^\infty \frac{\mathrm{d} r'}{ {r'}^4 f(r')}  \pqty{ A^{(2)}_{\mathrm{out}} + \int_{r_B}^{r'} \mathrm{d} r'' \, s^{(2)}_{\mathrm{out}} (r'') }.
\end{equation}

Now, near $r=r_B$, the function $\phi^{(2)}_{\mathrm{out}}$ can be expanded as,
\begin{equation}
\label{phi2sing}
\phi^{(2)}_{\mathrm{out}}= - \frac{\mi  \omega \dot{r}_h \phi^{(0)} }{18r_h (r-r_h)^2}+ \frac{ \omega^2 \phi^{(0)} }{9 r_h } \frac{\log (\frac{r-r_h}{r_h})}{r-r_h} - \frac{A^{(2)}_{\mathrm{out}} + \phi^{(0)} f_2 (r_B, r_h) }{6 r_h^2 (r-r_h)} + \cdots.
\end{equation}
Here, we have written the most important terms in $(r-r_h)/r_h$. There are some features in the above equation worth pointing out. First note the appearance of a more important (first) term in (\ref{phi2sing}), as compared to  eq.(\ref{phtoa}), which is, of course, dependent on $\dot{r}_h$.  Further, note that the next most important term, going like $\log(r-r_h)/(r-r_h)$ is independent of $\dot{r}_h$ --- this term is the same as eq.(\ref{phtoa}). Note also that the function $f_2(r_B, r_h)$ is a term of order $\epsilon^2$ --- it can be $\omega^2$ or $\omega \dot{r}_h$. 

Let us now turn to the near-horizon analysis, which will allow us to fix the constant $A^{(2)}_{\mathrm{out}}$. Note that the first and most important term in eq.(\ref{phi2sing}) is not required for the matching, since it cannot give rise to any $1/(r-r_h)$ term. Note further that written in the $\zeta$ variables, this term is actually of order $\mathcal{O}(\epsilon^0  )$:
\begin{equation}
\label{iorh}
\frac{\mi  \omega \dot{r}_h}{r_h (r-r_h)^2} \sim \pqty{\frac{\dot{r}_h}{\omega}} \zeta^2.
\end{equation}
In order to explain such terms in the insider region, we must look for an appropriate term in the source.

From the near-horizon analysis of the scalar differential equation, we have on the right hand side of the interior differential equation such a relevant term,
\begin{equation}
\label{odphin2v}
\frac{\del^2 \phi_{\mathrm{in}}}{\del {r^*}^2} - 2 \mi \omega \frac{\del \phi_{\mathrm{in}} }{\del {r^*}^2 } = -2 F(r)  \dot{r}_h \frac{\del  }{\del r_h} \frac{\del \phi^{(1)}_{\mathrm{in}} }{\del r} + \cdots.
\end{equation}
Using the form of $\phi^{(1)}_{\mathrm{in}}$ near $r=r_B$, eq.(\ref{phnh2}), we evaluate the leading term on the right hand side of eq.(\ref{odphin2v} to be,
\begin{equation}
\label{odphin2vs}
- \frac{4 \mi \omega \dot{r}_h \phi^{(0)} }{r_h} + \cdots.
\end{equation}
Performing the double integration from eq.(\ref{odphin2v}), we have near $r=r_B$, precisely the same term as the first term in (\ref{phi2sing}).

The  first term in eq.(\ref{phi2sing})  matches automatically with the outside in this manner.  The rest of the analysis of matching the two solutions  near $r_B$ then proceeds as before. 
As a result,  $A^{(2)}_{\mathrm{out}}$ has the same non-analytic structure as in the case with a constant $r_h$, eq.(\ref{a2out}),
\begin{equation}
\label{a2outcp}
A^{(2)}_{\mathrm{out}} \sim \omega^2 \log  \frac{\omega}{r_h} + \cdots,
\end{equation}
where the ellipsis indicates terms analytic in the derivatives.

\section{Near-Horizon Analysis of Metric and Gauge Field Components}\label{app-nhsm}

In this appendix, we show that the general principles outlined in the toy scalar field example in section \ref{sec-toy} continue to hold true for the actual gravitational and gauge field perturbations. Specifically, we consider the examples of the metric and gauge field components $h^y {}_x$, $h^y {}_v$ and $a_y$ to the first order in the $\epsilon$-expansion in linearised perturbation theory. It is sufficient to consider these components to exhibit the validity of our claims. These components correspond to the traceless symmetric tensor ($h^y {}_x$)  and vector ($h^y {}_v$ and $a_y$) of the $\mathrm{SO}(2)$.  It is in these sectors that the $\omega$-dependent terms become important towards  the horizon because of the scaling symmetry (\ref{scalads2}) mentioned in section \ref{sec-toy}.  For reasons mentioned in \S\ref{sss-scalar},  above eq.(\ref{gvrs1}), there are no such $\omega$--dependent terms in the scalar sector towards the horizon, in the first order analysis, and hence we do not discuss the scalar sector here.

For this appendix, we confine our attention to the zero-temperature case, so that $f(r)=f_0(r)$, eq.(\ref{f0r}), and $F(r) = 6(r-r_h)^2$. The metric and gauge field components scaling non-trivially under the scaling symmetry (\ref{scalads2}) are as follows,
\begin{align}
\pqty{ h^i {}_v, a_v } &\to \frac{1}{\lambda} \pqty{ h^i {}_v, a_v } , \nonumber \\ 
 h^i {}_r &\to \lambda h^i {}_r , \label{metscal}\\
  h_{vv} &\to \frac{1}{\lambda^2} h_{vv}. \nonumber
\end{align}  

Since this appendix concerns the interior region, we drop the subscript ``in'' from the metric and gaugef field perturbations.

\subsection{Tensor Sector}\label{appss-tensor}

Let us first discuss the mode $h^y {}_x$, because the discussion would be very similar for the scalar field discussed in section \ref{sec-toy}. The near-horizon form of the equation $\mathsf{E}_{xy}=0$, eq.(\ref{Exy}), gives, after taking the near-horizon form of the zeroth order metric and gauge field components eqs.(\ref{hir0}--\ref{av0}),
\begin{equation}
r_h^2 \frac{\mathrm{d}}{\mathrm{d} r} \pqty{ F(r) \frac{\mathrm{d} {h^y {}_x}^{(1)} }{\mathrm{d} r} } - 2 \mathrm{i} \omega r_h^2 \frac{\mathrm{d} {h^y {}_x}^{(1)} }{\mathrm{d} r}  = 0 \label{nhxy}.
\end{equation}
The near-horizon form of the full equation (\ref{Exy}) possesses the full scaling symmetry (\ref{scalads2}), if we take into account the transformations (\ref{metscal}). Now, however, we treat equation (\ref{nhxy}) as a dynamical equation in its own right.
Note that the form of the left hand side is the same as that of the scalar field considered in section \ref{sec-toy},see eq.(\ref{phi2}). We see that there is no source term at $\mathcal{O}(\epsilon)$.  We need to obtain an appropriate non-zero source at this order --- a source that would give rise to the  $\log (r-r_h)$ term we had seen near the matching region in the outside solution described in \S\ref{sss1-tensor}, see eqs.(\ref{hxy1full}) and (\ref{tenfun}).  Such a source term comes from considering the departure from the $\mathrm{AdS}_2 \times \mathbb{R}^2$ in the source. Recall that in the scalar toy model in section \ref{sec-toy}, too, we needed such a departure for having a non-zero source The relevant source term on the right hand side  of eq.(\ref{nhxy}) would be,
\begin{equation}
\label{srxy}
2 \mathrm{i} k_x r h^y {}_v^{(0)}.
\end{equation} 
Expanding this source in powers of $(r-r_h)/r_h$ and keeping the leading term, we obtain the dynamical equation,
\begin{equation}
\frac{\mathrm{d}}{\mathrm{d} r} \pqty{ F(r) \frac{\mathrm{d} {h^y {}_x}^{(1)} }{\mathrm{d} r} } - 2 \mathrm{i} \omega\frac{\mathrm{d} {h^y {}_x}^{(1)} }{\mathrm{d} r}  = -\frac{2\mi k_x \delta\beta_y}{r_h} \label{nhxy2}.
\end{equation} 
Note that this source term is consistent with the near-horizon source term (\ref{s1xy}), see eq.(\ref{sigmac}). With this source term, we get the logarithmic behaviour with the correct coefficient.

\subsection{Vector Sector}\label{appss-vector}
Let us next consider the near-horizon form of the Maxwell equation $\mathsf{M}^y = 0$, eq.(\ref{My}) and the Einstein equation $\mathsf{E}_{ry} = 0$, eq.(\ref{Eyr}). Using the zeroth order components (\ref{hir0}--\ref{av0})  and keeping appropriate terms arising due to the departure from the $\mathrm{AdS}_2 \times \mathbb{R}^2$ geometry, we have,
\begin{align}
\frac{\mathrm{d}}{\mathrm{d} r} \pqty{ F(r) \frac{\mathrm{d} a_y^{(1)} }{\mathrm{d} r} } + \sqrt{3} r_h^2 \frac{\mathrm{d} {h^y {}_v}^{(1)} }{\mathrm{d} r} - 2 \mathrm{i} \omega \frac{\mathrm{d} a_y^{(1)} }{\mathrm{d} r}  &=- \mathrm{i} \sqrt{3} \omega \, \delta \beta_y,  \label{nhmy} \\
r_h^2 \frac{\mathrm{d}^2 {h^y {}_v}^{(1)} }{\mathrm{d} r^2} + 4 \sqrt{3} \frac{\mathrm{d} a_y^{(1)} }{\mathrm{d} r}  &= \frac{2\mi \omega \delta \beta_y}{r_h}\label{nhery}.
\end{align}
The second equation above, (\ref{nhery}), immediately yields,
\begin{equation}
\label{nhyv}
r_h^2 \frac{\mathrm{d} {h^y {}_v}^{(1)} }{\mathrm{d} r} + 4 \sqrt{3} {a_y^{(1)}}  = \frac{C_1}{\sqrt{3}} +  \frac{2\mi \omega \delta \beta_y}{r_h} (r-r_h)  ,
\end{equation}
where $C_1$ is a constant. Plugging (\ref{nhyv}) in eq.(\ref{nhmy}) gives,
\begin{equation}
\frac{\mathrm{d}}{\mathrm{d} r} \pqty{ F(r) \frac{\mathrm{d} a_y^{(1)} }{\mathrm{d} r} }  - 2 \mathrm{i} \omega \frac{\mathrm{d} a_y^{(1)} }{\mathrm{d} r}  - 12 a_y^{(1)} = -\mathrm{i} \sqrt{3} \omega \, \delta \beta_y - C_1 - \frac{2\sqrt{3}\mi \omega \delta \beta_y}{r_h} (r-r_h) .  \label{nhay}
\end{equation}
Multiplying through by $F(r)$ and noting that,
\begin{equation}
F(r) = 6(r-r_h)^2 = \frac{1}{6 {r^*}^2}, \label{appfr}
\end{equation}
we have,
\begin{equation}
\frac{\mathrm{d}^2 a_y^{(1)} }{\mathrm{d} {r^*}^2 } - 2 \mathrm{i} \omega \frac{\mathrm{d} a_y^{(1)} }{\mathrm{d} {r^*}} - \frac{2}{{r^*}^2} a_y^{(1)} (r) = -\frac{\mathrm{i} \sqrt{3} \omega \, \delta \beta_y + C_1}{6{r^*}^2} + \frac{\mi  \sqrt{3} \omega \, \delta \beta_y}{18 r_h {r^*}^3 }. \label{deqay}
\end{equation}
The last term in the source above is sub-leading in $(r-r_h)$ or $1/r^*$ and can be thought of as arising due to a departure from the $\mathrm{AdS}_2 \times \mathbb{R}^2$ geometry. This source term will not be important for the leading order matching with the outside solution  --- this source term is of $\mathcal{O}(\epsilon^2)$ under a rescaling and expressing it in terms of the $\zeta$ variable (\ref{zeta}). Neglecting this source term, the equation (\ref{deqay}) has the solution, 
\begin{align}
a_y^{(1)} (r) &=  C_2 \pqty{r - r_h - \frac{\mathrm{i} \omega}{6} } + C_3 e^{2\mathrm{i} \omega r^*} \pqty{r - r_h + \frac{\mathrm{i} \omega}{6}  } +\frac{\mathrm{i} \sqrt{3} \omega \, \delta \beta_y + C_1}{12} .  \label{solay} 
\end{align}
We immediately see that imposing ingoing boundary conditions sets $C_3=0$. 

We can also solve for $h^y {}_v$ from  eq.(\ref{nhyv}), (we neglect the second term on the RHS of eq.(\ref{nhyv})),
\begin{equation}
{h^y {}_v}^{(1)} (r) = C_4  + \frac{\mathrm{i} \omega ( 2 \sqrt{3} C_2 - 3 \delta \beta_y) }{3r_h^2} (r-r_h) - \frac{2\sqrt{3}C_2}{r_h^2} (r-r_h)^2. \label{solhyv}
\end{equation}
The leading term of $a_y^{(1)}$ and leading and  first subleading terms of ${h^y {}_v}^{(1)}$  near $r=r_B$ are then given by,
\begin{align}
a_y^{(1)} (r) &= \frac{\mathrm{i} \sqrt{3} \omega \, \delta \beta_y + C_1}{12}  - \mi \frac{\omega}{6} C_2+ \mathcal{O} \pqty{  (r-r_h) \log( \frac{r-r_h}{\omega} )  } , \label{ayrb1} \\
{h^y {}_v}^{(1)} (r) &= C_4  + \frac{\mathrm{i} \omega ( 2 \sqrt{3} C_2 - 3 \delta \beta_y) }{3r_h^2} (r-r_h)  + \mathcal{O} \pqty{  (r-r_h)^2 \log( \frac{r-r_h}{\omega} )  } .  \label{hyvrb1} 
\end{align}

The remaining interior constants are easily determined by matching with the outer solution, given in  \S\ref{sss1-vector}. Using the explicit forms of these functions, eqs.(\ref{hyv1full}) and (\ref{ay1full}), we have, near $r=r_B$,
\begin{align}
{a_y}^{(1)} (r) &= \frac{\mathrm{i}  \omega \, \delta \beta_y}{2 \sqrt{3}} + \cdots,  \label{ayrb} \\
{h^y {}_v}^{(1)} (r) &= - \frac{\mathrm{i}  \omega \, \delta \beta_y}{r_h^2} (r-r_h) +\cdots. \label{hyvrb} 
\end{align}

We have, from comparison, 
\begin{align}
C_1 &= \mathrm{i}  \sqrt{3} \omega \, \delta \beta_y  , \label{C1} \\
C_2 &= 0 , \label{C2} \\
C_4 &= 0 \label{C3}.
\end{align}

Several comments are worth making at the end of this appendix. First note that, we have set the temperature to zero in the analysis above --- we could have kept $T$-dependent terms in the source, but as we saw in \S\ref{ss-ft}, for obtaining the outside solution the first order matching  can be carried out in the extremal limit itself. Keeping the $T$-dependent terms of course changes the interior solution in important ways.

Second, although we have explicitly considered the linearised analysis in this appendix, we can generalise the results for varying $r_h(x^\nu)$, $T(x^\nu)$, using the adiabatic approximation discussed in \S\ref{ss-varbg} --- there would,  be $\del_\nu r_h$-  and $\del_\nu T$-dependent source terms at the first order. These are analogous to the terms  we  saw in the previous appendix  which arise at second order in the scalar theory case. 

Finally, using arguments analogous to those in section \ref{sec-toy}, we can easily show that by choosing ingoing boundary conditions the interior solution for these metric and gauge field components  remain finite and well-behaved towards the future horizon  to higher orders as well.

\section{Conservation Equations and Time Evolution}\label{app-constime}

Let us start with the constitutive relations,
\begin{equation}
\label{t1mna}
T_{\mu \nu} = \frac{1}{2} \mathcal{E} \pqty{ \eta_{\mu \nu } + 3u_\mu u_\nu} - 2 \eta \sigma_{\mu \nu},
\end{equation}
and,
\begin{equation}
\label{jm1a}
J^\nu = \rho u^\nu - \chi_1 \mathfrak{a}^\nu -\chi_2 P^{\nu\lambda} \del_\lambda \pqty{ \mu + \frac{\sqrt{3} T}{2}  }    .
\end{equation}
The values of $\eta$, $\chi_1$ and $\chi_2$ are given in eqs.(\ref{etav}), (\ref{chi1}) and (\ref{chi2}) respectively. Using 
\begin{equation}
\label{defumu2}
u_\mu = \frac{1}{\sqrt{1-\bm{\beta}^2}} (-1, \beta_i),
\end{equation}
we can expand the constitutive relations up to quadratic order in $\beta$.  It is convenient to express the results in terms of the following quantities, (linearised forms of expansion (\ref{expansion}), shear tensor (\ref{sigma}), vorticity tensor\footnote{The vorticity tensor is defined as
$$
\omega^{\mu \nu} \equiv \frac{1}{2} P^{\mu \lambda} P^{\nu \kappa} ( \del_\kappa u_\lambda- \del_\lambda u_\kappa ). 
$$ } and acceleration vector (\ref{acc})).
\begin{align}
\theta_{(1)} &= \del_x \beta_x + \del_y \beta_y, \label{theta1} \\
\sigma^{\mu \nu}_{(1)} &= \begin{pmatrix}
0 & 0 & 0 \\
0 & \frac{1}{2} (\del_x \beta_x - \del_y \beta_y) & \frac{1}{2} (\del_x \beta_y + \del_y \beta_x) \\
0 & \frac{1}{2} (\del_x \beta_y + \del_y \beta_x)  & -\frac{1}{2} (\del_x \beta_x - \del_y \beta_y) 
\end{pmatrix}, \label{sigma1} \\
\omega^{\mu \nu}_{(1)} &= \begin{pmatrix}
0 & 0 & 0 \\
0  &0 &\frac{1}{2} (\del_y \beta_x - \del_x \beta_y) \\
0  & -\frac{1}{2} (\del_y \beta_x - \del_x \beta_y) &0
\end{pmatrix}, \label{omega1} \\
\mathfrak{a}_{(1)}^\nu &= \begin{pmatrix}
0 & \del_0 \beta_x & \del_0 \beta_y
\end{pmatrix}. \label{acc1}
\end{align}
Note that in the discussion below, we will not be very careful about the placement of the spatial indices $i,j$ etc. --- the results are unambiguous.

We have the following components of the stress tensor,
\begin{align}
T^{00} &= \mathcal{E} + \frac32 \mathcal{E} \bm{\beta}^2 , \label{t00} \\
T^{0i} &=\frac32 \mathcal{E} \beta_i - 2 \eta \beta_j \sigma^{ij}_{(1)} , \label{t0i} \\
T^{xx} &=   \frac{1}{2} \mathcal{E} + \frac32 \mathcal{E} \beta_x^2 - 2\eta \sigma^{xx}_{(1)} -  \eta  (\beta_x \mathfrak{a}^x_{(1)} - \beta_y \mathfrak{a}^y_{(1)}), \label{txx} \\
T^{yy} &=   \frac{1}{2} \mathcal{E} + \frac32 \mathcal{E} \beta_y^2 - 2\eta \sigma^{yy}_{(1)} + \eta (\beta_x \mathfrak{a}^x_{(1)} - \beta_y \mathfrak{a}^y_{(1)}), \label{tyy} \\
T^{xy} &=  \frac{3}{2} \mathcal{E} \beta_x \beta_y  - 2 \eta \sigma^{xy}_{(1)}  -\eta (\beta_x \mathfrak{a}^y_{(1)} + \beta_y \mathfrak{a}^x_{(1)}). \label{txy}
\end{align}
We also have the components of the current to be,
\begin{align}
J^0 &= \rho \pqty{1 + \frac{1}{2} \bm{\beta}^2 }  -\chi_1 \beta_i \mathfrak{a}^i_{(1)} - \chi_2 \bm{\beta}^2 \del_0 \pqty{\mu + \frac{\sqrt{3}}{2} T }  - \chi_2 \beta_i \del_i \pqty{\mu + \frac{\sqrt{3}}{2} T }, \label{j0comp}
\end{align}
and
\begin{align}
J^i &= \rho \beta_i -\chi_1 \mathfrak{a}^i_{(1)} - \frac{1}{2} \chi_1 \theta_{(1)} \beta_i -  \chi_1 \beta_j \pqty{  \sigma^{ij}_{(1)} + \omega^{ij}_{(1)} } \nonumber \\
&\quad - \chi_2 \beta_i \del_0 \pqty{\mu + \frac{\sqrt{3}}{2} T}- \chi_2 (\delta_{ij} + \beta_i \beta_j) \del_j \pqty{\mu + \frac{\sqrt{3}}{2} T}.\label{jicomp}
\end{align}

We now state the non-linear conservation equations in a simplified form. Note that in writing the following equations, we have set all terms containing an explicit factor of $\beta_i$ to zero, but we have retained terms involving a derivative in $\beta$.

We have,
\begin{align}
\del_\mu T^{\mu 0} &=  \del_0 \mathcal{E}  +\frac32 \mathcal{E}  \theta_{(1)}   -  2\eta \sigma^{ij}_{(1)} \sigma^{ij}_{(1)}, \label{pmtm0}
\end{align}
and
\begin{align}
\del_\mu T^{\mu i} &= \frac{3}{2} \mathcal{E} \mathfrak{a}^i_{(1)}  - 2 \eta \sigma^{ij}_{(1)} \mathfrak{a}^j_{(1)}  + \frac{1}{2} \del_i \mathcal{E} - 2 \sigma^{ij}_{(1)} \del_j \eta - 2 \eta \del_j \sigma^{ij}_{(1)} \nonumber \\
&\quad - \eta \theta_{(1)} \mathfrak{a}^i_{(1)} -2 \eta \omega^{ij}_{(1)}  \mathfrak{a}^j_{(1)} . \label{pmtm1}
\end{align}
The four-divergence of the current is given by,
\begin{align}
\del_\nu J^\nu &= \del_0 \rho  +  \rho \theta_{(1)}     - \chi_1 \mathfrak{a}^i_{(1)} \mathfrak{a}^i_{(1)} - \frac{1}{2} \chi_1 \theta_{(1)}^2  - \chi_1 \pqty{  \sigma^{ij}_{(1)} \sigma^{ij}_{(1)}  - \omega^{ij}_{(1)} \omega^{ij}_{(1)} } -\del_i \chi_1 \mathfrak{a}^i_{(1)}  \nonumber \\
&\quad - \chi_1 \del_i \mathfrak{a}^i_{(1)} - \mathfrak{a}^i_{(1)} \chi_2  \del_i \pqty{ \mu + \frac{\sqrt{3} T}{2}  }   - \theta_{(1)}  \chi_2  \del_0 \pqty{ \mu + \frac{\sqrt{3} T}{2}  }  - \chi_2  \del_i \del_i \pqty{ \mu + \frac{\sqrt{3} T}{2}  } . \label{pmjm1}
\end{align}
We have used the fact that in our approximation, $\chi_2$ is a constant, eq.(\ref{chi2}).

Note that in the charge conservation equation the acceleration appears as parts of $\mathcal{O} (\epsilon^2)$ terms. It is possible to trade this time-derivative for a spatial derivative using $\del_\mu T^{\mu i} = 0$, eq.(\ref{pmtm1}). We need to keep only the $\mathcal{O} (\epsilon)$ term for this purpose:
\begin{equation}
\label{aieni}
\mathfrak{a}^i_{(1)}  = -\frac{1}{3} \frac{ \del_i \mathcal{E}}{\mathcal{E}} .
\end{equation}

We also need to obtain the term $\del_i \mathfrak{a}^i_{(1)}$. For that purpose, we need to be careful about keeping the various $\beta_i$'s again. We have, from the $\del_\mu T^{\mu i}=0$ equation, keeping only the perfect fluid part because only this part gives the relevant $\mathcal{O}(\epsilon^2)$ expression:
\begin{equation}
\label{aini1}
\frac{3}{2} \del_0 (\mathcal{E} \beta_i) + \del_j \bqty{ \frac{\mathcal{E}}{2}  ( \delta_{ij} + 3 \beta_i \beta_j)  }= 0 .
\end{equation}
This leads to, upon taking a spatial derivative of $\del_0 \beta_i$
\begin{equation}
\label{aini2}
\del_i  \mathfrak{a}^i_{(1)}= - \theta_{(1)} \frac{\del_0 \mathcal{E} }{\mathcal{E}} - \frac{1}{3} \frac{\del_i \del_i \mathcal{E}}{\mathcal{E}} + \frac{1}{3} \frac{\del_i \mathcal{E} \del_i \mathcal{E}}{\mathcal{E}^2} -  \frac{3}{2}\theta_{(1)}^2 - \pqty{  \sigma^{ij}_{(1)} \sigma^{ij}_{(1)}  - \omega^{ij}_{(1)} \omega^{ij}_{(1)} } .
\end{equation}

The current four-divergence (\ref{pmjm1}) then reads,
\begin{align}
\del_\nu J^\nu &= \del_0 \rho + \rho \theta_{(1)} -  \chi_1 \bqty{ \frac{4}{9} \frac{\del^i  \mathcal{E}\del_i  \mathcal{E}}{\mathcal{E}^2} - \frac{1}{3} \frac{\del_i \del^i \mathcal{E}}{\mathcal{E}}  - \theta_{(1)}^2 } + \chi_1 \theta_{(1)} \frac{\del_0 \mathcal{E}}{\mathcal{E}} + \frac{1}{3} \del_i \chi_1  \frac{\del_i \mathcal{E}}{\mathcal{E}}   \nonumber \\
&\quad + \frac{1}{3} \frac{\del_i \mathcal{E}}{\mathcal{E}}  \chi_2   \del_i \pqty{ \mu + \frac{\sqrt{3} T}{2}  } - \theta_{(1)}  \chi_2  \del_0 \pqty{ \mu + \frac{\sqrt{3} T}{2}  }  - \chi_2  \del_i \del_i \pqty{ \mu + \frac{\sqrt{3} T}{2}  } . \label{pmjm2}
\end{align}

We can now write these equations in a form in which the  time derivative is on the left hand side and the rest of the terms, involving only spatial derivatives is on the right hand side.

The energy conservation equation is written as,
\begin{align}
2 \pqty{ \frac{\mu^2}{\sqrt{3}} + \mu T} \del_0 \mu + \pqty{ \mu^2 + 2 \sqrt{3} \mu T } \del_0 T  = 4\pi G\bqty{-  \frac32 \mathcal{E}  \theta_{(1)}   +  2\eta \sigma^{ij}_{(1)}  \sigma_{{ij},(1)} } \equiv B_1. \label{defb1}
\end{align}
The momentum conservation equations can be written from eq,(\ref{pmtm1}) as,
\begin{equation}
\del_0 \beta_i =- \frac{1}{3} \frac{ \del_i \mathcal{E}}{\mathcal{E}} - \frac29 \frac{ \del_j \mathcal{E}}{\mathcal{E}^2} \eta \pqty{2  \sigma^{ij}_{(1)}   + 2 \omega^{ij}_{(1)} +  \theta_{(1)} \delta_{ij} }   + \frac{4}{3\mathcal{E}} \sigma^{ij}_{(1)} \del_j \eta + \frac{4}{3\mathcal{E}} \eta \del_j \sigma^{ij}_{(1)} \equiv B_{i+1} . \label{defb2}
\end{equation}
The current conservation equation can be stated as,
\begin{align}
&\quad \pqty{ \frac{2 \mu}{\sqrt{3}} + T +  2\theta_{(1)}   } \del_0 \mu + \pqty{ \mu + \sqrt{3} T + \sqrt{3} \theta_{(1)}   } \del_0 T  \nonumber  \\
&=4\pi G \Bigg[ - \rho \theta_{(1)} +  \chi_1 \bqty{ \frac{4}{9} \frac{\del^i  \mathcal{E}\del_i  \mathcal{E}}{\mathcal{E}^2} - \frac{1}{3} \frac{\del_i \del^i \mathcal{E}}{\mathcal{E}}  - \theta_{(1)}^2 } - \frac{1}{3} \del_i \chi_1  \frac{\del_i \mathcal{E}}{\mathcal{E}}   \nonumber \\
&\quad - \frac{1}{3} \frac{\del_i \mathcal{E}}{\mathcal{E}}  \chi_2   \del_i \pqty{ \mu + \frac{\sqrt{3} T}{2}  } + \chi_2  \del_i \del_i \pqty{ \mu + \frac{\sqrt{3} T}{2}  } \Bigg] \equiv B_4. \label{defb4}
\end{align}

The matrix  describing the time evolution can then be written as,
\begin{equation}
A =  \begin{pmatrix}
2 \pqty{ \frac{\mu^2}{\sqrt{3}} + \mu T}  &0 &0  & \pqty{ \mu^2 + 2 \sqrt{3} \mu T } \\
0 & 1 & 0 & 0 \\
0 & 0 & 1 & 0 \\
\pqty{ \frac{2 \mu}{\sqrt{3}} + T +  2\theta_{(1)}   } & 0 &0 &  \pqty{ \mu + \sqrt{3} T +  \sqrt{3} \theta_{(1)}  }
\end{pmatrix}. \label{adef}
\end{equation}
The time evolution equation as a matrix equation reads,
\begin{equation}
\label{timeevoleq}
A \begin{pmatrix}
\dot \mu  \\ \dot \beta^1 \\ \dot \beta^2 \\ \dot T
\end{pmatrix} = \begin{pmatrix}
B_1 \\ B_2 \\ B_3 \\ B_4
\end{pmatrix},
\end{equation}
thus agreeing with eq.(\ref{tevo}).

Let us now analyse the equations in more detail. We consider eq.(\ref{defb1}) and a linear combination of eqs.(\ref{defb1}) and (\ref{defb4}), namely:  (\ref{defb1}) $- \mu \times$  (\ref{defb4}). We also write the right hand sides of this equations, in a perturbative series in $\epsilon$, where $\epsilon$ generically includes factors of $T$ or spatial derivatives:
\begin{align}
B_1 &= B_1^{(1)} \epsilon + B_1^{(2)} \epsilon^2, \label{b1ex} \\ 
B_1 - \mu B_4 \equiv \hat{B} &= \hat{B}^{(2)} \epsilon^2. \label{b2exp}
\end{align}
Note that we have used the fact that the $\mathcal{O}(\epsilon)$ term in $\hat{B}$ vanishes having used the definitions of $B_1$, eq.(\ref{defb1}) and $B_2$ eq.(\ref{defb4}) and using the explicit expressions for the energy density (\ref{elint}) and charge density (\ref{rholint}): the $\mathcal{O} (\epsilon)$ term in $\hat{B}$ is given by,
\begin{equation}
\theta_{(1)} \pqty{ - \frac{3}{2} 4\pi G \mathcal{E} (T=0) + \mu \, 4\pi G  \rho (T=0)   } =  \theta_{(1)} \pqty{ - \frac{3}{2} \frac{2\mu^2}{3\sqrt{3}} + \mu  \frac{\mu}{\sqrt{3}} } = 0. 
\end{equation} 
We can thus write these equations as,

\begin{align}
 \frac{\mu^2}{\sqrt{3}} \pqty{ 2\dot{\mu} + \sqrt{3} \dot{T} }+ 2\mu T \pqty{ \dot{\mu} + \sqrt{3} \dot{T} } &= B_1^{(1)} \epsilon + B_1^{(2)} \epsilon^2,  \label{linc1}\\
\mu T \pqty{ \dot{\mu} + \sqrt{3}  \dot{T} } - \mu \theta_{(1)} \pqty{ 2\dot{\mu} + \sqrt{3} \dot{T} } &= \hat{B}^{(2)} \epsilon^2. \label{linc2}
\end{align}

Solving the first equation perturbatively, we have at the leading order,
\begin{equation}
\label{2m3t1}
2\dot{\mu} + \sqrt{3} \dot{T} = \frac{\sqrt{3}}{\mu^2} B_1^{(1)} \epsilon.
\end{equation} 
Inputting this into the second equation (\ref{linc2}), we have,
\begin{equation}
\dot{\mu} +\sqrt{3}  \dot{T} = \frac{\hat{B}^{(2)} \epsilon^2}{\mu T} + \frac{\sqrt{3}}{\mu^2 T} \theta_{(1)} B_1^{(1)} \epsilon. \label{2m3t2}
\end{equation}

Before we end this appendix, let us comment on the nature of approximations. While it is true that, as mentioned in section \ref{sec-deo}, we count any derivative generically as $\mathcal{O}(\epsilon)$, we should actually treat a spatial derivative of $T$ --- $\del_i T$ as $\mathcal{O} (\epsilon^2)$, from the point of view of the time development considered here. The reason is that the spatial derivative $\del_i T$ should be considered to be part of the initial value data. If we require the initial system to lie in our regime (\ref{condtwo}), we should not allow the variation of the temperature to be bigger than the temperature itself.

\bibliographystyle{JHEP}
\bibliography{nefl_ref}

\providecommand{\href}[2]{#2}\begingroup\raggedright\begin{thebibliography}{100}

\bibitem{Aharony:1999ti}
O.~Aharony, S.~S. Gubser, J.~M. Maldacena, H.~Ooguri and Y.~Oz, \emph{{Large N
  field theories, string theory and gravity}},
  \href{https://doi.org/10.1016/S0370-1573(99)00083-6}{\emph{Phys. Rept.}
  {\bfseries 323} (2000) 183}
  [\href{https://arxiv.org/abs/hep-th/9905111}{{\ttfamily hep-th/9905111}}].

\bibitem{Bhattacharyya:2008jc}
S.~Bhattacharyya, V.~E. Hubeny, S.~Minwalla and M.~Rangamani, \emph{{Nonlinear
  Fluid Dynamics from Gravity}},
  \href{https://doi.org/10.1088/1126-6708/2008/02/045}{\emph{JHEP} {\bfseries
  02} (2008) 045} [\href{https://arxiv.org/abs/0712.2456}{{\ttfamily
  0712.2456}}].

\bibitem{Bhattacharyya:2007vs}
S.~Bhattacharyya, S.~Lahiri, R.~Loganayagam and S.~Minwalla, \emph{{Large
  rotating AdS black holes from fluid mechanics}},
  \href{https://doi.org/10.1088/1126-6708/2008/09/054}{\emph{JHEP} {\bfseries
  09} (2008) 054} [\href{https://arxiv.org/abs/0708.1770}{{\ttfamily
  0708.1770}}].

\bibitem{Loganayagam:2008is}
R.~Loganayagam, \emph{{Entropy Current in Conformal Hydrodynamics}},
  \href{https://doi.org/10.1088/1126-6708/2008/05/087}{\emph{JHEP} {\bfseries
  05} (2008) 087} [\href{https://arxiv.org/abs/0801.3701}{{\ttfamily
  0801.3701}}].

\bibitem{Bhattacharyya:2008xc}
S.~Bhattacharyya, V.~E. Hubeny, R.~Loganayagam, G.~Mandal, S.~Minwalla,
  T.~Morita et~al., \emph{{Local Fluid Dynamical Entropy from Gravity}},
  \href{https://doi.org/10.1088/1126-6708/2008/06/055}{\emph{JHEP} {\bfseries
  06} (2008) 055} [\href{https://arxiv.org/abs/0803.2526}{{\ttfamily
  0803.2526}}].

\bibitem{Bhattacharyya:2008ji}
S.~Bhattacharyya, R.~Loganayagam, S.~Minwalla, S.~Nampuri, S.~P. Trivedi and
  S.~R. Wadia, \emph{{Forced Fluid Dynamics from Gravity}},
  \href{https://doi.org/10.1088/1126-6708/2009/02/018}{\emph{JHEP} {\bfseries
  02} (2009) 018} [\href{https://arxiv.org/abs/0806.0006}{{\ttfamily
  0806.0006}}].

\bibitem{Banerjee:2008th}
N.~Banerjee, J.~Bhattacharya, S.~Bhattacharyya, S.~Dutta, R.~Loganayagam and
  P.~Surowka, \emph{{Hydrodynamics from charged black branes}},
  \href{https://doi.org/10.1007/JHEP01(2011)094}{\emph{JHEP} {\bfseries 01}
  (2011) 094} [\href{https://arxiv.org/abs/0809.2596}{{\ttfamily 0809.2596}}].

\bibitem{Bhattacharyya:2008mz}
S.~Bhattacharyya, R.~Loganayagam, I.~Mandal, S.~Minwalla and A.~Sharma,
  \emph{{Conformal Nonlinear Fluid Dynamics from Gravity in Arbitrary
  Dimensions}},
  \href{https://doi.org/10.1088/1126-6708/2008/12/116}{\emph{JHEP} {\bfseries
  12} (2008) 116} [\href{https://arxiv.org/abs/0809.4272}{{\ttfamily
  0809.4272}}].

\bibitem{Bhattacharyya:2008kq}
S.~Bhattacharyya, S.~Minwalla and S.~R. Wadia, \emph{{The Incompressible
  Non-Relativistic Navier-Stokes Equation from Gravity}},
  \href{https://doi.org/10.1088/1126-6708/2009/08/059}{\emph{JHEP} {\bfseries
  08} (2009) 059} [\href{https://arxiv.org/abs/0810.1545}{{\ttfamily
  0810.1545}}].

\bibitem{Hubeny:2009zz}
V.~E. Hubeny, M.~Rangamani, S.~Minwalla and M.~Van~Raamsdonk, \emph{{The
  fluid-gravity correspondence: The membrane at the end of the universe}},
  \href{https://doi.org/10.1142/S0218271808014084}{\emph{Int. J. Mod. Phys. D}
  {\bfseries 17} (2009) 2571}.

\bibitem{Bhattacharya:2011eea}
J.~Bhattacharya, S.~Bhattacharyya and S.~Minwalla, \emph{{Dissipative
  Superfluid dynamics from gravity}},
  \href{https://doi.org/10.1007/JHEP04(2011)125}{\emph{JHEP} {\bfseries 04}
  (2011) 125} [\href{https://arxiv.org/abs/1101.3332}{{\ttfamily 1101.3332}}].

\bibitem{Bhattacharya:2011tra}
J.~Bhattacharya, S.~Bhattacharyya, S.~Minwalla and A.~Yarom, \emph{{A Theory of
  first order dissipative superfluid dynamics}},
  \href{https://doi.org/10.1007/JHEP05(2014)147}{\emph{JHEP} {\bfseries 05}
  (2014) 147} [\href{https://arxiv.org/abs/1105.3733}{{\ttfamily 1105.3733}}].

\bibitem{Rangamani:2009xk}
M.~Rangamani, \emph{{Gravity and Hydrodynamics: Lectures on the fluid-gravity
  correspondence}},
  \href{https://doi.org/10.1088/0264-9381/26/22/224003}{\emph{Class. Quant.
  Grav.} {\bfseries 26} (2009) 224003}
  [\href{https://arxiv.org/abs/0905.4352}{{\ttfamily 0905.4352}}].

\bibitem{Hubeny:2011hd}
V.~E. Hubeny, S.~Minwalla and M.~Rangamani, \emph{{The fluid/gravity
  correspondence}},  in \emph{{Black holes in higher dimensions}} (G.~T.
  Horowitz, ed.), pp.~348--386.
\newblock Cambridge University Press, Cambridge, 2012.
\newblock \href{https://arxiv.org/abs/1107.5780}{{\ttfamily 1107.5780}}.

\bibitem{Kanitscheider:2009as}
I.~Kanitscheider and K.~Skenderis, \emph{{Universal hydrodynamics of
  non-conformal branes}},
  \href{https://doi.org/10.1088/1126-6708/2009/04/062}{\emph{JHEP} {\bfseries
  04} (2009) 062} [\href{https://arxiv.org/abs/0901.1487}{{\ttfamily
  0901.1487}}].

\bibitem{David:2009np}
J.~R. David, M.~Mahato and S.~R. Wadia, \emph{{Hydrodynamics from the
  D1-brane}}, \href{https://doi.org/10.1088/1126-6708/2009/04/042}{\emph{JHEP}
  {\bfseries 04} (2009) 042} [\href{https://arxiv.org/abs/0901.2013}{{\ttfamily
  0901.2013}}].

\bibitem{Erdmenger:2008rm}
J.~Erdmenger, M.~Haack, M.~Kaminski and A.~Yarom, \emph{{Fluid dynamics of
  R-charged black holes}},
  \href{https://doi.org/10.1088/1126-6708/2009/01/055}{\emph{JHEP} {\bfseries
  01} (2009) 055} [\href{https://arxiv.org/abs/0809.2488}{{\ttfamily
  0809.2488}}].

\bibitem{Son:2007vk}
D.~T. Son and A.~O. Starinets, \emph{{Viscosity, Black Holes, and Quantum Field
  Theory}},
  \href{https://doi.org/10.1146/annurev.nucl.57.090506.123120}{\emph{Ann.\
  Rev.\ Nucl.\ Part.\ Sci.} {\bfseries 57} (2007) 95}
  [\href{https://arxiv.org/abs/0704.0240}{{\ttfamily 0704.0240}}].

\bibitem{Erdmenger:2007cm}
J.~Erdmenger, N.~Evans, I.~Kirsch and E.~Threlfall, \emph{{Mesons in
  Gauge/Gravity Duals - A Review}},
  \href{https://doi.org/10.1140/epja/i2007-10540-1}{\emph{Eur. Phys. J. A}
  {\bfseries 35} (2008) 81} [\href{https://arxiv.org/abs/0711.4467}{{\ttfamily
  0711.4467}}].

\bibitem{Gubser:2009md}
S.~S. Gubser and A.~Karch, \emph{{From gauge-string duality to strong
  interactions: A Pedestrian's Guide}},
  \href{https://doi.org/10.1146/annurev.nucl.010909.083602}{\emph{Ann. Rev.
  Nucl. Part. Sci.} {\bfseries 59} (2009) 145}
  [\href{https://arxiv.org/abs/0901.0935}{{\ttfamily 0901.0935}}].

\bibitem{Herzog:2009xv}
C.~P. Herzog, \emph{{Lectures on Holographic Superfluidity and
  Superconductivity}},
  \href{https://doi.org/10.1088/1751-8113/42/34/343001}{\emph{J. Phys. A}
  {\bfseries 42} (2009) 343001}
  [\href{https://arxiv.org/abs/0904.1975}{{\ttfamily 0904.1975}}].

\bibitem{Hartnoll:2009sz}
S.~A. Hartnoll, \emph{{Lectures on holographic methods for condensed matter
  physics}}, \href{https://doi.org/10.1088/0264-9381/26/22/224002}{\emph{Class.
  Quant. Grav.} {\bfseries 26} (2009) 224002}
  [\href{https://arxiv.org/abs/0903.3246}{{\ttfamily 0903.3246}}].

\bibitem{McGreevy:2009xe}
J.~McGreevy, \emph{{Holographic duality with a view toward many-body physics}},
  \href{https://doi.org/10.1155/2010/723105}{\emph{Adv. High Energy Phys.}
  {\bfseries 2010} (2010) 723105}
  [\href{https://arxiv.org/abs/0909.0518}{{\ttfamily 0909.0518}}].

\bibitem{Horowitz:2010gk}
G.~T. Horowitz, \emph{{Introduction to Holographic Superconductors}},
  \href{https://doi.org/10.1007/978-3-642-04864-7\_10}{\emph{Lect. Notes Phys.}
  {\bfseries 828} (2011) 313}
  [\href{https://arxiv.org/abs/1002.1722}{{\ttfamily 1002.1722}}].

\bibitem{CasalderreySolana:2011us}
J.~Casalderrey-Solana, H.~Liu, D.~Mateos, K.~Rajagopal and U.~A. Wiedemann,
  \emph{{Gauge/String Duality, Hot QCD and Heavy Ion Collisions}}. Cambridge
  University Press, 2014,
  \href{https://doi.org/10.1017/CBO9781139136747}{10.1017/CBO9781139136747},
  [\href{https://arxiv.org/abs/1101.0618}{{\ttfamily 1101.0618}}].

\bibitem{Hartnoll:2016apf}
S.~A. Hartnoll, A.~Lucas and S.~Sachdev, \emph{{Holographic quantum matter}},
  \href{https://arxiv.org/abs/1612.07324}{{\ttfamily 1612.07324}}.

\bibitem{Maldacena:2016upp}
J.~Maldacena, D.~Stanford and Z.~Yang, \emph{{Conformal symmetry and its
  breaking in two dimensional Nearly Anti-de-Sitter space}},
  \href{https://doi.org/10.1093/ptep/ptw124}{\emph{PTEP} {\bfseries 2016}
  (2016) 12C104} [\href{https://arxiv.org/abs/1606.01857}{{\ttfamily
  1606.01857}}].

\bibitem{Teitelboim:1983ux}
C.~Teitelboim, \emph{{Gravitation and Hamiltonian Structure in Two Space-Time
  Dimensions}}, \href{https://doi.org/10.1016/0370-2693(83)90012-6}{\emph{Phys.
  Lett. B} {\bfseries 126} (1983) 41}.

\bibitem{Jackiw:1984je}
R.~Jackiw, \emph{{Lower Dimensional Gravity}},
  \href{https://doi.org/10.1016/0550-3213(85)90448-1}{\emph{Nucl. Phys. B}
  {\bfseries 252} (1985) 343}.

\bibitem{Sachdev:1992fk}
S.~Sachdev and J.~Ye, \emph{{Gapless spin fluid ground state in a random,
  quantum Heisenberg magnet}},
  \href{https://doi.org/10.1103/PhysRevLett.70.3339}{\emph{Phys. Rev. Lett.}
  {\bfseries 70} (1993) 3339}
  [\href{https://arxiv.org/abs/cond-mat/9212030}{{\ttfamily
  cond-mat/9212030}}].

\bibitem{Kitaev}
A.~Y. Kitaev, ``{\it A simple model of quantum holography}.'' Talks
  (\href{http://online.kitp.ucsb.edu/online/entangled15/kitaev/}{I},
  \href{http://online.kitp.ucsb.edu/online/entangled15/kitaev2/}{II}) presented
  at the programme {\it Entanglement in Strongly-Correlated Quantum Matter},
  Kavli Intitute for Theoretical Physics, University of California, Santa
  Barbara, 2015.

\bibitem{Maldacena:2016hyu}
J.~Maldacena and D.~Stanford, \emph{{Remarks on the Sachdev-Ye-Kitaev model}},
  \href{https://doi.org/10.1103/PhysRevD.94.106002}{\emph{Phys. Rev. D}
  {\bfseries 94} (2016) 106002}
  [\href{https://arxiv.org/abs/1604.07818}{{\ttfamily 1604.07818}}].

\bibitem{Davison:2016ngz}
R.~A. Davison, W.~Fu, A.~Georges, Y.~Gu, K.~Jensen and S.~Sachdev,
  \emph{{Thermoelectric transport in disordered metals without quasiparticles:
  The Sachdev-Ye-Kitaev models and holography}},
  \href{https://doi.org/10.1103/PhysRevB.95.155131}{\emph{Phys. Rev. B}
  {\bfseries 95} (2017) 155131}
  [\href{https://arxiv.org/abs/1612.00849}{{\ttfamily 1612.00849}}].

\bibitem{Nayak:2018qej}
P.~Nayak, A.~Shukla, R.~M. Soni, S.~P. Trivedi and V.~Vishal, \emph{{On the
  Dynamics of Near-Extremal Black Holes}},
  \href{https://doi.org/10.1007/JHEP09(2018)048}{\emph{JHEP} {\bfseries 09}
  (2018) 048} [\href{https://arxiv.org/abs/1802.09547}{{\ttfamily
  1802.09547}}].

\bibitem{Moitra:2018jqs}
U.~Moitra, S.~P. Trivedi and V.~Vishal, \emph{{Extremal and near-extremal black
  holes and near-CFT$_{1}$}},
  \href{https://doi.org/10.1007/JHEP07(2019)055}{\emph{JHEP} {\bfseries 07}
  (2019) 055} [\href{https://arxiv.org/abs/1808.08239}{{\ttfamily
  1808.08239}}].

\bibitem{Moitra:2019bub}
U.~Moitra, S.~K. Sake, S.~P. Trivedi and V.~Vishal, \emph{{Jackiw-Teitelboim
  Gravity and Rotating Black Holes}},
  \href{https://doi.org/10.1007/JHEP11(2019)047}{\emph{JHEP} {\bfseries 11}
  (2019) 047} [\href{https://arxiv.org/abs/1905.10378}{{\ttfamily
  1905.10378}}].

\bibitem{Moitra:2019xoj}
U.~Moitra, S.~K. Sake, S.~P. Trivedi and V.~Vishal, \emph{{Jackiw-Teitelboim
  Model Coupled to Conformal Matter in the Semi-Classical Limit}},
  \href{https://arxiv.org/abs/1908.08523}{{\ttfamily 1908.08523}}.

\bibitem{Sarosi:2017ykf}
G.~Sárosi, \emph{{AdS$_{2}$ holography and the SYK model}},
  \href{https://doi.org/10.22323/1.323.0001}{\emph{PoS} {\bfseries Modave2017}
  (2018) 001} [\href{https://arxiv.org/abs/1711.08482}{{\ttfamily
  1711.08482}}].

\bibitem{Rosenhaus:2018dtp}
V.~Rosenhaus, \emph{{An introduction to the SYK model}},
  \href{https://doi.org/10.1088/1751-8121/ab2ce1}{\emph{J. Phys. A} {\bfseries
  52} (2019) 323001} [\href{https://arxiv.org/abs/1807.03334}{{\ttfamily
  1807.03334}}].

\bibitem{Denef:2009yy}
F.~Denef, S.~A. Hartnoll and S.~Sachdev, \emph{{Quantum oscillations and black
  hole ringing}},
  \href{https://doi.org/10.1103/PhysRevD.80.126016}{\emph{Phys.\ Rev.\ D}
  {\bfseries 80} (2009) 126016}
  [\href{https://arxiv.org/abs/0908.1788}{{\ttfamily 0908.1788}}].

\bibitem{Edalati:2009bi}
M.~Edalati, J.~I. Jottar and R.~G. Leigh, \emph{{Transport Coefficients at Zero
  Temperature from Extremal Black Holes}},
  \href{https://doi.org/10.1007/JHEP01(2010)018}{\emph{JHEP} {\bfseries 01}
  (2010) 018} [\href{https://arxiv.org/abs/0910.0645}{{\ttfamily 0910.0645}}].

\bibitem{Edalati:2010hk}
M.~Edalati, J.~I. Jottar and R.~G. Leigh, \emph{{Shear Modes, Criticality and
  Extremal Black Holes}},
  \href{https://doi.org/10.1007/JHEP04(2010)075}{\emph{JHEP} {\bfseries 04}
  (2010) 075} [\href{https://arxiv.org/abs/1001.0779}{{\ttfamily 1001.0779}}].

\bibitem{Edalati:2010pn}
M.~Edalati, J.~I. Jottar and R.~G. Leigh, \emph{{Holography and the sound of
  criticality}}, \href{https://doi.org/10.1007/JHEP10(2010)058}{\emph{JHEP}
  {\bfseries 10} (2010) 058} [\href{https://arxiv.org/abs/1005.4075}{{\ttfamily
  1005.4075}}].

\bibitem{Davison:2011uk}
R.~A. Davison and N.~K. Kaplis, \emph{{Bosonic excitations of the $AdS_4$
  Reissner-Nordstrom black hole}},
  \href{https://doi.org/10.1007/JHEP12(2011)037}{\emph{JHEP} {\bfseries 12}
  (2011) 037} [\href{https://arxiv.org/abs/1111.0660}{{\ttfamily 1111.0660}}].

\bibitem{Davison:2013bxa}
R.~A. Davison and A.~Parnachev, \emph{{Hydrodynamics of cold holographic
  matter}}, \href{https://doi.org/10.1007/JHEP06(2013)100}{\emph{JHEP}
  {\bfseries 06} (2013) 100} [\href{https://arxiv.org/abs/1303.6334}{{\ttfamily
  1303.6334}}].

\bibitem{Erdmenger:2016wyp}
J.~Erdmenger, D.~Fernandez, P.~Goulart and P.~Witkowski, \emph{{Conductivities
  from attractors}}, \href{https://doi.org/10.1007/JHEP03(2017)147}{\emph{JHEP}
  {\bfseries 03} (2017) 147}
  [\href{https://arxiv.org/abs/1611.09381}{{\ttfamily 1611.09381}}].

\bibitem{Oh:2010jp}
J.-H. Oh, \emph{{Small Amplitude Forced Fluid Dynamics from Gravity at T = 0}},
  \href{https://doi.org/10.1140/epjc/s10052-011-1841-9}{\emph{Eur. Phys. J. C}
  {\bfseries 71} (2011) 1841}
  [\href{https://arxiv.org/abs/1012.1040}{{\ttfamily 1012.1040}}].

\bibitem{Policastro:2001yc}
G.~Policastro, D.~T. Son and A.~O. Starinets, \emph{{The Shear viscosity of
  strongly coupled N=4 supersymmetric Yang-Mills plasma}},
  \href{https://doi.org/10.1103/PhysRevLett.87.081601}{\emph{Phys.\ Rev.\
  Lett.} {\bfseries 87} (2001) 081601}
  [\href{https://arxiv.org/abs/hep-th/0104066}{{\ttfamily hep-th/0104066}}].

\bibitem{Policastro:2002se}
G.~Policastro, D.~T. Son and A.~O. Starinets, \emph{{From AdS / CFT
  correspondence to hydrodynamics}},
  \href{https://doi.org/10.1088/1126-6708/2002/09/043}{\emph{JHEP} {\bfseries
  09} (2002) 043} [\href{https://arxiv.org/abs/hep-th/0205052}{{\ttfamily
  hep-th/0205052}}].

\bibitem{Son:2002sd}
D.~T. Son and A.~O. Starinets, \emph{{Minkowski space correlators in AdS / CFT
  correspondence: Recipe and applications}},
  \href{https://doi.org/10.1088/1126-6708/2002/09/042}{\emph{JHEP} {\bfseries
  09} (2002) 042} [\href{https://arxiv.org/abs/hep-th/0205051}{{\ttfamily
  hep-th/0205051}}].

\bibitem{Herzog:2002fn}
C.~P. Herzog, \emph{{The Hydrodynamics of M theory}},
  \href{https://doi.org/10.1088/1126-6708/2002/12/026}{\emph{JHEP} {\bfseries
  12} (2002) 026} [\href{https://arxiv.org/abs/hep-th/0210126}{{\ttfamily
  hep-th/0210126}}].

\bibitem{Policastro:2002tn}
G.~Policastro, D.~T. Son and A.~O. Starinets, \emph{{From AdS / CFT
  correspondence to hydrodynamics. 2. Sound waves}},
  \href{https://doi.org/10.1088/1126-6708/2002/12/054}{\emph{JHEP} {\bfseries
  12} (2002) 054} [\href{https://arxiv.org/abs/hep-th/0210220}{{\ttfamily
  hep-th/0210220}}].

\bibitem{Herzog:2002pc}
C.~Herzog and D.~Son, \emph{{Schwinger-Keldysh propagators from AdS/CFT
  correspondence}},
  \href{https://doi.org/10.1088/1126-6708/2003/03/046}{\emph{JHEP} {\bfseries
  03} (2003) 046} [\href{https://arxiv.org/abs/hep-th/0212072}{{\ttfamily
  hep-th/0212072}}].

\bibitem{Nunez:2003eq}
A.~Nunez and A.~O. Starinets, \emph{{AdS / CFT correspondence, quasinormal
  modes, and thermal correlators in N=4 SYM}},
  \href{https://doi.org/10.1103/PhysRevD.67.124013}{\emph{Phys. Rev. D}
  {\bfseries 67} (2003) 124013}
  [\href{https://arxiv.org/abs/hep-th/0302026}{{\ttfamily hep-th/0302026}}].

\bibitem{Herzog:2003ke}
C.~P. Herzog, \emph{{The Sound of M theory}},
  \href{https://doi.org/10.1103/PhysRevD.68.024013}{\emph{Phys.\ Rev.\ D}
  {\bfseries 68} (2003) 024013}
  [\href{https://arxiv.org/abs/hep-th/0302086}{{\ttfamily hep-th/0302086}}].

\bibitem{Kovtun:2003wp}
P.~Kovtun, D.~T. Son and A.~O. Starinets, \emph{{Holography and hydrodynamics:
  Diffusion on stretched horizons}},
  \href{https://doi.org/10.1088/1126-6708/2003/10/064}{\emph{JHEP} {\bfseries
  10} (2003) 064} [\href{https://arxiv.org/abs/hep-th/0309213}{{\ttfamily
  hep-th/0309213}}].

\bibitem{Buchel:2003tz}
A.~Buchel and J.~T. Liu, \emph{{Universality of the shear viscosity in
  supergravity}},
  \href{https://doi.org/10.1103/PhysRevLett.93.090602}{\emph{Phys.\ Rev.\
  Lett.} {\bfseries 93} (2004) 090602}
  [\href{https://arxiv.org/abs/hep-th/0311175}{{\ttfamily hep-th/0311175}}].

\bibitem{Kovtun:2004de}
P.~Kovtun, D.~T. Son and A.~O. Starinets, \emph{{Viscosity in strongly
  interacting quantum field theories from black hole physics}},
  \href{https://doi.org/10.1103/PhysRevLett.94.111601}{\emph{Phys.\ Rev.\
  Lett.} {\bfseries 94} (2005) 111601}
  [\href{https://arxiv.org/abs/hep-th/0405231}{{\ttfamily hep-th/0405231}}].

\bibitem{Buchel:2004hw}
A.~Buchel, \emph{{N=2* hydrodynamics}},
  \href{https://doi.org/10.1016/j.nuclphysb.2004.11.039}{\emph{Nucl.\ Phys.\ B}
  {\bfseries 708} (2005) 451}
  [\href{https://arxiv.org/abs/hep-th/0406200}{{\ttfamily hep-th/0406200}}].

\bibitem{Buchel:2004di}
A.~Buchel, J.~T. Liu and A.~O. Starinets, \emph{{Coupling constant dependence
  of the shear viscosity in N=4 supersymmetric Yang-Mills theory}},
  \href{https://doi.org/10.1016/j.nuclphysb.2004.11.055}{\emph{Nucl.\ Phys.\ B}
  {\bfseries 707} (2005) 56}
  [\href{https://arxiv.org/abs/hep-th/0406264}{{\ttfamily hep-th/0406264}}].

\bibitem{Buchel:2004qq}
A.~Buchel, \emph{{On universality of stress-energy tensor correlation functions
  in supergravity}},
  \href{https://doi.org/10.1016/j.physletb.2005.01.052}{\emph{Phys.\ Lett.\ B}
  {\bfseries 609} (2005) 392}
  [\href{https://arxiv.org/abs/hep-th/0408095}{{\ttfamily hep-th/0408095}}].

\bibitem{Kovtun:2005ev}
P.~K. Kovtun and A.~O. Starinets, \emph{{Quasinormal modes and holography}},
  \href{https://doi.org/10.1103/PhysRevD.72.086009}{\emph{Phys.\ Rev.\ D}
  {\bfseries 72} (2005) 086009}
  [\href{https://arxiv.org/abs/hep-th/0506184}{{\ttfamily hep-th/0506184}}].

\bibitem{Benincasa:2005iv}
P.~Benincasa, A.~Buchel and A.~O. Starinets, \emph{{Sound waves in strongly
  coupled non-conformal gauge theory plasma}},
  \href{https://doi.org/10.1016/j.nuclphysb.2005.11.005}{\emph{Nucl.\ Phys.\ B}
  {\bfseries 733} (2006) 160}
  [\href{https://arxiv.org/abs/hep-th/0507026}{{\ttfamily hep-th/0507026}}].

\bibitem{Shuryak:2005ia}
E.~Shuryak, S.-J. Sin and I.~Zahed, \emph{{A Gravity dual of RHIC collisions}},
  \href{https://doi.org/10.3938/jkps.50.384}{\emph{J. Korean Phys. Soc.}
  {\bfseries 50} (2007) 384}
  [\href{https://arxiv.org/abs/hep-th/0511199}{{\ttfamily hep-th/0511199}}].

\bibitem{Janik:2005zt}
R.~A. Janik and R.~B. Peschanski, \emph{{Asymptotic perfect fluid dynamics as a
  consequence of Ads/CFT}},
  \href{https://doi.org/10.1103/PhysRevD.73.045013}{\emph{Phys. Rev. D}
  {\bfseries 73} (2006) 045013}
  [\href{https://arxiv.org/abs/hep-th/0512162}{{\ttfamily hep-th/0512162}}].

\bibitem{Benincasa:2006fu}
P.~Benincasa, A.~Buchel and R.~Naryshkin, \emph{{The Shear viscosity of gauge
  theory plasma with chemical potentials}},
  \href{https://doi.org/10.1016/j.physletb.2006.12.030}{\emph{Phys.\ Lett.\ B}
  {\bfseries 645} (2007) 309}
  [\href{https://arxiv.org/abs/hep-th/0610145}{{\ttfamily hep-th/0610145}}].

\bibitem{Mas:2006dy}
J.~Mas, \emph{{Shear viscosity from R-charged AdS black holes}},
  \href{https://doi.org/10.1088/1126-6708/2006/03/016}{\emph{JHEP} {\bfseries
  03} (2006) 016} [\href{https://arxiv.org/abs/hep-th/0601144}{{\ttfamily
  hep-th/0601144}}].

\bibitem{Son:2006em}
D.~T. Son and A.~O. Starinets, \emph{{Hydrodynamics of r-charged black holes}},
  \href{https://doi.org/10.1088/1126-6708/2006/03/052}{\emph{JHEP} {\bfseries
  03} (2006) 052} [\href{https://arxiv.org/abs/hep-th/0601157}{{\ttfamily
  hep-th/0601157}}].

\bibitem{Saremi:2006ep}
O.~Saremi, \emph{{The Viscosity bound conjecture and hydrodynamics of M2-brane
  theory at finite chemical potential}},
  \href{https://doi.org/10.1088/1126-6708/2006/10/083}{\emph{JHEP} {\bfseries
  10} (2006) 083} [\href{https://arxiv.org/abs/hep-th/0601159}{{\ttfamily
  hep-th/0601159}}].

\bibitem{Maeda:2006by}
K.~Maeda, M.~Natsuume and T.~Okamura, \emph{{Viscosity of gauge theory plasma
  with a chemical potential from AdS/CFT}},
  \href{https://doi.org/10.1103/PhysRevD.73.066013}{\emph{Phys.\ Rev.\ D}
  {\bfseries 73} (2006) 066013}
  [\href{https://arxiv.org/abs/hep-th/0602010}{{\ttfamily hep-th/0602010}}].

\bibitem{Herzog:2006gh}
C.~Herzog, A.~Karch, P.~Kovtun, C.~Kozcaz and L.~Yaffe, \emph{{Energy loss of a
  heavy quark moving through N=4 supersymmetric Yang-Mills plasma}},
  \href{https://doi.org/10.1088/1126-6708/2006/07/013}{\emph{JHEP} {\bfseries
  07} (2006) 013} [\href{https://arxiv.org/abs/hep-th/0605158}{{\ttfamily
  hep-th/0605158}}].

\bibitem{Liu:2006ug}
H.~Liu, K.~Rajagopal and U.~A. Wiedemann, \emph{{Calculating the jet quenching
  parameter from AdS/CFT}},
  \href{https://doi.org/10.1103/PhysRevLett.97.182301}{\emph{Phys. Rev. Lett.}
  {\bfseries 97} (2006) 182301}
  [\href{https://arxiv.org/abs/hep-ph/0605178}{{\ttfamily hep-ph/0605178}}].

\bibitem{Gubser:2006bz}
S.~S. Gubser, \emph{{Drag force in AdS/CFT}},
  \href{https://doi.org/10.1103/PhysRevD.74.126005}{\emph{Phys. Rev. D}
  {\bfseries 74} (2006) 126005}
  [\href{https://arxiv.org/abs/hep-th/0605182}{{\ttfamily hep-th/0605182}}].

\bibitem{Janik:2006gp}
R.~A. Janik and R.~B. Peschanski, \emph{{Gauge/gravity duality and
  thermalization of a boost-invariant perfect fluid}},
  \href{https://doi.org/10.1103/PhysRevD.74.046007}{\emph{Phys. Rev. D}
  {\bfseries 74} (2006) 046007}
  [\href{https://arxiv.org/abs/hep-th/0606149}{{\ttfamily hep-th/0606149}}].

\bibitem{Nakamura:2006ih}
S.~Nakamura and S.-J. Sin, \emph{{A Holographic dual of hydrodynamics}},
  \href{https://doi.org/10.1088/1126-6708/2006/09/020}{\emph{JHEP} {\bfseries
  09} (2006) 020} [\href{https://arxiv.org/abs/hep-th/0607123}{{\ttfamily
  hep-th/0607123}}].

\bibitem{Janik:2006ft}
R.~A. Janik, \emph{{Viscous plasma evolution from gravity using AdS/CFT}},
  \href{https://doi.org/10.1103/PhysRevLett.98.022302}{\emph{Phys. Rev. Lett.}
  {\bfseries 98} (2007) 022302}
  [\href{https://arxiv.org/abs/hep-th/0610144}{{\ttfamily hep-th/0610144}}].

\bibitem{Friess:2006kw}
J.~J. Friess, S.~S. Gubser, G.~Michalogiorgakis and S.~S. Pufu,
  \emph{{Expanding plasmas and quasinormal modes of anti-de Sitter black
  holes}}, \href{https://doi.org/10.1088/1126-6708/2007/04/080}{\emph{JHEP}
  {\bfseries 04} (2007) 080}
  [\href{https://arxiv.org/abs/hep-th/0611005}{{\ttfamily hep-th/0611005}}].

\bibitem{Nakamura:2006xk}
S.~Nakamura, Y.~Seo, S.-J. Sin and K.~Yogendran, \emph{{A New Phase at Finite
  Quark Density from AdS/CFT}},
  \href{https://doi.org/10.3938/jkps.52.1734}{\emph{J. Korean Phys. Soc.}
  {\bfseries 52} (2008) 1734}
  [\href{https://arxiv.org/abs/hep-th/0611021}{{\ttfamily hep-th/0611021}}].

\bibitem{Liu:2006he}
H.~Liu, K.~Rajagopal and U.~A. Wiedemann, \emph{{Wilson loops in heavy ion
  collisions and their calculation in AdS/CFT}},
  \href{https://doi.org/10.1088/1126-6708/2007/03/066}{\emph{JHEP} {\bfseries
  03} (2007) 066} [\href{https://arxiv.org/abs/hep-ph/0612168}{{\ttfamily
  hep-ph/0612168}}].

\bibitem{Herzog:2007ij}
C.~P. Herzog, P.~Kovtun, S.~Sachdev and D.~T. Son, \emph{{Quantum critical
  transport, duality, and M-theory}},
  \href{https://doi.org/10.1103/PhysRevD.75.085020}{\emph{Phys.\ Rev.\ D}
  {\bfseries 75} (2007) 085020}
  [\href{https://arxiv.org/abs/hep-th/0701036}{{\ttfamily hep-th/0701036}}].

\bibitem{Heller:2007qt}
M.~P. Heller and R.~A. Janik, \emph{{Viscous hydrodynamics relaxation time from
  AdS/CFT}}, \href{https://doi.org/10.1103/PhysRevD.76.025027}{\emph{Phys. Rev.
  D} {\bfseries 76} (2007) 025027}
  [\href{https://arxiv.org/abs/hep-th/0703243}{{\ttfamily hep-th/0703243}}].

\bibitem{Hartnoll:2007ai}
S.~A. Hartnoll and P.~Kovtun, \emph{{Hall conductivity from dyonic black
  holes}}, \href{https://doi.org/10.1103/PhysRevD.76.066001}{\emph{Phys.\ Rev.\
  D} {\bfseries 76} (2007) 066001}
  [\href{https://arxiv.org/abs/0704.1160}{{\ttfamily 0704.1160}}].

\bibitem{Gubser:2007ga}
S.~S. Gubser, S.~S. Pufu and A.~Yarom, \emph{{Sonic booms and diffusion wakes
  generated by a heavy quark in thermal AdS/CFT}},
  \href{https://doi.org/10.1103/PhysRevLett.100.012301}{\emph{Phys. Rev. Lett.}
  {\bfseries 100} (2008) 012301}
  [\href{https://arxiv.org/abs/0706.4307}{{\ttfamily 0706.4307}}].

\bibitem{Kats:2007mq}
Y.~Kats and P.~Petrov, \emph{{Effect of curvature squared corrections in AdS on
  the viscosity of the dual gauge theory}},
  \href{https://doi.org/10.1088/1126-6708/2009/01/044}{\emph{JHEP} {\bfseries
  01} (2009) 044} [\href{https://arxiv.org/abs/0712.0743}{{\ttfamily
  0712.0743}}].

\bibitem{Brigante:2007nu}
M.~Brigante, H.~Liu, R.~C. Myers, S.~Shenker and S.~Yaida, \emph{{Viscosity
  Bound Violation in Higher Derivative Gravity}},
  \href{https://doi.org/10.1103/PhysRevD.77.126006}{\emph{Phys.\ Rev.\ D}
  {\bfseries 77} (2008) 126006}
  [\href{https://arxiv.org/abs/0712.0805}{{\ttfamily 0712.0805}}].

\bibitem{Baier:2007ix}
R.~Baier, P.~Romatschke, D.~T. Son, A.~O. Starinets and M.~A. Stephanov,
  \emph{{Relativistic viscous hydrodynamics, conformal invariance, and
  holography}},
  \href{https://doi.org/10.1088/1126-6708/2008/04/100}{\emph{JHEP} {\bfseries
  04} (2008) 100} [\href{https://arxiv.org/abs/0712.2451}{{\ttfamily
  0712.2451}}].

\bibitem{Gubser:2008px}
S.~S. Gubser, \emph{{Breaking an Abelian gauge symmetry near a black hole
  horizon}}, \href{https://doi.org/10.1103/PhysRevD.78.065034}{\emph{Phys. Rev.
  D} {\bfseries 78} (2008) 065034}
  [\href{https://arxiv.org/abs/0801.2977}{{\ttfamily 0801.2977}}].

\bibitem{Brigante:2008gz}
M.~Brigante, H.~Liu, R.~C. Myers, S.~Shenker and S.~Yaida, \emph{{The Viscosity
  Bound and Causality Violation}},
  \href{https://doi.org/10.1103/PhysRevLett.100.191601}{\emph{Phys.\ Rev.\
  Lett.} {\bfseries 100} (2008) 191601}
  [\href{https://arxiv.org/abs/0802.3318}{{\ttfamily 0802.3318}}].

\bibitem{Hartnoll:2008vx}
S.~A. Hartnoll, C.~P. Herzog and G.~T. Horowitz, \emph{{Building a Holographic
  Superconductor}},
  \href{https://doi.org/10.1103/PhysRevLett.101.031601}{\emph{Phys. Rev. Lett.}
  {\bfseries 101} (2008) 031601}
  [\href{https://arxiv.org/abs/0803.3295}{{\ttfamily 0803.3295}}].

\bibitem{Gubser:2008yx}
S.~S. Gubser, A.~Nellore, S.~S. Pufu and F.~D. Rocha, \emph{{Thermodynamics and
  bulk viscosity of approximate black hole duals to finite temperature quantum
  chromodynamics}},
  \href{https://doi.org/10.1103/PhysRevLett.101.131601}{\emph{Phys. Rev. Lett.}
  {\bfseries 101} (2008) 131601}
  [\href{https://arxiv.org/abs/0804.1950}{{\ttfamily 0804.1950}}].

\bibitem{Son:2008ye}
D.~Son, \emph{{Toward an AdS/cold atoms correspondence: A Geometric realization
  of the Schrodinger symmetry}},
  \href{https://doi.org/10.1103/PhysRevD.78.046003}{\emph{Phys. Rev. D}
  {\bfseries 78} (2008) 046003}
  [\href{https://arxiv.org/abs/0804.3972}{{\ttfamily 0804.3972}}].

\bibitem{Balasubramanian:2008dm}
K.~Balasubramanian and J.~McGreevy, \emph{{Gravity duals for non-relativistic
  CFTs}}, \href{https://doi.org/10.1103/PhysRevLett.101.061601}{\emph{Phys.
  Rev. Lett.} {\bfseries 101} (2008) 061601}
  [\href{https://arxiv.org/abs/0804.4053}{{\ttfamily 0804.4053}}].

\bibitem{Gubser:2008pc}
S.~S. Gubser, S.~S. Pufu and A.~Yarom, \emph{{Entropy production in collisions
  of gravitational shock waves and of heavy ions}},
  \href{https://doi.org/10.1103/PhysRevD.78.066014}{\emph{Phys. Rev. D}
  {\bfseries 78} (2008) 066014}
  [\href{https://arxiv.org/abs/0805.1551}{{\ttfamily 0805.1551}}].

\bibitem{Gubser:2008sz}
S.~S. Gubser, S.~S. Pufu and F.~D. Rocha, \emph{{Bulk viscosity of strongly
  coupled plasmas with holographic duals}},
  \href{https://doi.org/10.1088/1126-6708/2008/08/085}{\emph{JHEP} {\bfseries
  08} (2008) 085} [\href{https://arxiv.org/abs/0806.0407}{{\ttfamily
  0806.0407}}].

\bibitem{Myers:2008yi}
R.~C. Myers, M.~F. Paulos and A.~Sinha, \emph{{Quantum corrections to eta/s}},
  \href{https://doi.org/10.1103/PhysRevD.79.041901}{\emph{Phys.\ Rev.\ D}
  {\bfseries 79} (2009) 041901}
  [\href{https://arxiv.org/abs/0806.2156}{{\ttfamily 0806.2156}}].

\bibitem{Karch:2008fa}
A.~Karch, D.~Son and A.~Starinets, \emph{{Zero Sound from Holography}},
  \href{https://arxiv.org/abs/0806.3796}{{\ttfamily 0806.3796}}.

\bibitem{Haack:2008cp}
M.~Haack and A.~Yarom, \emph{{Nonlinear viscous hydrodynamics in various
  dimensions using AdS/CFT}},
  \href{https://doi.org/10.1088/1126-6708/2008/10/063}{\emph{JHEP} {\bfseries
  10} (2008) 063} [\href{https://arxiv.org/abs/0806.4602}{{\ttfamily
  0806.4602}}].

\bibitem{Herzog:2008wg}
C.~P. Herzog, M.~Rangamani and S.~F. Ross, \emph{{Heating up Galilean
  holography}},
  \href{https://doi.org/10.1088/1126-6708/2008/11/080}{\emph{JHEP} {\bfseries
  11} (2008) 080} [\href{https://arxiv.org/abs/0807.1099}{{\ttfamily
  0807.1099}}].

\bibitem{Adams:2008wt}
A.~Adams, K.~Balasubramanian and J.~McGreevy, \emph{{Hot Spacetimes for Cold
  Atoms}}, \href{https://doi.org/10.1088/1126-6708/2008/11/059}{\emph{JHEP}
  {\bfseries 11} (2008) 059} [\href{https://arxiv.org/abs/0807.1111}{{\ttfamily
  0807.1111}}].

\bibitem{Buchel:2008ae}
A.~Buchel, R.~C. Myers, M.~F. Paulos and A.~Sinha, \emph{{Universal holographic
  hydrodynamics at finite coupling}},
  \href{https://doi.org/10.1016/j.physletb.2008.10.003}{\emph{Phys.\ Lett.\ B}
  {\bfseries 669} (2008) 364}
  [\href{https://arxiv.org/abs/0808.1837}{{\ttfamily 0808.1837}}].

\bibitem{Iqbal:2008by}
N.~Iqbal and H.~Liu, \emph{{Universality of the hydrodynamic limit in AdS/CFT
  and the membrane paradigm}},
  \href{https://doi.org/10.1103/PhysRevD.79.025023}{\emph{Phys. Rev. D}
  {\bfseries 79} (2009) 025023}
  [\href{https://arxiv.org/abs/0809.3808}{{\ttfamily 0809.3808}}].

\bibitem{Herzog:2008he}
C.~Herzog, P.~Kovtun and D.~Son, \emph{{Holographic model of superfluidity}},
  \href{https://doi.org/10.1103/PhysRevD.79.066002}{\emph{Phys.\ Rev.\ D}
  {\bfseries 79} (2009) 066002}
  [\href{https://arxiv.org/abs/0809.4870}{{\ttfamily 0809.4870}}].

\bibitem{Hartnoll:2008kx}
S.~A. Hartnoll, C.~P. Herzog and G.~T. Horowitz, \emph{{Holographic
  Superconductors}},
  \href{https://doi.org/10.1088/1126-6708/2008/12/015}{\emph{JHEP} {\bfseries
  12} (2008) 015} [\href{https://arxiv.org/abs/0810.1563}{{\ttfamily
  0810.1563}}].

\bibitem{Rangamani:2008gi}
M.~Rangamani, S.~F. Ross, D.~Son and E.~G. Thompson, \emph{{Conformal
  non-relativistic hydrodynamics from gravity}},
  \href{https://doi.org/10.1088/1126-6708/2009/01/075}{\emph{JHEP} {\bfseries
  01} (2009) 075} [\href{https://arxiv.org/abs/0811.2049}{{\ttfamily
  0811.2049}}].

\bibitem{Buchel:2008vz}
A.~Buchel, R.~C. Myers and A.~Sinha, \emph{{Beyond eta/s = 1/4 pi}},
  \href{https://doi.org/10.1088/1126-6708/2009/03/084}{\emph{JHEP} {\bfseries
  03} (2009) 084} [\href{https://arxiv.org/abs/0812.2521}{{\ttfamily
  0812.2521}}].

\bibitem{Cai:2009zv}
R.-G. Cai, Z.-Y. Nie, N.~Ohta and Y.-W. Sun, \emph{{Shear Viscosity from
  Gauss-Bonnet Gravity with a Dilaton Coupling}},
  \href{https://doi.org/10.1103/PhysRevD.79.066004}{\emph{Phys. Rev. D}
  {\bfseries 79} (2009) 066004}
  [\href{https://arxiv.org/abs/0901.1421}{{\ttfamily 0901.1421}}].

\bibitem{Myers:2009ij}
R.~C. Myers, M.~F. Paulos and A.~Sinha, \emph{{Holographic Hydrodynamics with a
  Chemical Potential}},
  \href{https://doi.org/10.1088/1126-6708/2009/06/006}{\emph{JHEP} {\bfseries
  06} (2009) 006} [\href{https://arxiv.org/abs/0903.2834}{{\ttfamily
  0903.2834}}].

\bibitem{Chesler:2009cy}
P.~M. Chesler and L.~G. Yaffe, \emph{{Boost invariant flow, black hole
  formation, and far-from-equilibrium dynamics in N = 4 supersymmetric
  Yang-Mills theory}},
  \href{https://doi.org/10.1103/PhysRevD.82.026006}{\emph{Phys. Rev. D}
  {\bfseries 82} (2010) 026006}
  [\href{https://arxiv.org/abs/0906.4426}{{\ttfamily 0906.4426}}].

\bibitem{deBoer:2009pn}
J.~de~Boer, M.~Kulaxizi and A.~Parnachev, \emph{{AdS(7)/CFT(6), Gauss-Bonnet
  Gravity, and Viscosity Bound}},
  \href{https://doi.org/10.1007/JHEP03(2010)087}{\emph{JHEP} {\bfseries 03}
  (2010) 087} [\href{https://arxiv.org/abs/0910.5347}{{\ttfamily 0910.5347}}].

\bibitem{Goldstein:2009cv}
K.~Goldstein, S.~Kachru, S.~Prakash and S.~P. Trivedi, \emph{{Holography of
  Charged Dilaton Black Holes}},
  \href{https://doi.org/10.1007/JHEP08(2010)078}{\emph{JHEP} {\bfseries 08}
  (2010) 078} [\href{https://arxiv.org/abs/0911.3586}{{\ttfamily 0911.3586}}].

\bibitem{Rebhan:2011vd}
A.~Rebhan and D.~Steineder, \emph{{Violation of the Holographic Viscosity Bound
  in a Strongly Coupled Anisotropic Plasma}},
  \href{https://doi.org/10.1103/PhysRevLett.108.021601}{\emph{Phys. Rev. Lett.}
  {\bfseries 108} (2012) 021601}
  [\href{https://arxiv.org/abs/1110.6825}{{\ttfamily 1110.6825}}].

\bibitem{Bredberg:2011jq}
I.~Bredberg, C.~Keeler, V.~Lysov and A.~Strominger, \emph{{From Navier-Stokes
  To Einstein}}, \href{https://doi.org/10.1007/JHEP07(2012)146}{\emph{JHEP}
  {\bfseries 07} (2012) 146} [\href{https://arxiv.org/abs/1101.2451}{{\ttfamily
  1101.2451}}].

\bibitem{Davison:2011ek}
R.~A. Davison and A.~O. Starinets, \emph{{Holographic zero sound at finite
  temperature}}, \href{https://doi.org/10.1103/PhysRevD.85.026004}{\emph{Phys.
  Rev. D} {\bfseries 85} (2012) 026004}
  [\href{https://arxiv.org/abs/1109.6343}{{\ttfamily 1109.6343}}].

\bibitem{Donos:2014cya}
A.~Donos and J.~P. Gauntlett, \emph{{Thermoelectric DC conductivities from
  black hole horizons}},
  \href{https://doi.org/10.1007/JHEP11(2014)081}{\emph{JHEP} {\bfseries 11}
  (2014) 081} [\href{https://arxiv.org/abs/1406.4742}{{\ttfamily 1406.4742}}].

\bibitem{Henningson:1998gx}
M.~Henningson and K.~Skenderis, \emph{{The Holographic Weyl anomaly}},
  \href{https://doi.org/10.1088/1126-6708/1998/07/023}{\emph{JHEP} {\bfseries
  07} (1998) 023} [\href{https://arxiv.org/abs/hep-th/9806087}{{\ttfamily
  hep-th/9806087}}].

\bibitem{Balasubramanian:1999re}
V.~Balasubramanian and P.~Kraus, \emph{{A Stress tensor for Anti-de Sitter
  gravity}}, \href{https://doi.org/10.1007/s002200050764}{\emph{Commun.\ Math.\
  Phys.} {\bfseries 208} (1999) 413}
  [\href{https://arxiv.org/abs/hep-th/9902121}{{\ttfamily hep-th/9902121}}].

\bibitem{Kraus:1999di}
P.~Kraus, F.~Larsen and R.~Siebelink, \emph{{The gravitational action in
  asymptotically AdS and flat space-times}},
  \href{https://doi.org/10.1016/S0550-3213(99)00549-0}{\emph{Nucl.\ Phys.\ B}
  {\bfseries 563} (1999) 259}
  [\href{https://arxiv.org/abs/hep-th/9906127}{{\ttfamily hep-th/9906127}}].

\bibitem{Skenderis:2002wp}
K.~Skenderis, \emph{{Lecture notes on holographic renormalization}},
  \href{https://doi.org/10.1088/0264-9381/19/22/306}{\emph{Class.\ Quant.\
  Grav.} {\bfseries 19} (2002) 5849}
  [\href{https://arxiv.org/abs/hep-th/0209067}{{\ttfamily hep-th/0209067}}].

\bibitem{LANDAU}
L.~D. Landau and E.~M. Lifshitz, \emph{Fluid Mechanics (Second Edition): Volume
  6 of Course of Theoretical Physics}. Pergamon, 1987.

\bibitem{AShandbook}
M.~Abramowitz and I.~A. Stegun, eds., \emph{Handbook of Mathematical Functions,
  With Formulas, Graphs, and Mathematical Tables,}. United States Department of
  Commerce, National Bureau of Standards, USA, 1972.

\bibitem{Ferrara:1995ih}
S.~Ferrara, R.~Kallosh and A.~Strominger, \emph{{N=2 extremal black holes}},
  \href{https://doi.org/10.1103/PhysRevD.52.R5412}{\emph{Phys. Rev. D}
  {\bfseries 52} (1995) 5412}
  [\href{https://arxiv.org/abs/hep-th/9508072}{{\ttfamily hep-th/9508072}}].

\bibitem{Ferrara:1996dd}
S.~Ferrara and R.~Kallosh, \emph{{Supersymmetry and attractors}},
  \href{https://doi.org/10.1103/PhysRevD.54.1514}{\emph{Phys. Rev. D}
  {\bfseries 54} (1996) 1514}
  [\href{https://arxiv.org/abs/hep-th/9602136}{{\ttfamily hep-th/9602136}}].

\bibitem{Sen:2005wa}
A.~Sen, \emph{{Black hole entropy function and the attractor mechanism in
  higher derivative gravity}},
  \href{https://doi.org/10.1088/1126-6708/2005/09/038}{\emph{JHEP} {\bfseries
  09} (2005) 038} [\href{https://arxiv.org/abs/hep-th/0506177}{{\ttfamily
  hep-th/0506177}}].

\bibitem{Goldstein:2005hq}
K.~Goldstein, N.~Iizuka, R.~P. Jena and S.~P. Trivedi,
  \emph{{Non-supersymmetric attractors}},
  \href{https://doi.org/10.1103/PhysRevD.72.124021}{\emph{Phys. Rev. D}
  {\bfseries 72} (2005) 124021}
  [\href{https://arxiv.org/abs/hep-th/0507096}{{\ttfamily hep-th/0507096}}].

\bibitem{Astefanesei:2006dd}
D.~Astefanesei, K.~Goldstein, R.~P. Jena, A.~Sen and S.~P. Trivedi,
  \emph{{Rotating attractors}},
  \href{https://doi.org/10.1088/1126-6708/2006/10/058}{\emph{JHEP} {\bfseries
  10} (2006) 058} [\href{https://arxiv.org/abs/hep-th/0606244}{{\ttfamily
  hep-th/0606244}}].

\bibitem{Andrianopoli:2006ub}
L.~Andrianopoli, R.~D'Auria, S.~Ferrara and M.~Trigiante, \emph{{Extremal black
  holes in supergravity}}.
\newblock 2008.
\newblock \href{https://arxiv.org/abs/hep-th/0611345}{{\ttfamily
  hep-th/0611345}}.

\bibitem{Larsen:2006xm}
F.~Larsen, \emph{{The Attractor Mechanism in Five Dimensions}}, {\emph{Lect.
  Notes Phys.} {\bfseries 755} (2008) 249}
  [\href{https://arxiv.org/abs/hep-th/0608191}{{\ttfamily hep-th/0608191}}].

\bibitem{Sen:2007qy}
A.~Sen, \emph{{Black Hole Entropy Function, Attractors and Precision Counting
  of Microstates}}, \href{https://doi.org/10.1007/s10714-008-0626-4}{\emph{Gen.
  Rel. Grav.} {\bfseries 40} (2008) 2249}
  [\href{https://arxiv.org/abs/0708.1270}{{\ttfamily 0708.1270}}].

\bibitem{hawking_ellis_1973}
S.~W. Hawking and G.~F.~R. Ellis, \emph{The Large Scale Structure of
  Space-Time}, Cambridge Monographs on Mathematical Physics. Cambridge
  University Press, 1973,
  \href{https://doi.org/10.1017/CBO9780511524646}{10.1017/CBO9780511524646}.

\bibitem{Wald:1984rg}
R.~M. Wald, \emph{{General Relativity}}. Chicago Univ. Pr., Chicago, USA, 1984,
  \href{https://doi.org/10.7208/chicago/9780226870373.001.0001}{10.7208/chicago/9780226870373.001.0001}.

\bibitem{MMST}
S.~Minwalla, U.~Moitra, S.~K. Sake, S.~P. Trivedi et~al. {\it In preparation}.

\bibitem{Brito:2015oca}
R.~Brito, V.~Cardoso and P.~Pani, \emph{{Superradiance}: {Energy Extraction,
  Black-Hole Bombs and Implications for Astrophysics and Particle Physics}},
  vol.~906. Springer, 2015,
  \href{https://doi.org/10.1007/978-3-319-19000-6}{10.1007/978-3-319-19000-6},
  [\href{https://arxiv.org/abs/1501.06570}{{\ttfamily 1501.06570}}].

\end{thebibliography}\endgroup
\end{document}